\documentclass[showpacs, amsmath,amsfonts,amssymb,aps]{revtex4}

\usepackage{tabularx}
\usepackage[counter=equation]{glossaries}
\usepackage{enumitem}
\usepackage{graphicx}
\usepackage{float}
\usepackage{titlesec}
\usepackage{dcolumn}
\usepackage{bm}
\usepackage{mathrsfs}
\usepackage{amsmath}

\usepackage{amsfonts} 

\DeclareMathOperator\arctanh{arctanh}

\numberwithin{equation}{section}

\begin{document}

\title{New vistas on the Laplace-Runge-Lenz vector}
\author{Davide Batic}
\email{davide.batic@ku.ac.ae}
\affiliation{
Department of Mathematics,\\  Khalifa University of Science and Technology,\\ Main Campus, Abu Dhabi,\\ United Arab Emirates}
\author{M. Nowakowski}
\email{mnowakos@uniandes.edu.co}
\affiliation{
Departamento de Fisica,\\ Universidad de los Andes, Cra.1E
No.18A-10, Bogota, Colombia
}
\author{Aya Mohammad Abdelhaq}
\email{1000538296@ku.ac.ae}
\affiliation{
Department of Physics,\\  Khalifa University of Science and Technology,\\ Main Campus, Abu Dhabi,\\ United Arab Emirates}

\date{\today}

\begin{abstract}
Scalar, vector and tensor conserved quantities are essential tools
in solving different problems in physics and complex, nonlinear differential 
equations in mathematics. In many guises they enter our understanding of nature:
charge, lepton, baryon numbers conservation accompanied with constant energy,
linear or angular total momenta and the conservation of energy-momentum/angular momentum tensors in field theories due to Noether theorem which is based on the 
translational and Lorentz symmetry of the Lagrangians. One of the oldest discovered conserved
quantities is the Laplace-Runge-Lenz vector for the $1/r$-potential. Its different aspects have been discussed
many times in the literature. But explicit generalisations to other spherically symmetric potentials are still rare.
Here, we attempt to fill this gap by constructing explicit examples of a conserved vector perpendicular to the angular momentum
for a class of phenomenologically relevant potentials. Hereby, we maintain the nomenclature and keep calling these constant vectors
Laplace-Runge-Lenz vectors.    
\end{abstract}
\pacs{xyz}
\maketitle

\tableofcontents
\section{Introduction}
The progress of science is seldom a smooth linear process. Unsolved
problems and incomplete solutions are often left aside to proceed to
new directions, areas, problems and pastures motivated either by
experiments/observations, discoveries or deeper theoretical
insight. Hereby, it is almost guaranteed that scientists equipped with
better technology and/or mathematical formalism will return to the old
questions, often re-discover them independently going one or two steps
ahead. The older the area of physics, the more examples can be found
to exemplify this aspect of the history of science. We mention here
the famous three-body-problem \cite{three} and the Laplace-Runge-Lenz (LRL)
vector \cite{Goldstein, Arnold, Landau}. Since this conserved vector is the subject of the
present article, it is somewhat instructive to have a glimpse at its
history spread over five centuries \cite{Gold2, Gold3}. It appears
that as soon as classical mechanics was put into a mathematical
framework, the conservation of the magnitude of the LRL-vector was
discovered by Jakob Hermann (1678-1732) \cite{Herrmann} who was a
contemporary of Newton (1643-1727). Bernoulli (1667-1748) gave it its
modern appearance making it the first conserved vector in classical mechanics
\cite{Bernoulli}. Subsequent, independent re-discoveries were made by
two famous mathematicians: Pierre Simon Laplace (1749-1827)
\cite{Laplace} and
William Rowan Hamilton (1805-1865) \cite{Hamilton}. In the year 1901, 
the LRL-vector has
been discussed in a book on vector analysis by Gibbs and Wilson
\cite{Gibbs} and in 1919 by Runge (1856-1927) also in a book on
similar topics in German \cite{Runge}. Finally, Wilhelm Lenz (1888-1957)
mentioned it in his treatise on the hydrogen atom \cite{Lenz}. The
quantum mechanical aspect of the LRL-vector was picked up by Wolfgang Pauli (1900-1958)
to obtain the full spectrum of the hydrogen atom relying on purely
algebraic method \cite{Pauli}. This has found its entry in several textbooks on
quantum mechanics \cite{Bohm, Weinberg, Shankar}. In view of the many
contributors, it does not come as
a surprise that \cite{Sub} suggests a new name for the
vector, namely the  Herrmann-Bernoulli-Laplace-Hamilton-Runge-Lenz vector.
To appreciate its role in the hydrogen atom we could add the name of
Pauli making it the Herrmann-Bernoulli-Laplace-Hamilton-Runge-Lenz-Pauli vector.

This historical
overview outlined above reveals some aspects of the working of science that are 
worth to be mentioned. One of them is that knowledge can get lost albeit
published and
needs therefore an efficient catalogization made nowadays possible by internet
and search engines. Secondly, in spite of the many returns, the
explicit construction was always limited to the
$1/r$-potential. Indeed, the many facets of this potential have been
discussed at length equally by physicists \cite{Itzy} and by
mathematicians \cite{Morse}.  The second half of the twentieth century
brought some advancement  in connecting the conserved vector to a
symmetry of a spherically symmetric potential \cite{Fradkin}. Hence, it
became obvious that other potentials $V(r)$ should also have  a
corresponding LRL-vector. Some general ansatz has been made by 
\cite{Peres, Yoshida} where the LRL-vector for the three dimensional harmonic oscillator with $V(r) \propto r^2$ were 
found. But the ansatz remained in general unexploited.  A conserved
scalar quantity has been found in the problem of the $1/r$-potential in
the presence of a constant magnetic field \cite{Redmond} while a
conserved vector has been derived for a class of time-dependent Kepler
potentials \cite{Ritter}. For another class of forces including the
drag forces it has been constructed by \cite{Gorringe}. This is summarized in the review article \cite{Leach} where the authors study and construct the equivalent to the
LRL-vector for problems described by an equation of motion of the form
$\ddot{\mathbf{r}}+f\mathbf{i}+g\mathbf{j}=0$ with $f$ and $g$
functions and $(\mathbf{i}, \mathbf{j})$ different vectors, e.g.,
$(\hat{\mathbf{r}}, \hat{\mathbf{\theta}})$,
$(\mathbf{L}, \hat{\mathbf{r}})$, $(\dot{\mathbf{r}}, \mathbf{r})$
etc.  Evidently these conserved vectors do also exist for
non-spherically symmetric forces. Surprising, especially from the point of
view of a symmetry underlying a conserved quantity,  
is also the existence of a conserved quantity in cases with friction 
where not even the energy is conserved.

In this article, we restrict ourselves mostly to central potentials $V(r)$
for which the motion takes place in a plane perpendicular to the angular
momentum $\mathbf{L}$. With a general ansatz for a vector in this
plane
we demand its conservation under the condition of an equation of
motion
$\dot{\mathbf{r}}^2 +V_{eff}(r)=E=const$. A set of differential
equations is derived and solved explicitly for specific cases. The
Kepler problem and the isotropic harmonic potential in three dimensions
served as a check of the correctness of the formalism. We proceed to
construct the LRL-vector for cases that are relevant to gravitation: a
slightly deformed 
spheroid, Kepler plus the isotropic harmonic potential and the Coulomb
potential with the Cornell confining one. We demonstrate in an
appendix that combining the Newtonian potential with the harmonic
one allows to cover several interesting cases. Among them not only the  non-relativistic
de Sitter or any de Sitter scenario in General Relativity \cite{mine1} but also the potential of Mercury emerging from an approximation of the contributions due to the Sun and the other planets.
The calculations are extensive as they use different special
functions. With the help of Maple, we used these LRL-vectors to plot
the corresponding particle trajectories relying on the same method used in the Kepler problem with the corresponding LRL-vector. 

{\bf{A technical and at the same time non-technical (when used verbally) argument for the existence of a conserved  LRL-vector is usually sought in the symmetry of the potential. For instance, every central potential has the symmetry of the rotation group $SO(3)$. Indeed, in \cite{Fradkin} the author establishes a theorem which ensures the existence of a conserved vector for central potential problems. As expected, the proof relies on symmetry arguments. This, however, works only in one direction, i.e., from a symmetry to a conservation law. As can be seen from the examples discussed in \cite{Leach}, a symmetry is not a necessary ingredient for the system to posses a conserved vector. Indeed, one would not expect to encounter a symmetry in mechanical problems with friction. Yet, a sub-class of frictional problems in classical mechanics has a conserved vector as explicitly demonstrated in Section G. Such state of affair does not diminish the power of arguments relying on symmetry, but, on the other hand, shows that we must be ready for caveats considering conservation laws.}}

\section{Prolegomena}
For central force problems where the force acting upon a particle with mass $m$ is given by
\begin{equation}
{\bf{F}}=\frac{f(r)}{r}{\bf{r}}=f(r)\widehat{{\bf{r}}},
\end{equation}
the equation of motion controlling the particle dynamics is represented by Newton' second law
\begin{equation}\label{NL}
m\ddot{{\bf{r}}}=f(r)\widehat{\bf{r}}.
\end{equation}
Here, the double dot denotes the second derivative with respect to time while $\widehat{{\bf{r}}}={\bf{r}}/r$ is the usual unit vector in the radial direction. It is known that due to homogeneity in time as seen from the point of view of a Lagrangian, the total energy 
\begin{equation}\label{EC0}
E=\frac{1}{2}m\dot{{\bf{r}}}^2+U(r)
\end{equation}
is conserved. In addition, if the force field is conservative, the potential energy $U(r)$  is linked to the force acting on the particle by the relation
\begin{equation}
{\bf{F}}=-\frac{dU}{dr}\widehat{{\bf{r}}}.
\end{equation}
We recall that in the Kepler problem the potential energy is
\begin{equation}\label{KP}
    U(r)=-G_N\frac{Mm}{r},
\end{equation}
where $G_N$ is Newton's gravitational constant and $M$ is the mass of the central gravitational object. For later convenience, we rescale the conservation equation by $m$ as in \cite{Arnold}. To this purpose, let $\mathcal{E}=E/m$ and $V(r)=U(r)/m$ be the corresponding potential, then (\ref{EC}) becomes
\begin{equation}\label{EC}
\mathcal{E}=\frac{\dot{{\bf{r}}}^2}{2}+V(r)
\end{equation}
with $V(r)=-k_N/r$ and $k_N=G_N M$ in the case of the Kepler problem. It is well-known that due to isotropy in space, the angular momentum 
\begin{equation}\label{L}
{\bf{L}}={\bf{r}}\times{\bf{p}},\quad{\bf{p}}=m\dot{{\bf{r}}}
\end{equation}
is conserved. Hence, by differentiating (\ref{L}) with respect to the time variable, it is straightforward to verify that
\begin{equation}
\frac{d{\bf{L}}}{dt}={\bf{0}}.
\end{equation}
Note that the conservation of angular momentum implies that the motion of the test particle takes place on a plane. This is due to the fact that ${\bf{L}}$ is conserved together with the fact that ${\bf{L}}\cdot{\bf{r}}=0$. Hence, the position of the particle is on a plane orthogonal to ${\bf{L}}$. Similarly we have that ${\bf{L}}\cdot\dot{{\bf{r}}}=0$ signalizing that the velocity of the particle belongs to the same plane as well. To show that 
\begin{equation}
\frac{d\mathcal{E}}{dt}=0,
\end{equation}
we recall that in polar coordinates
\begin{equation}\label{rpol}
\dot{{\bf{r}}}=\dot{r}\widehat{{\bf{r}}}+r\dot{\varphi}\widehat{{\bm{\varphi}}}.
\end{equation}
Moreover, $\dot{\widehat{{\bf{r}}}}=\dot{\varphi}\widehat{{\bm{\varphi}}}$ and $\dot{\widehat{{\bm{\varphi}}}}=-\dot{\varphi}\widehat{{\bf{r}}}$. Hence, the left hand side of (\ref{NL}) can be rewritten as 
\begin{equation}\label{exprM}
m\ddot{{\bf{r}}}=m(\ddot{r}-r\dot{\varphi}^2)\widehat{{\bf{r}}}+m(r\ddot{\varphi}+2\dot{r}\dot{\varphi})\widehat{{\bm{\varphi}}}
\end{equation}
and for a conservative force, we obtain the equation of motion
\begin{equation}\label{moto}
\ddot{r}-r\dot{\varphi}^2=\frac{f(r)}{m}=-\frac{dV}{dr},\quad
r\ddot{\varphi}+2\dot{r}\dot{\varphi}=0.
\end{equation}
By means of the second equation above it can be verified that
\begin{equation}\label{aiutino}
\frac{1}{2}\frac{d}{dt}\left(\dot{{\bf{r}}}\cdot\dot{{\bf{r}}}\right)=\dot{r}\ddot{r}-r\dot{r}\dot{\varphi}^2.
\end{equation}
Multiplying the first equation in (\ref{moto}) by $\dot{r}$ and using (\ref{aiutino}) shows that 
\begin{equation}
\frac{d}{dt}\left[\frac{\dot{{\bf{r}}}^2}{2}+V(r)\right]=0
\end{equation}
from which the energy conservation follows. Essentially the homogeneity of time and isotropy of space are symmetries which lead to conservation laws \cite{Landau}. Moreover, it can be also shown that any central potential in a classical dynamic problem admits also an $O(4)$ and $SU(3)$ symmetry related to other conserved quantities \cite{Fradkin}. Such a quantity in case of the Kepler problem is known as Laplace-Runge-Lenz vector. We refer to \cite{Goldstein} for history and the choice of name. Let us introduce the vector ${\bf{r}}_{\bot}=r\dot{\varphi}\widehat{{\bm{\varphi}}}$ which can be rewritten in terms of (\ref{rpol}) as follows
\begin{equation}\label{rbot}
{\bf{r}}_{\bot}= \dot{{\bf{r}}}-\frac{\dot{r}}{r}{{\bf{r}}}.
\end{equation}
The benefit of using the above representation for ${\bf{r}}_{\bot}$ will become apparent in Section~\ref{GenAn}. Here, it suffices to observe that by construction ${\bf{r}}\cdot{\bf{r}}_{\bot}=0$. This signalizes that ${\bf{r}}_{\bot}$ represents a vector which is orthogonal to the radial direction. Moreover, it is straightforward to verify that the velocity vector and its magnitude admit the representations
\begin{equation}\label{rap}
\dot{\bf{r}}={\bf{r}}_{\bot}+\frac{\dot{r}}{r}{{\bf{r}}},\quad
\dot{{\bf{r}}}^2={\bf{r}}_{\bot}^2+\dot{r}^2,
\end{equation}
while replacing the first equation in (\ref{rap}) into the expression for the angular momentum leads to
\begin{equation}
{\bf{L}}=m{\bf{r}}\times{\bf{r}}_{\bot}.
\end{equation}
Note that using the orthogonality of the vectors ${\bf{r}}$ and ${\bf{r}}_{\bot}$ the magnitude of ${\bf{L}}$ is
\begin{equation}
L= mrr_\bot.
\end{equation}
Solving the above equation for $r_\bot$ yields
\begin{equation}\label{15}
r_\bot =\frac{\ell}{r},\quad\ell=\frac{L}{m}.
\end{equation}
which allows to rewrite the second equation in (\ref{rap}) as
\begin{equation}\label{idI}
\dot{{\bf{r}}}^2=\frac{\ell^2}{r^2}+\dot{r}^2.
\end{equation}
As a consequence the energy conservation equation (\ref{EC}) takes the form
\begin{equation}\label{CE2}
\mathcal{E}=\frac{\dot{r}^2}{2}+V_{eff}(r),\quad
V_{eff}(r)=\frac{\ell^2}{2r^2}+V(r),
\end{equation}
where $V_{eff}(r)$ is the so-called effective potential. Hence, a three dimensional problem for a central potential $V$ has been transformed into an equivalent one dimensional problem where $V$ is replaced by $V_{eff}$. We end this section by shortly review the standard derivation of the LRL-vector in the case of the Kepler problem. First of all, we observe that by means of Newton' second law we can rewrite the rate of change of the momentum as
\begin{equation}
\dot{{\bf{p}}}=\frac{f(r)}{r}{\bf{r}}.
\end{equation}
Moreover, taking into account the conservation of the angular momentum, the rate of change of the vector ${\bf{p}}\times{\bf{L}}$ is computed to be
\begin{equation}
\frac{d}{dt}\left({\bf{p}}\times{\bf{L}}\right)=m\frac{f(r)}{r}{\bf{r}}\times\left({\bf{r}}\times\dot{{\bf{r}}}\right)=m\frac{f(r)}{r}\left[\left({\bf{r}}\cdot\dot{{\bf{r}}}\right){\bf{r}}-r^2\dot{{\bf{r}}}\right]=m\frac{f(r)}{r}\left[r\dot{r}{\bf{r}}-r^2\dot{{\bf{r}}}\right], 
\end{equation}
where in the last step we used the identity ${\bf{r}}\cdot\dot{{\bf{r}}}=r\dot{r}$. The above expression can be further manipulated until we end up with 
\begin{equation}
\frac{d}{dt}\left({\bf{p}}\times{\bf{L}}\right)=-m r^2f(r)\left(\frac{\dot{{\bf{r}}}}{r}-\frac{\dot{r}}{r^2}{\bf{r}}\right)=-m r^2f(r)\frac{d}{dt}\left(\frac{{\bf{r}}}{r}\right).
\end{equation}
Finally, for the Kepler potential energy (\ref{KP}) we have $f(r)=-dU/dr=-G_N Mm/r^2$ and the above expression becomes
\begin{equation}\label{pp}
\frac{d}{dt}\left({\bf{p}}\times{\bf{L}}\right)=G_N Mm^2\frac{d}{dt}\left(\frac{{\bf{r}}}{r}\right).
\end{equation}
It follows from (\ref{pp}) that the LRL-vector ${\bf{A}}$ is conserved because 
\begin{equation}\label{Ap}
\frac{d{\bf{A}}}{dt}=\frac{d}{dt}\left({\bf{p}}\times{\bf{L}}-G_N Mm^2\frac{{\bf{r}}}{r}\right)=0
\end{equation}
At this point a comment is in order. First of all, the expression in the bracket has SI units J$\cdot$Kg$\cdot$m$=$Kg$^2\cdot$m$^3/$s$^2$ as one would expect for the LRL-vector as defined in \cite{Goldstein}. However, in the next section it will be more convenient to work with its rescaled version. To this purpose, let $\bm{\pi}={\bf{p}}/m$ and $\bm{\ell}={\bf{L}}/m$. Then, we define the LRL vector as
\begin{equation}\label{ourA}
{\bm{\mathcal{A}}}=\frac{{\bf{A}}}{m^2}=\bm{\pi}\times\bm{\ell}-\frac{k_N}{r}{\bf{r}}
\end{equation}
with $k_N=G_N M$. Note that $\bm{\mathcal{A}}$ is perpendicular to ${\bm{\ell}}$ since both vectors ${\bm{\pi}}\times{\bm{\ell}}$ and ${\bf{r}}$ are orthogonal to ${\bm{\ell}}$. This observation signalizes that ${\bm{\mathcal{A}}}$ belongs to the plane of motion and as a consequence, the particle orbit $r(\vartheta)$ is simple to calculate. To this purpose, we observe that combining the dot product definition with (\ref{ourA}) yields
\begin{equation}
{\bm{\mathcal{A}}}\cdot{\bf{r}}=\mathcal{A}r\cos{\varphi}={\bf{r}}\cdot({\bm{\pi}}\times{\bm{\ell}})-k_N r.
\end{equation}
Taking into account that ${\bf{r}}\cdot({\bm{\pi}}\times{\bm{\ell}})=({\bf{r}}\times{\bm{\pi}})\cdot{\bm{\ell}}=\ell^2$ and solving the above expression for $r$, we end up with  
\begin{equation}\label{orbit}
\frac{1}{r(\varphi)}=\frac{k_N}{\ell^2}\left(1+\frac{\mathcal{A}}{k_N}\cos{\varphi}\right),
\end{equation}
which agrees with the formula for a conic section with eccentricity
\begin{equation}\label{ecc}
e=\frac{\mathcal{A}}{k_N}=\sqrt{1+\frac{2\mathcal{E}\ell^2}{k_N^2}}.
\end{equation}
Formula (\ref{ecc}) allows also to express the magnitude of the LRL-vector as
\begin{equation}
\mathcal{A}^2=k_N^2+2\mathcal{E}\ell^2
\end{equation}
and to rewrite the trajectory (\ref{orbit}) as
\begin{equation}
\frac{1}{r(\varphi)}=\frac{k_N}{\ell^2}\left[1+\sqrt{1+\frac{2\mathcal{E}\ell^2}{k^2_N}}\cos{\varphi}\right].
\end{equation}
One would like to copy such steps for other central potentials also. Although the existence of ${\bm{\mathcal{A}}}$ for other central potentials is guaranteed, its functional form is often not known.

\section{A different point of view on the LRL-vector for the Kepler problem}
We first express the vector ${\bm{\pi}}\times{\bm{\ell}}$ in terms of the vector ${\bf{r}}_\bot$. To this purpose, we observe that by means of the identity ${\bf{r}}\cdot\dot{{\bf{r}}}=r\dot{r}$ and (\ref{rbot}) we find that
\begin{equation}
{\bm{\pi}}\times{\bm{\ell}}=\dot{{\bf{r}}}\times({\bf{r}}\times\dot{{\bf{r}}})=\dot{{\bf{r}}}^2{\bf{r}}-r\dot{r}\dot{{\bf{r}}}=\left[(\dot{{\bf{r}}})^2-\dot{r}^2\right]{\bf{r}}-r\dot{r}{\bf{r}}_\bot.
\end{equation}
Moreover, using (\ref{idI}) yields
\begin{equation}
{\bm{\pi}}\times{\bm{\ell}}=\frac{\ell^2}{r^2}{\bf{r}}-r\dot{r}{\bf{r}}_\bot.
\end{equation}
If we replace the above expression in the formula (\ref{ourA}) for the LRL-vector and reintroduce there the potential $V(r)=-k_N/r$ together with (\ref{CE2}), we find that
\begin{equation}\label{Am}
\bm{\mathcal{A}}=\left[\frac{\ell^2}{2r^2}+V_{eff}(r)\right]{\bf{r}}-r\dot{r}{\bf{r}}_\bot.
\end{equation}
Differentiating the above expression with respect to time leads to
\begin{equation}\label{square}
\frac{d\bm{\mathcal{A}}}{dt}=\left[\frac{\ell^2}{2r^2}+V_{eff}(r)\right]\dot{{\bf{r}}}+\dot{r}\frac{d}{dr}\left[\frac{\ell^2}{2r^2}+V_{eff}(r)\right]{\bf{r}}-\dot{r}^2{\bf{r}}_\bot-r\ddot{r}{\bf{r}}_\bot-r\dot{r}\dot{{\bf{r}}}_\bot.
\end{equation}
In order to further simplify the above expression, we observe that solving the energy conservation equation (\ref{CE2}) with respect to $\dot{r}$ gives
\begin{equation}
\dot{r}=\pm\sqrt{2\left[\mathcal{E}-V_{eff}(r)\right]}
\end{equation}
from which it can be easily verified that
\begin{equation}\label{2star}
\ddot{r}=-\frac{dV_{eff}}{dr}.
\end{equation}
If we differentiate the first equation in (\ref{rap}) with respect to the time variable and use (\ref{moto}) together with the first relation in (\ref{rap}), we find that
\begin{equation}
\dot{{\bf{r}}}_\bot=\left[\frac{f(r)}{m}-\ddot{r}\right]\frac{{\bf{r}}}{r}-\frac{\dot{r}}{r}{\bf{r}}_\bot.
\end{equation}
At this point, a straightforward application of (\ref{2star}) gives that
\begin{equation}\label{ascia}
\dot{{\bf{r}}}_\bot=-\frac{\dot{r}}{r}{\bf{r}}_\bot-\frac{\ell^2}{r^4}{\bf{r}}.
\end{equation}
Finally, if we insert the first equation in (\ref{rap}), (\ref{2star}) and (\ref{ascia}) into (\ref{square}), we find that (\ref{square}) can be expressed as a linear combination with respect to the vectors ${\bf{r}}$ and ${\bf{r}}_\bot$, namely
\begin{equation}\label{37}
\frac{d\bm{\mathcal{A}}}{dt}=\left[\frac{dV_{eff}}{dr}+\frac{V_{eff}(r)}{r}+\frac{\ell^2}{2r^3}\right]\left(\dot{r}{\bf{r}}+r{\bf{r}}_\bot\right).
\end{equation}
At this point a comment is in order. Since the conservation of the LRL-vector ${\bm{\mathcal{A}}}$ ensures that $d\bm{\mathcal{A}}/dt=0$ and moreover, the vectors ${\bf{r}}$ and ${\bf{r}}_\bot$ are perpendicular to each other thus spanning the plane in which the trajectory of the particle takes place, we immediately conclude that (\ref{37}) can also be used to derive an equation for the potential. It is interesting to observe that the effective Kepler potential is indeed a solution of the differential equation
\begin{equation}\label{eq1}
    \frac{dV_{eff}}{dr}+\frac{V_{eff}(r)}{r}=-\frac{\ell^2}{2r^3}.
\end{equation}
If we use (\ref{CE2}) to rewrite the above equation in terms of the Kepler potential $V(r)$, it is not difficult to check that such a potential satisfies the equation
\begin{equation}
    \frac{dV}{dr}+\frac{V(r)}{r}=0,
\end{equation}
beside also being a solution of the Laplace equation
\begin{equation}\label{Laplace}
    \Delta V=\frac{d^2 V}{dr^2}+\frac{2}{r}\frac{dV}{dr}=0.
\end{equation}
Moreover, if we first rewrite (\ref{Am}) by means of (\ref{eq1}) as follows
\begin{equation}\label{finale}
{\bm{\mathcal{A}}}=\left(\frac{\ell^2}{r^2}-r\frac{dV}{dr}\right){\bf{r}}-r\dot{r}{\bf{r}}_\bot
\end{equation}
and then, we repeat the same steps from before, i.e. we differentiate ${\bm{\mathcal{A}}}$ with respect to time, we will find that it is conserved, whenever $V(r)$ satisfies the Laplace equation. Hence, we found that the conservation of a LRL-vector can be encoded in a differential equation for the effective potential. In case of the Kepler problem, this is the Laplace equation. A legitimate question is then, if we can generalise the above result for other classes of central potentials. If it turns to be possible, such a procedure should give us the corresponding LRL-vector for a certain family of potentials.

\section{General ansatz for a general LRL-vector}\label{GenAn}
Inspired by equation (\ref{37}), we demand that
\begin{equation}
\bm{\mathcal{A}}=S(r)\frac{{\bf{r}}}{r}+ P(r)\frac{{\bf{r}}_\bot}{r_\bot}
\end{equation}
and we impose the conservation condition
\begin{equation}\label{ansatz}
\frac{d\bm{\mathcal{A}}}{dt}=0, 
\end{equation}
where $S(r)$ and $P(r)$ are two unknown functions that depend implicitly on the time variable through $r$. Furthermore, we suppose they are both at least once continuously differentiable with respect to time. If we compute the above derivative, we get
\begin{equation}\label{cucu}
\frac{d\bm{\mathcal{A}}}{dt}=
\frac{1}{r}\left(\dot{{\bf{r}}}-\frac{\dot{r}}{r}{\bf{r}}\right)S(r)+\dot{r}\frac{dS}{dr}\frac{{\bf{r}}}{r}+\frac{1}{r_\bot}\left(\dot{{\bf{r}}}_\bot-\frac{\dot{r}_\bot}{r_\bot}{\bf{r}}_\bot\right)P(r)+\dot{r}_\bot\frac{dP}{dr}\frac{{\bf{r}}_\bot}{r_\bot}.
\end{equation}
The above expression can be further simplified if we apply the first relation in (\ref{rap}) and equation (\ref{ascia}) rewritten in the following equivalent form by means of (\ref{15}), namely
\begin{equation}
\dot{{\bf{r}}}_\bot=-r^2_\bot\left(\frac{{\bf{r}}}{r^2}+\frac{\dot{r}}{r}\frac{{\bf{r}}_\bot}{r^2_\bot}\right).
\end{equation}
Then, (\ref{cucu}) becomes
\begin{equation}\label{haha}
\frac{d\bm{\mathcal{A}}}{dt}=\frac{{\bf{r}}_\bot}{r}S(r)+\dot{r}\frac{dS}{dr}\frac{{\bf{r}}}{r}-r_\bot\left(\frac{{\bf{r}}}{r^2}+\frac{\dot{r}}{r}\frac{{\bf{r}}_\bot}{r^2_\bot}\right)P(r)-\frac{\dot{r}_\bot}{r_\bot}\frac{{\bf{r}}_\bot}{r_\bot}P(r)+\frac{\dot{r}}{r_\bot}\frac{dP}{dr}{\bf{r}}_\bot.
\end{equation}
If we differentiate (\ref{15}) with respect to $t$, we find that
\begin{equation}\label{pic}
\dot{r}_\bot=-\frac{\ell}{r^2}\dot{r}=-r_\bot\frac{\dot{r}}{r},
\end{equation}
where in the last step we used again (\ref{15}). Finally, substituting (\ref{pic}) into (\ref{haha}) yields
\begin{equation}
\frac{d\bm{\mathcal{A}}}{dt}=\left[\frac{\dot{r}}{r}\frac{dS}{dr}-\frac{r_\bot}{r^2}P(r)\right]{\bf{r}}+\left[\frac{\dot{r}}{r_\bot}\frac{dP}{dr}+\frac{S(r)}{r}\right]{\bf{r}}_\bot.
\end{equation}
Since the vector ${\bf{r}}$ and ${\bf{r}}_\bot$ are linearly independent, we conclude that equation (\ref{ansatz}) is equivalent to the following first order system for the unknown functions $S(r)$ and $P(r)$, that is
\begin{eqnarray}
\frac{dS}{dr}&=&\frac{r_\bot}{r\dot{r}}P(r)=\frac{\ell}{r^2\dot{r}}P(r),\label{eqS}\\
\frac{dP}{dr}&=&-\frac{r_\bot}{r\dot{r}}S(r)=-\frac{\ell}{r^2\dot{r}}S(r),\label{eqp}
\end{eqnarray}
where in the last step we used (\ref{15}). Note that the conservation of $\bm{\mathcal{A}}$ implies that the modulus of this vector is constant in time and hence, there must be some combination of the unknown functions $S(r)$ and $P(r)$ which remains constant. More precisely, if we multiply (\ref{eqS}) by $S(r)$, (\ref{eqp}) by $P(r)$ and sum them together, after integration we end up with the following constraint
\begin{equation}\label{constraint}
S^2(r)+P^2(r)=const.
\end{equation}
However, integrating directly the system (\ref{eqS}) and (\ref{eqp}) is not an efficient procedure for finding a LRL-vector because at some stage for solving the above system of differential equations we would have to integrate $(r^2\dot r)^{-1}$ over the variable $r$. It is then simpler to integrate directly 
\begin{equation}
dt=\pm\frac{dr}{\sqrt{2\left[\mathcal{E}-V_{eff}(r)\right]}},
\end{equation}
which solves the problem. Despite this problem, we can try to copy some of the successful steps used in finding the LRL-vector for the Kepler potential. More precisely, we assume that $S(r)$ and $P(r)$ are proportional to $r$ and $r\dot r$, respectively. Hence, we introduce the following educated guess
\begin{equation}\label{educguess}
S(r)=rg(r),\quad
P(r)= r\dot{r}r_\bot h(r)= \ell\dot{r} h(r).
\end{equation}
Substitution of (\ref{educguess}) into (\ref{eqS}) and (\ref{eqp}) leads to the system
\begin{eqnarray}
\dot{r}^2\frac{dh}{dr}+\frac{1}{2}\frac{d\dot{r}^2}{dr}h(r)&=&-\frac{g(r)}{r},\label{A}\\
\frac{d}{dr}\left(rg(r)\right)&=&\frac{\ell^2}{r^2}h(r).\label{B}
\end{eqnarray}
On the other hand, from the conservation equation (\ref{CE2}) we find that
\begin{equation}
\frac{d\dot{r}^2}{dr}=-2\frac{dV_{eff}}{dr},
\end{equation}
which replaced into (\ref{A}) allows to represent the differential system (\ref{A}) and (\ref{B}) as follows
\begin{eqnarray}
&&2\left[\mathcal{E}-V_{eff}(r)\right]\frac{dh}{dr}-\frac{dV_{eff}}{dr}h(r)+\frac{g(r)}{r}=0,\label{e1}\\
&&\frac{d}{dr}\left(rg(r)\right)-\frac{\ell^2}{r^2}h(r)=0.\label{e2}
\end{eqnarray}
Note that the constraint (\ref{constraint}) reads now
\begin{equation}\label{con2}
r^2g^2(r)+\ell^2\dot{r}^2h^2(r)=const.
\end{equation}
Using the conservation equation (\ref{CE2}) allows to rewrite (\ref{con2}) as 
\begin{equation}
r^2 g^2(r)+2\ell^2 h^2(r)\left[\mathcal{E}-V_{eff}(r)\right]=const.
\end{equation}
If we differentiate the above equation with respect to $r$, we obtain
\begin{equation}\label{aiuto}
2h\frac{dh}{dr}\left[\mathcal{E}-V_eff(r)\right]-\frac{dV_{eff}}{dr}h^2(r)=-\frac{1}{2\ell^2}\frac{d}{dr}\left(r^2 g^2(r)\right).
\end{equation}
At this point, we can multiply (\ref{e1}) by $h(r)$ and apply (\ref{aiuto}) in order to eliminate the dependence on $V_{eff}(r)$. This leads to the following system for the unknown functions $h$ and $g$, namely
\begin{eqnarray}
&&\frac{h(r)g(r)}{r}-\frac{1}{2\ell^2}\frac{d}{dr}\left(r^2 g^2(r)\right)=0,\label{e3}\\
&&\frac{d}{dr}\left(rg(r)\right)-\frac{\ell^2}{r^2}h(r)=0.\label{e4}
\end{eqnarray}
However, the above equations are not independent because the combination $(\ell^2/(r g(r)))\cdot$(\ref{e3})$+$(\ref{e4}) vanishes. Hence, we end up with the following underdetermined system
\begin{eqnarray}
&&\frac{d}{dr}(rg(r))=\frac{\ell^2}{r^2}h(r),\label{e5}\\
&&\frac{dV_{eff}}{dr}+\frac{1}{h^2(r)}\frac{dh^2}{dr}V_{eff}(r)=\frac{\mathcal{E}}{h^2(r)}\frac{dh^2}{dr}+\frac{1}{2\ell^2 h^2(r)}\frac{d}{dr}\left(r^2 g^2(r)\right),\label{e6}
\end{eqnarray}
where the second equation is simply (\ref{aiuto}) rewritten in an equivalent form. At this point, we see that two strategies are possible. We can assign $h$, solve (\ref{e5}) for $g$ and then, recover the effective potential from (\ref{e6}) or we can select $V_{eff}$ and solve the differential system (\ref{e5}) and (\ref{e6}) for $h$ and $g$. Either cases the LRL-vector is given by
\begin{equation}\label{newLRL}
\bm{\mathcal{A}}=g(r){\bf{r}}+r\dot{r}h(r){\bf{r}}_\bot
\end{equation}
and by the construction outlined in the present section such a vector is conserved. Note that in the case we assign the effective potential, equations (\ref{e5}) and (\ref{e6}) can be decoupled. More precisely, $h(r)$ should be a solution of the second order differential equation
\begin{equation}\label{equah}
\frac{d^2 h}{dr^2}+P_1(r)\frac{dh}{dr}+P_2(r)h(r)=0
\end{equation}
with
\begin{equation}\label{P1P2}
P_1(r)=\frac{4[\mathcal{E}-V_{eff}(r)]-3r V^{'}_{eff}(r)}{2r[\mathcal{E}-V_{eff}(r)]},\quad
P_2(r)=-\frac{r^4\Delta V_{eff}-\ell^2}{2r^4[\mathcal{E}-V_{eff}(r)]},
\end{equation}
where $\Delta$ is the Laplace operator introduced in (\ref{Laplace}) and the prime denotes differentiation with respect to the variable $r$ while $g(r)$ can be retrieved from (\ref{e1})  according to
\begin{equation}\label{equaG}
g(r)=rh(r)\frac{dV_{eff}}{dr}-2r\frac{dh}{dr}[\mathcal{E}-V_{eff}(r)]
\end{equation}
once $h(r)$ has been found from (\ref{equah}). By means of the variable transformation $r=x/(1-x)$ one may be tempted to try to find the most general potential such that the transformed equation (\ref{equah}) becomes the hypergeometric equation or some other equation relevant to applications. Unfortunately, this strategy is equivalent to impose that the effective potential satisfies at the same time two distinct differential equations whose solution spaces in general do not necessarily need to intersect all though there are special cases where they admit a common particular solution. Regarding this aspect we refer to the last part of the present section. From a computational point of view, it turns out to be more convenient to assign effective potentials relevant in physics, solve the corresponding differential equations for $h(r)$ and $g(r)$ and construct the corresponding LRL-vector. As a consistency check we consider the special case $h(r)=a=const$. Then, equation (\ref{e5}) can be immediately integrated and we obtain
\begin{equation}\label{gr}
g(r)=-\frac{a\ell^2}{r^2}+\frac{b}{r},
\end{equation}\
where $b$ is an arbitrary integration constant. If we replace $h(r)=a$ and $g(r)$ as given by (\ref{gr}) into (\ref{e6}), we end up with the following differential equation for the effective potential
\begin{equation}
\frac{dV_{eff}}{dr}=\frac{b}{ar^2}-\frac{\ell^2}{r^3},
\end{equation}
whose integration is straightforward and gives
\begin{equation}\label{Vex}
V_{eff}(r)=\frac{\ell^2}{2r^2}-\frac{b}{ar}.
\end{equation}
Note that we set the integration constant equal to zero because the potential is in any case defined up to an arbitrary additive constant. Furthermore, if substitute $h(r)=a$, (\ref{gr}) and (\ref{Vex}) into the constraint (\ref{constraint}), it can be easily verified that the combination
\begin{equation}
r^2 g^2(r)+2\ell^2 h^2(r)\left[\mathcal{E}-V_{eff}(r)\right]=b^2+2\mathcal{E}a^2\ell^2
\end{equation}
is indeed constant. Finally, if we compute the LRL-vector according to (\ref{newLRL}), it is not difficult to check that it agrees with the corresponding LRL-vector (\ref{Am}) for the Kepler problem whenever $a=-1$ and $b=-k_N$. Finally, in Table~\ref{overview} we present some potentials emerging from different choices of the function $h(r)$.
\begin{table}[!ht]
\caption{Analytic formulae of the potential $V_{eff}(r)$ and the function $g(r)$ for different choices of $h(r)$. Here, $a$ is a real parameter while $c_1$ and $c_2$ are arbitrary integration constants.}
\begin{center}
\begin{tabular}{ | l | l | l | l|l|}
\hline
$h(r)$&  $g(r)$  & $V_{eff}(r)$    \\ \hline
$ar$  &  $\frac{c_1}{r}+a\ell^2\frac{\ln{r}}{r}$ & $\frac{\ell^2}{2r^2}+\frac{\ell^2}{2}\frac{\ln^2{r}}{r^2}+\frac{c_1}{a}\frac{\ln{r}}{r^2}+\frac{c_2}{r^2}$            \\ \hline
$ar^2$ &  $a\ell^2+\frac{c_1}{r}$ & $\frac{\ell^2}{2r^2}+\frac{c_1}{ar^3}+\frac{c_2}{r^4}$          \\ \hline
$ar^3$ &  $\frac{a\ell^2}{2}r+\frac{c_1}{r}$ & $\frac{\ell^2}{2r^2}-\frac{3\ell^2}{8r^2}+\frac{c_1}{2ar^4}+\frac{c_2}{r^6}$            \\ \hline
$a\sqrt{r}$ & $\frac{c_1}{r}-\frac{2a\ell^2}{\sqrt{r}}$ & $\frac{\ell^2}{2r^2}+\frac{3\ell^2}{2mr^2}-\frac{2c_1}{ar\sqrt{r}}+\frac{c_2}{r}$\\ \hline
\end{tabular}
\label{overview}
\end{center}
\end{table}
We conclude this section with the explicit construction of the LRL-vector for several classes of potentials. The first two examples serve the purpose of verifying that our general construction of the LRL-vector for a central potential is sound in the sense that it reproduces correctly the known trajectories for the Coulomb and the harmonic potentials.

\subsection{The Kepler potential $V(r)=-k/r$ with $k>0$}
\begin{figure}[!ht]\label{uno}
\centering
    \includegraphics[width=0.4\textwidth]{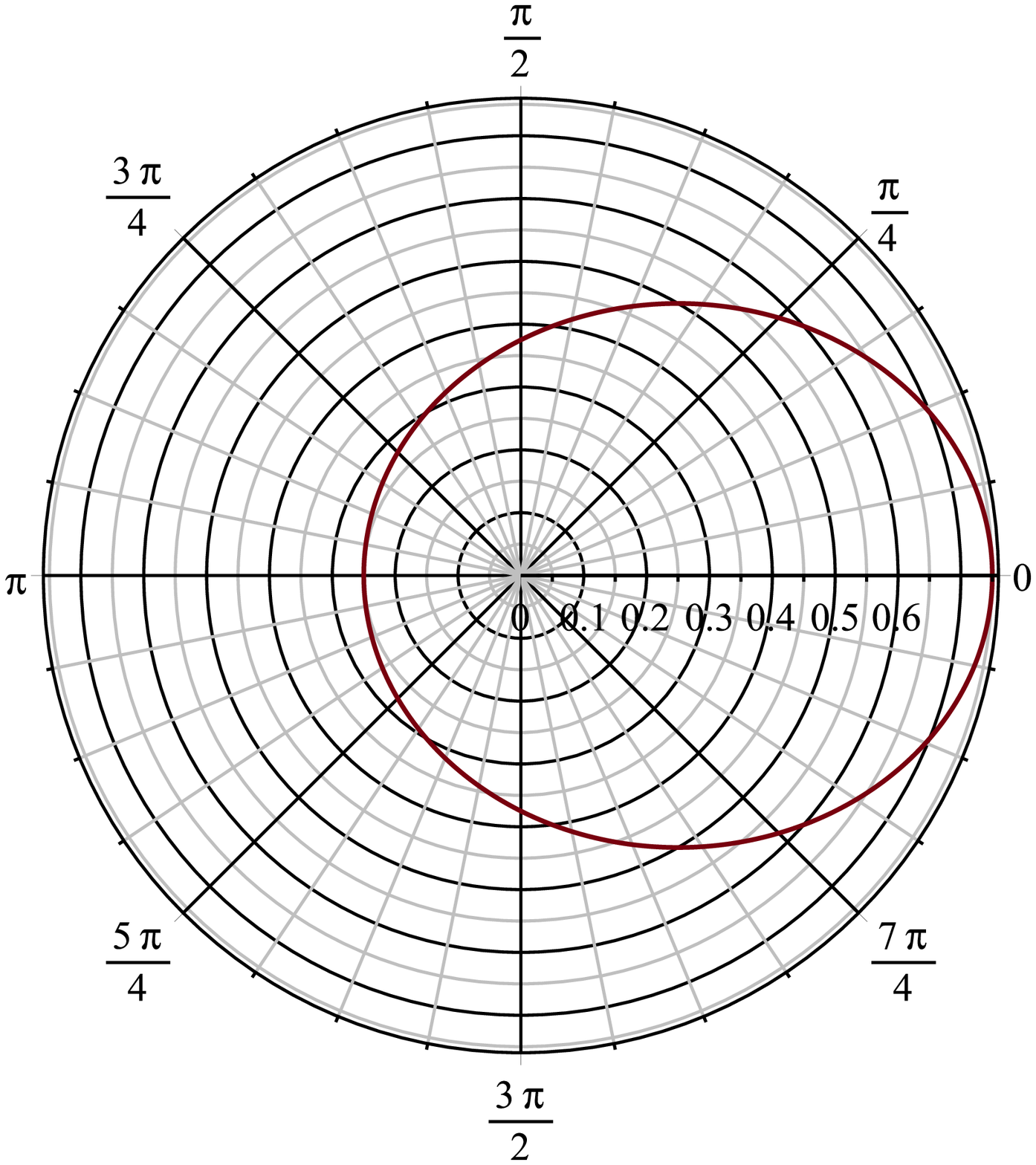}
    \includegraphics[width=0.4\textwidth]{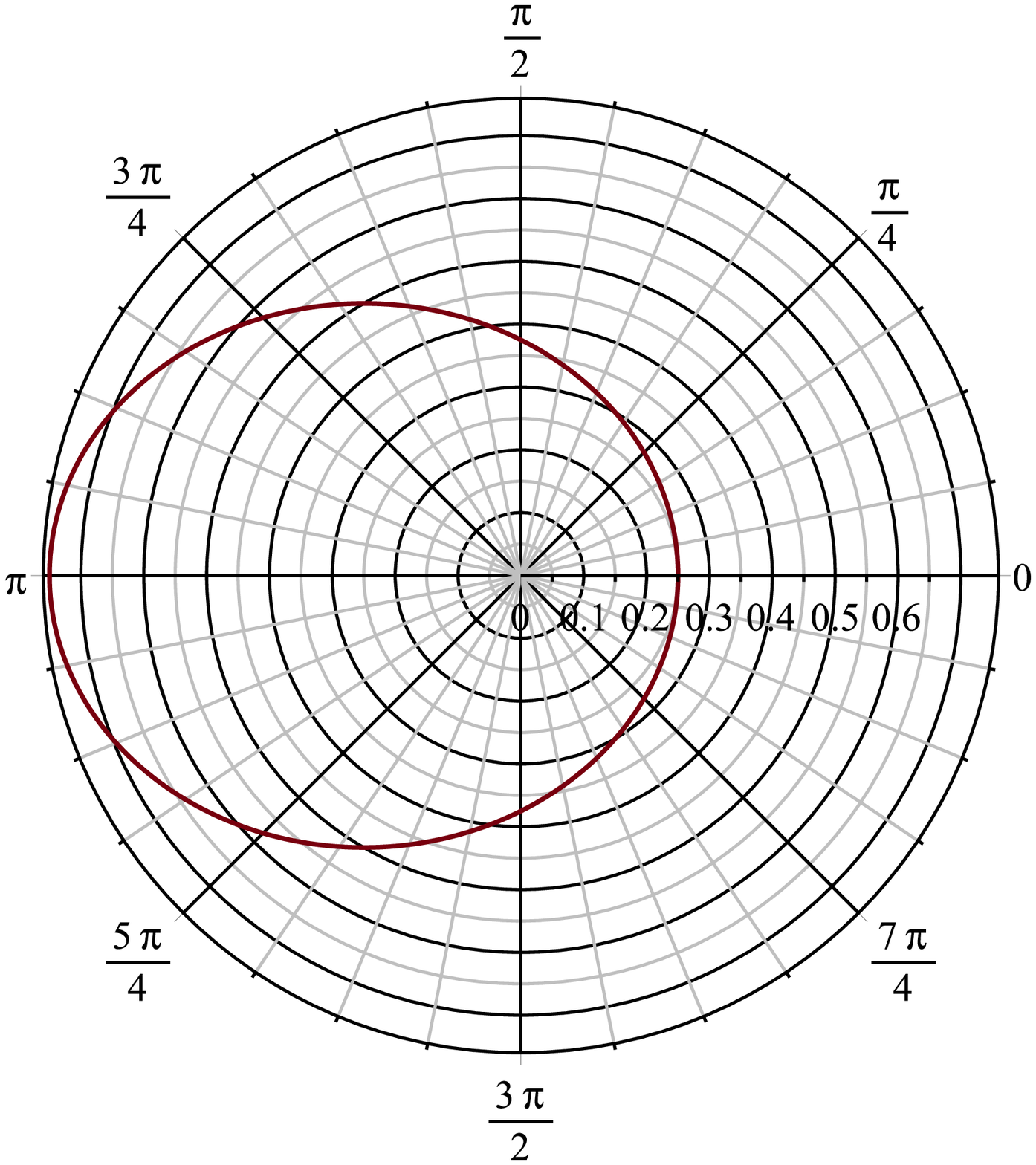}
\caption{\label{figure1}
Plots of the particle trajectories with eccentricity $e=0.5$ for the case $\mathcal{E}=-1$, $k=1$ and $\ell=\sqrt{3/8}$. The figure on the left displays $r^{(1)}_{-}(\varphi)$ while the one on the right $r^{(1)}_{+}(\varphi)$.}
\end{figure}

The corresponding effective potential is
\begin{equation}
V_{eff}(r)=\frac{\ell^2}{2r^2}-\frac{k}{r}
\end{equation}
and has a minimum at $r_{min}=\ell^2/k$ where $V_{eff}(r_{min})=-k^2/2\ell^2$. From (\ref{CE2}) we obtain the motion reality condition 
\begin{equation}
\mathcal{E}-V_{eff}(r)=\frac{2\mathcal{E}r^2+2kr-\ell^2}{2r^2}\geq 0
\end{equation}
and closed trajectories take place exhibiting two turning points when $V_{eff}(r_{min})<\mathcal{E}<0$. Note that for $\mathcal{E}=V_{eff}(r_{min})$ we have $r=k/|\mathcal{E}|$ corresponding to a circular path. To construct the LRL-vector we need to determine the unknown functions $h(r)$ and $g(r)$ which are obtained by solving (\ref{equah}) with
\begin{equation}
P_1(r)=\frac{4\mathcal{E}r^2+kr+\ell^2}{r(2\mathcal{E}r^2+2kr-\ell^2)},\quad
P_2(r)=0
\end{equation}
and equation (\ref{equaG}), respectively. First of all, we observe that (\ref{equah}) admits the particular solution $h_1(r)=c_1$ with $c_1$ an arbitrary integration constant. To find a second linearly independent particular solution $h_2(r)$ we set $\Phi(r)=h^{'}(r)$ so that  equation (\ref{equah}) reduces to the first order separable equation  $\Phi^{'}(r)+P_1(r)\Phi(r)=0$. With the help of $2.172$ and $2.175.1$ in \cite{Grad} we find that a particular solution is $\Phi(r)=r/(2\mathcal{E}r^2+2kr-\ell^2)^{3/2}$ which integrated once more leads to
\begin{equation}
h_2(r)=\frac{kr-\ell^2}{\sqrt{2\mathcal{E}r^2+2kr-\ell^2}}.
\end{equation}
Hence, the general solution reads
\begin{equation}
h(r)=c_1+c_2\frac{kr-\ell^2}{\sqrt{2\mathcal{E}r^2+2kr-\ell^2}}.
\end{equation}
Let us first consider the case $h(r)=c_1$. By means of (\ref{equaG}) we find
\begin{equation}
g(r)=c_1\left(\frac{k}{r}-\frac{\ell^2}{r^2}\right).
\end{equation}
If we rewrite (\ref{newLRL}) in the equivalent form $\bm{\mathcal{A}}=rg(r){\widehat{\bf{r}}}+\ell r\dot{r}h(r){\widehat{\bf{r}}}_\bot$ and use (\ref{CE2}), the square of the modulus of the LRL-vector is evaluated to be
\begin{equation}\label{Amodulus}
\mathcal{A}^2=\bm{\mathcal{A}}\cdot\bm{\mathcal{A}}=r^2g^2(r)+2\ell^2 h^2(r)\left[\mathcal{E}-V_{eff}(r)\right]=c_1^2\left(k^2+2\mathcal{E}\ell^2\right).
\end{equation}
Note that the motion reality condition ensures that $A^2$ cannot become negative. To derive the particle trajectory we observe that $\bm{\mathcal{A}}\cdot\bm{r}=\mathcal{A}r\cos{\varphi}$ and $\bm{\mathcal{A}}\cdot\bm{r}=r^2 g(r)$. Combining them yields the equation
\begin{equation}\label{equag}
\mathcal{A}\frac{\cos{\varphi}}{r}=g(r).
\end{equation}
If we solve for $r$ the above equation, we end up with
\begin{equation}\label{traj1}
r^{(1)}_\pm(\varphi)=\frac{\ell^2/k}{1\pm e\cos{\varphi}},\quad
e=\sqrt{1+\frac{2\mathcal{E}\ell^2}{k^2}}
\end{equation}
where $e$ is the eccentricity. From Fig.~\ref{figure1} we immediately see that $r^{(1)}_{-}(\varphi)$ gives the polar representation of an ellipse with one focus coinciding with the origin of the coordinate system while the second focus is located on the positive $x$-axis. In the case the positive sign is chosen, as in $r^{(1)}_{+}(\varphi)$, the position of the first focus is the same as before but the second focus is now on the negative $x$-xis.
We conclude the analysis by considering the case when $h(r)=c_2 h_2(r)$ for which
\begin{equation}
g(r)=c_2\ell^2\frac{\sqrt{2\mathcal{E}r^2+2kr-\ell^2}}{r^2}.
\end{equation}
By means of (\ref{Amodulus}) we find that $\mathcal{A}^2=c_2^2\ell^2(k^2+2\mathcal{E}\ell^2)$. Moreover, solving (\ref{equag}) gives
\begin{equation}\label{cv}
r_\pm(\varphi)=\ell^2\frac{k\pm\sqrt{k^2+2\mathcal{E}\ell^2}\sin{\varphi}}{(2\mathcal{E}\ell^2+k^2)\cos^2{\varphi}-2\mathcal{E}\ell^2}.
\end{equation}
If we introduce the eccentricity $e$ defined in (\ref{traj1}), it possible to rewrite (\ref{cv}) as
\begin{equation}
r^{(2)}_\pm(\varphi)=\frac{\ell^2/k}{1\pm e\sin{\varphi}}.
\end{equation}
\begin{figure}[!ht]\label{due}
\centering
    \includegraphics[width=0.4\textwidth]{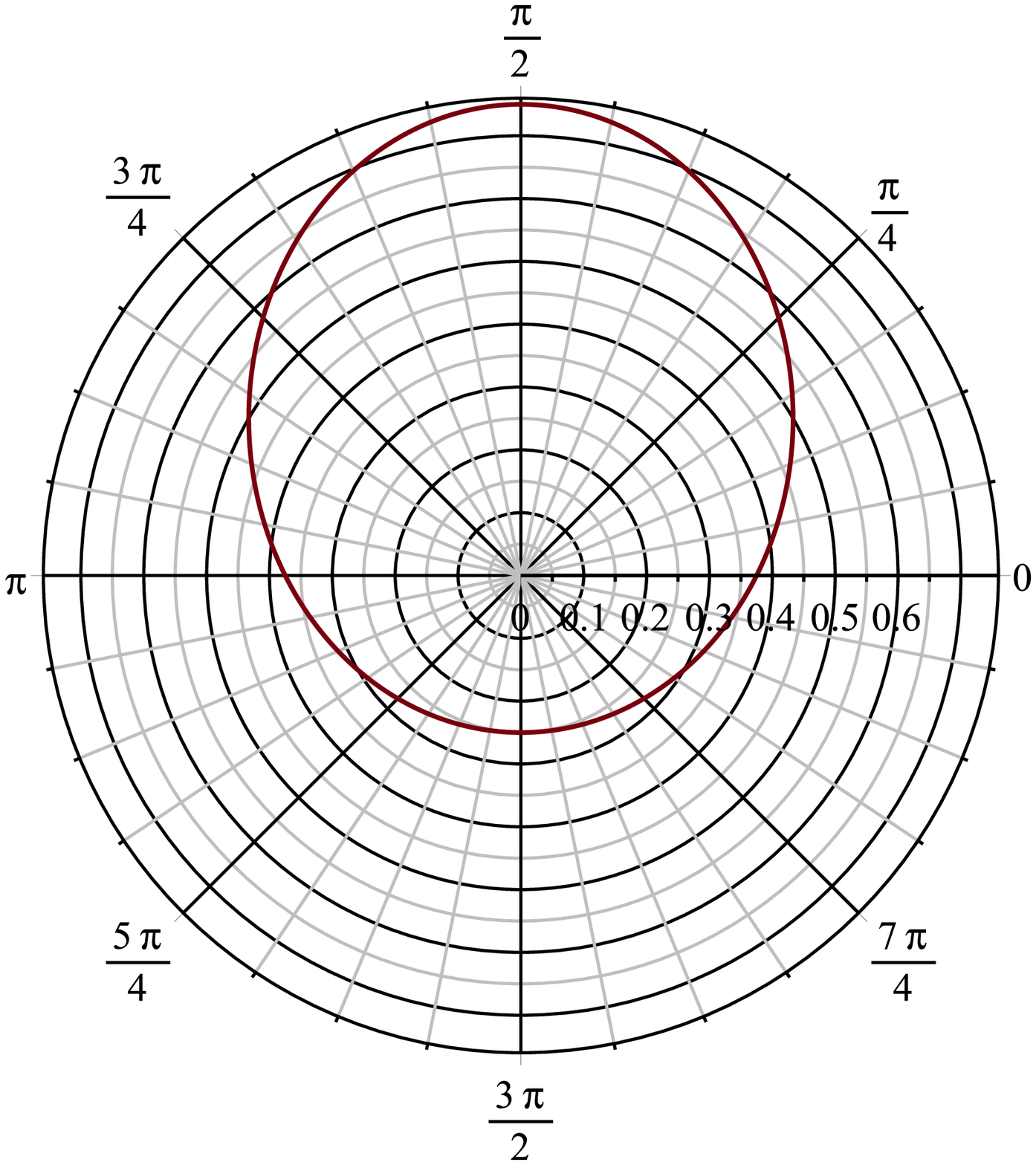}
    \includegraphics[width=0.4\textwidth]{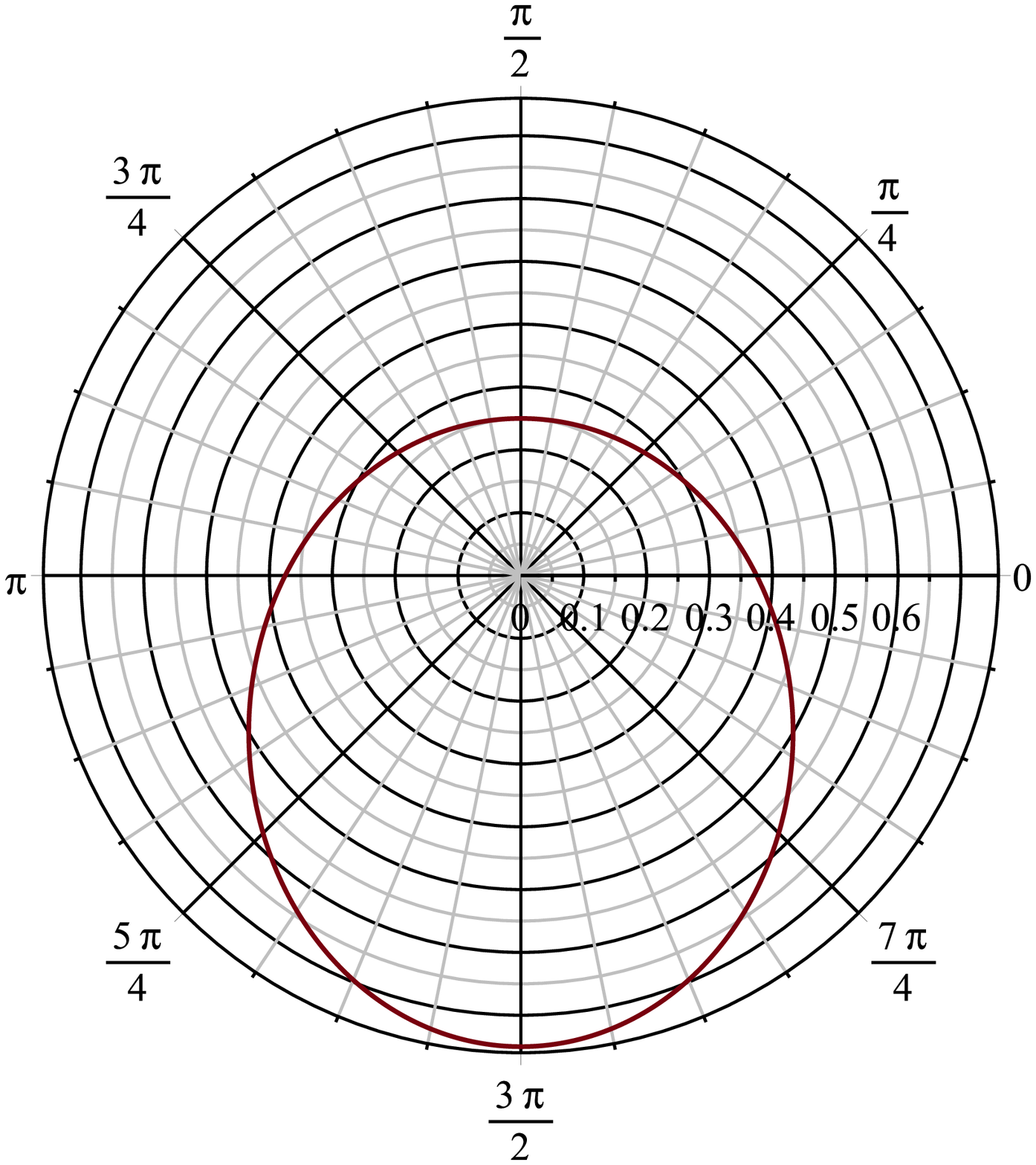}
\caption{\label{figure2}
Plots of the particle trajectories with eccentricity $e=0.5$ for the case $\mathcal{E}=-1$, $k=1$ and $\ell=\sqrt{3/8}$. The figure on the left displays $r^{(2)}_{-}(\varphi)$ while the one on the right $r^{(2)}_{+}(\varphi)$.}
\end{figure}
By looking at Fig.~\ref{figure2} we realize that $r^{(2)}_{-}(\varphi)$ is the polar representation of an ellipse with one focus on the origin of the coordinate system while the second focus is located on the positive $y$-axis. In the case of $r^{(2)}_{+}(\varphi)$, the position of the first focus is the same as before but the second focus is now on the negative $y$-xis.

{\bf{Finally, in order to enable the reader to extract new insights and potential applications of our result, the restored expression of the generalised LRL vector for the nontrivial example studied here is presented below in explicit closed form }}
 \begin{equation}
 \boxed{\bm{\mathcal{A}}=rg(r){\widehat{\bf{r}}}+\ell r\dot{r}h(r){\widehat{\bf{r}}}_\bot,\quad\dot{r}=\pm\sqrt{2\left[\mathcal{E}-V_{eff}(r)\right]},\quad V_{eff}(r)=\frac{\ell^2}{2r^2}-\frac{k}{r}},
 \end{equation}
 \begin{equation}
 \boxed{g(r)=c_1\left(\frac{k}{r}-\frac{\ell^2}{r^2}\right)+c_2\ell^2\frac{\sqrt{2\mathcal{E}r^2+2kr-\ell^2}}{r^2},\quad
 h(r)=c_1+c_2\frac{kr-\ell^2}{\sqrt{2\mathcal{E}r^2+2kr-\ell^2}}.
 }
\end{equation}

\subsection{The isotropic harmonic potential $V(r)=kr^2$ with $k>0$}
The corresponding effective potential is
\begin{equation}
V_{eff}(r)=\frac{\ell^2}{2r^2}+kr^2
\end{equation}
and has a global minimum at $r_{min}=\sqrt{\ell/\sqrt{2k}}$ where $V_{eff}(r_{min})=\sqrt{2k}\ell$. If we impose the motion reality condition 
\begin{equation}\label{quartic}
\mathcal{E}-V_{eff}(r)=-\frac{2kr^4-2\mathcal{E}r^2+\ell^2}{2r^2}\geq 0,
\end{equation}
it is not difficult to verify that the above inequality is satisfied whenever $\mathcal{E}\geq V_{eff}(r_{min})$ and $r_1\leq r\leq r_2$ where 
\begin{equation}
r_1=\sqrt{\frac{\mathcal{E}-\sqrt{\mathcal{E}^2-2k\ell^2}}{2k}},\quad
r_2=\sqrt{\frac{\mathcal{E}+\sqrt{\mathcal{E}^2-2k\ell^2}}{2k}}
\end{equation}
are the turning points in the particle trajectory. Note that the quartic polynomial in (\ref{quartic}) has two additional roots given by $r_3=-r_1$ and $r_4=-r_2$. This remark allows us to introduce the following factorisation 
\begin{equation}
\mathcal{E}-V_{eff}(r)=\frac{k}{r^2}(r^2-r_1^2)(r_2^2-r^2)
\end{equation}
which is especially useful in solving the differential equation (\ref{equah}) with
\begin{equation}
P_1(r)=-\frac{1}{r}+\frac{6r(2kr^2-\mathcal{E})}{2kr^4-2\mathcal{E}r^2+\ell^2},\quad
P_2(r)=\frac{6kr^2}{2kr^4-2\mathcal{E}r^2+\ell^2}.
\end{equation}
To construct the LRL-vector we need to determine the unknown functions $h(r)$ and $g(r)$ which are obtained by solving (\ref{equah}) and (\ref{equaG}), respectively. The corresponding expression for the LRL-vector is then computed according to (\ref{newLRL}). The general solution of (\ref{equah}) is $h(r)=c_1 h_1(r)+c_2 h_2(r)$ with
\begin{equation}
h_1(r)=\frac{1}{\sqrt{r^2-r_1^2}},\quad
h_2(r)=\frac{1}{\sqrt{r_2^2-r^2}}.
\end{equation}
Moreover the functions $g(r)$ generated by the above solutions are
\begin{equation}
g_1(r)=(2\mathcal{E}-2kr_1^2)\frac{\sqrt{r^2-r_1^2}}{r^2},\quad
g_2(r)=(2\mathcal{E}-2kr_2^2)\frac{\sqrt{r^2_2-r^2}}{r^2}.
\end{equation}
Let us first consider the case $h(r)=h_1(r)$. Proceeding as before, the square of the modulus of the LRL-vector is given by
\begin{equation}\label{Amodulusis}
\mathcal{A}^2=4kc_1^2 r_2^2\sqrt{\mathcal{E}^2-2k\ell^2}
\end{equation}
while the trajectory can be computed according to (\ref{equag}) and is given by the following formula
\begin{equation}\label{traj2}
r^{(1)}(\varphi)=\frac{r_1}{\sqrt{1-B_1\cos^2{\varphi}}},\quad B_1=\frac{2\sqrt{\mathcal{E}^2-2k\ell^2}}{\mathcal{E}+\sqrt{\mathcal{E}^2-2k\ell^2}}.
\end{equation}
If we consider the case $h(r)=h_2(r)$, we find that $\mathcal{A}^2=4kc_2^2 r_1^2\sqrt{\mathcal{E}^2-2k\ell^2}$ and in this case the trajectory is 
\begin{equation}\label{traj3}
r^{(2)}(\varphi)=\frac{r_2}{\sqrt{1+B_2\cos^2{\varphi}}},\quad B_2=\frac{2\sqrt{\mathcal{E}^2-2k\ell^2}}{\mathcal{E}-\sqrt{\mathcal{E}^2-2k\ell^2}}.
\end{equation}
\begin{figure}[!ht]\label{tre}
\centering
    \includegraphics[width=0.4\textwidth]{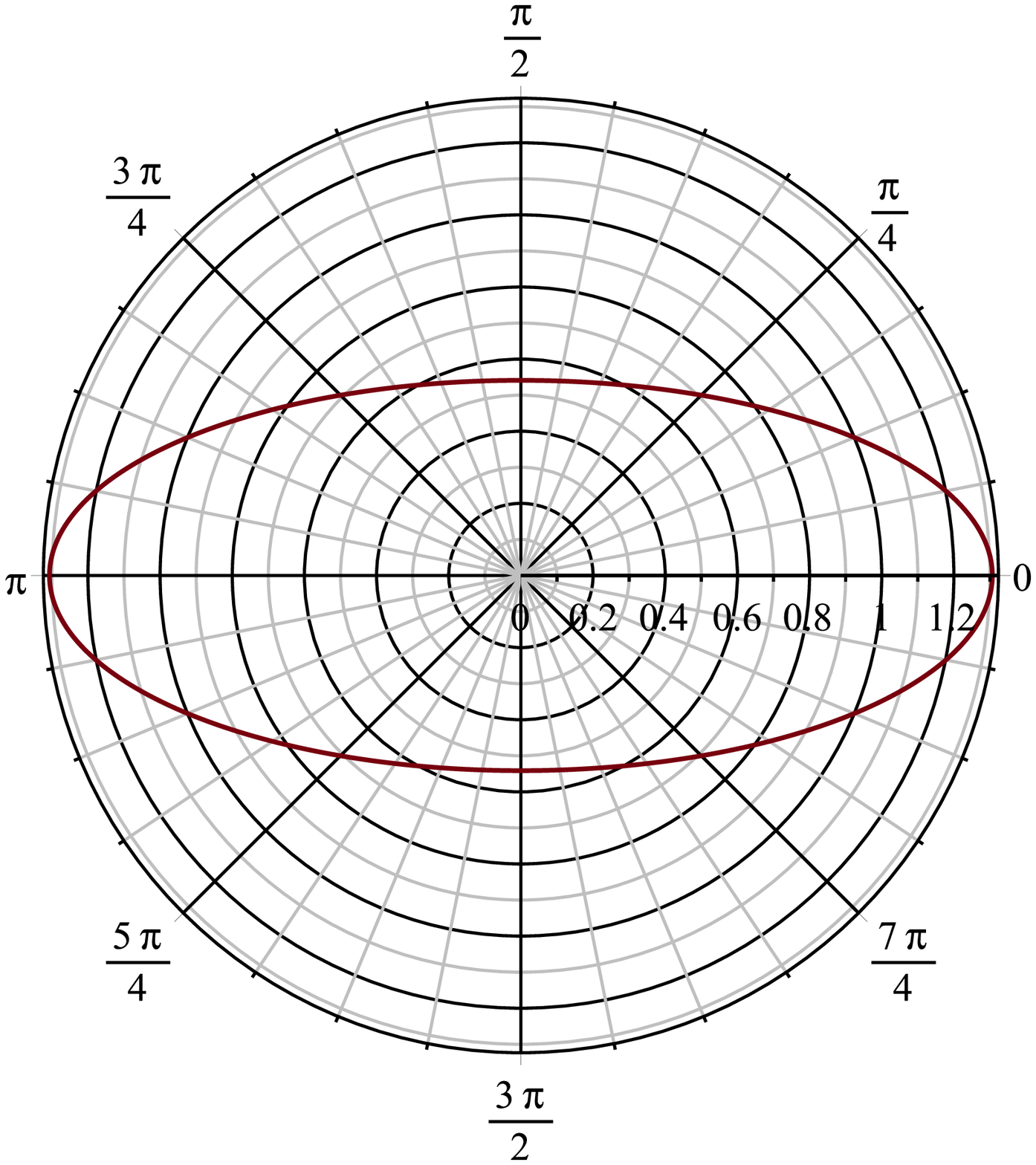}
    \includegraphics[width=0.4\textwidth]{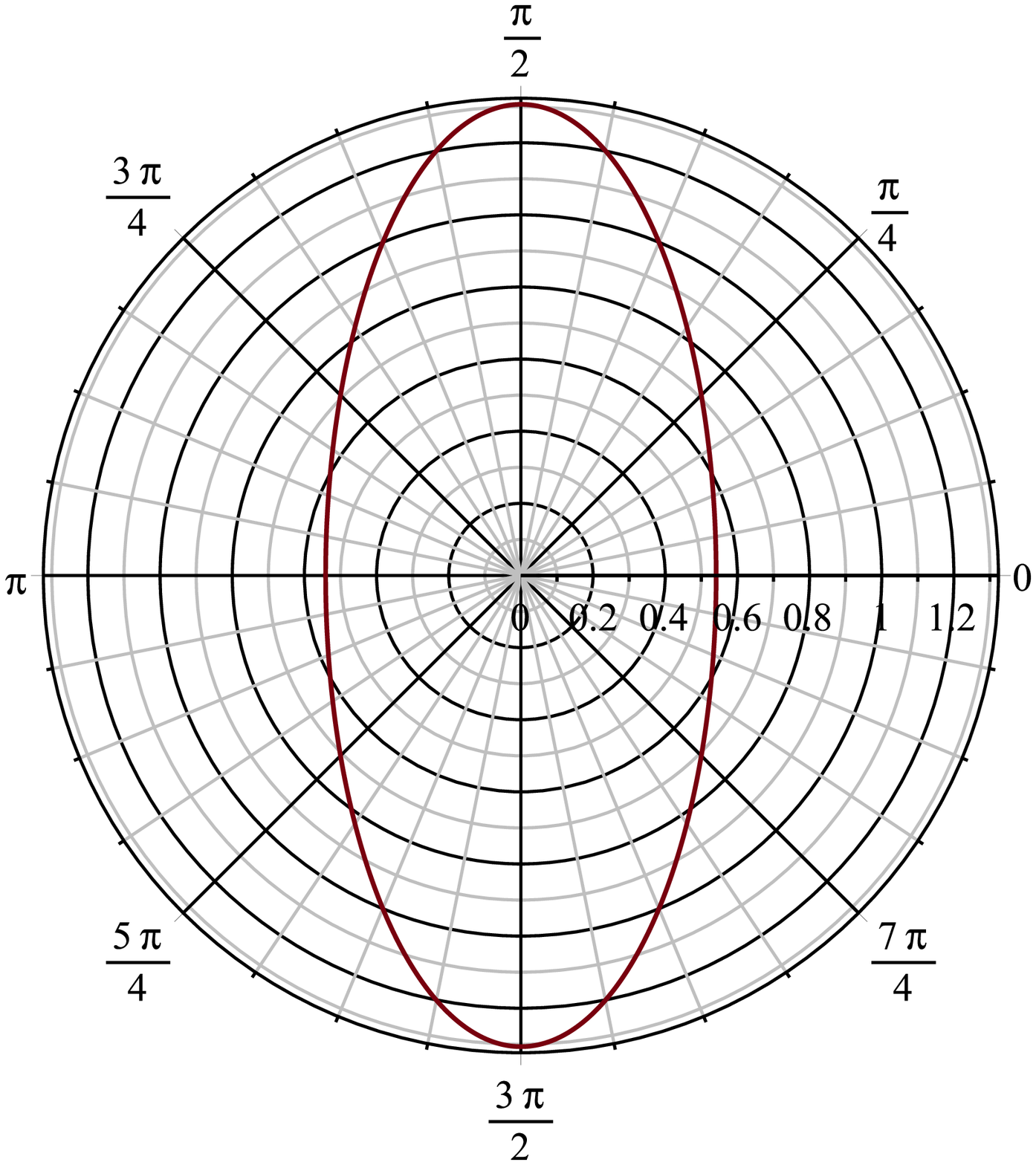}
\caption{\label{figure3}
Plots of the particle trajectories with eccentricity for $k=1=\ell$ and $\mathcal{E}=2$. The figure on the left shows $r^{(1)}_{-}(\varphi)$ with $r_1=0.5412$ and $B_1=0.8284$ while the one on the right $r^{(2)}_{+}(\varphi)$ with $r_2=1.3066$ and $B_2=4.8284$.}
\end{figure}
It is gratifying to observe in Fig.~\ref{figure3} that the trajectories (\ref{traj2}) and (\ref{traj3}) are again ellipses as in the case of the Kepler problem but with the difference now that the origin of the coordinate system is located at the midpoint of the line segment joining the foci of the ellipse.
{\bf{Finally, in order to facilitate the reader to extract new insights and potential applications of our result, the restored expression of the generalised LRL vector for the nontrivial example studied here is presented below in explicit closed form }}
 \begin{equation}
 \boxed{\bm{\mathcal{A}}=rg(r){\widehat{\bf{r}}}+\ell r\dot{r}h(r){\widehat{\bf{r}}}_\bot,\quad\dot{r}=\pm\sqrt{2\left[\mathcal{E}-V_{eff}(r)\right]},\quad V_{eff}(r)=\frac{\ell^2}{2r^2}+kr^2},
 \end{equation}
 \begin{equation}
 \boxed{g(r)=c_1(2\mathcal{E}-2kr_1^2)\sqrt{1-\left(\frac{r_1}{r}\right)^2}+c_2(2\mathcal{E}-2kr_2^2)\sqrt{\left(\frac{r_2}{r}\right)^2-1},\quad
 h(r)=\frac{c_1}{\sqrt{r^2-r_1^2}}+\frac{c_2}{\sqrt{r_2^2-r^2}},
 }
\end{equation}
\begin{equation}
\boxed{
r_1=\sqrt{\frac{\mathcal{E}-\sqrt{\mathcal{E}^2-2k\ell^2}}{2k}},\quad
r_2=\sqrt{\frac{\mathcal{E}+\sqrt{\mathcal{E}^2-2k\ell^2}}{2k}}.
}
\end{equation}

\subsection{The potential $V(r)=-k/r-B/r^3$ with $k>0$ and $B>0$}\label{tukaj}
This is the potential considered by \cite{Greiner,Garavaglia,Eliseo,Wayne} in connection with the problem of determining the perihelion rotation of a planet moving in the gravitational field of the Sun. The $1/r^3$ term takes into account the flattening of the Sun at the poles. A similar contribution appears also in the general relativistic case for the Schwarzschild metric but it has of course a different origin. The effective potential of our problem reads
\begin{equation}\label{GRpotG}
V_{eff}(r)=\frac{\ell^2}{2r^2}-\frac{k}{r}-\frac{B}{r^3}.
\end{equation}
If $\ell>\sqrt[4]{12kB}$, the above potential admits a local minimum and a global maximum at
\begin{equation}
r_{max}=\frac{\ell^2-\sqrt{\ell^4-12kB}}{2k},\quad
r_{min}=\frac{\ell^2+\sqrt{\ell^4-12kB}}{2k}
\end{equation}
with $r_{max}<r_{min}$. If we introduce the parameter
\begin{equation}
\alpha=\frac{\ell^4}{12kB}>1,
\end{equation}
then
\begin{eqnarray}
V_{eff}(r_{max})&=&\frac{\ell^6-18kB\ell^2+\sqrt{(\ell^4-12kB)^3}}{108B^2}=\frac{\ell^6}{108B^2}\left[1-\frac{3}{2\alpha}+\sqrt{\left(1-\frac{1}{\alpha}\right)^3}\right],\\
V_{eff}(r_{min})&=&\frac{\ell^6-18kB\ell^2-\sqrt{(\ell^4-12kB)^3}}{108B^2}=\frac{\ell^6}{108B^2}\left[1-\frac{3}{2\alpha}-\sqrt{\left(1-\frac{1}{\alpha}\right)^3}\right].
\end{eqnarray}
At this point a comment is in order. Taking into account that the intersection of the effective potential with the $r$-axis are
\begin{equation}
r_{I,II}=\frac{\ell^2}{4k}\left[1\pm\sqrt{1-\frac{4}{3\alpha}}\right],
\end{equation}
the following scenarios emerge
\begin{enumerate}
\item
if $1<\alpha<4/3$ both maximum and minimum are negative because $r_{I,II}$ become complex.
\item
If $0<\alpha<1$ both $r_{I,II}$ and $r_{max,min}$ become complex. In this case, the effective potential is monotonically increasing on the interval $(0,+\infty)$ and vanishes as $r\to+\infty$. 
\item
If $\alpha>4/3$, there is a maximum and minimum with $V_{eff}(r_{max})>0$ and $V_{eff}(r_{min})<0$.
\end{enumerate}
Hence, bounded trajectories are allowed only in the case 1. and 2. As before we introduce the motion reality condition $\mathcal{E}-V_{eff}(r)\geq 0$. From the behaviour of the effective potential we deduce that closed trajectories are possible whenever $V_{eff}(r_{min})\leq\mathcal{E}<0$. Imposing such a condition is equivalent to request that the cubic polynomial appearing in the expression below
\begin{equation}\label{cubic}
\mathcal{E}-V_{eff}(r)=\frac{2kr^2-2|\mathcal{E}|r^3-\ell^2 r+2B}{2r^3}=\frac{|\mathcal{E}|}{r^3}(r-r_0)(r-r_1)(r_2-r)
\end{equation}
admits three real distinct roots here denoted by $0<r_0<r_1<r_2$. Clearly, the motion reality condition is satisfied between the turning points $r_1$ and $r_2$ and in what follows, we will always consider the case when $r_1<r<r_2$. To find the aforementioned roots we proceed as in \cite{Bron}. More precisely, we divide the cubic in (\ref{cubic}) by $-2|\mathcal{E}|$ and introduce the variable transformation
\begin{equation}
r=\frac{k}{3|\mathcal{E}|}+y
\end{equation}
in order to get the corresponding reduced cubic
\begin{equation}
y^3+3py+2q=0,\quad
p=\frac{3|\mathcal{E}|\ell^2-2k^2}{18\mathcal{E}^2},\quad
q=\frac{9k|\mathcal{E}|\ell^2-54B\mathcal{E}^2-4k^3}{108|\mathcal{E}|^3}.
\end{equation}
There will be three different real roots when the discriminant $D=q^2+p^3$ is negative. After some lengthy but straightforward algebra we find that
\begin{equation}
r_0=\frac{k}{3|\mathcal{E}|}-2\sqrt{|p|}\cos{\left(\frac{\pi-\psi}{3}\right)},\quad
r_1=\frac{k}{3|\mathcal{E}|}-2\sqrt{|p|}\cos{\left(\frac{\pi+\psi}{3}\right)},\quad
r_2=\frac{k}{3|\mathcal{E}|}+2\sqrt{|p|}\cos{\left(\frac{\psi}{3}\right)}
\end{equation}
with $\cos{\psi}=|q|/|p|^{3/2}$. As a consistency check we verified that in the limit of $B\to 0$ the formulae for the turning points $r_1$ and $r_2$ agree with the corresponding ones in the Kepler problem. In the same limit $r_0$ vanishes but this does not pose any problem because the Kepler effective potential diverges at $r=0$. The next step is the construction of the LRL-vector associated to bounded trajectories with $r_1<r<r_2$. First of all, we compute the solution of the differential equation (\ref{equah}) for $h(r)$ with 
\begin{equation}
P_1(r)=-\frac{5}{2r}+\frac{12kr-18|\mathcal{E}|r^2-3\ell^2}{2kr^2-2|\mathcal{E}|r^3-\ell^2 r+2B},\quad
P_2(r)=\frac{3\ell^2}{2Br}+\frac{3}{r^2}+\frac{6|\mathcal{E}|\ell^2 r^2-3(2k\ell^2-4|\mathcal{E}|B)r-3(4kB-\ell^4)}{2B(2kr^2-2|\mathcal{E}|r^3-\ell^2 r+2B)},
\end{equation}
where $P_1$ and $P_2$ have been computed according to (\ref{P1P2}). If we try the ansatz
\begin{equation}\label{ansatzo}
h(r)=\frac{r^{3/2}e^{\int_r^{r_2} w(u)du}}{\sqrt{2kr^2-2|\mathcal{E}|r^3-\ell^2 r+2B}},
\end{equation}
we end up with the following nonlinear first order differential equation for the unknown function $w(r)$, namely
\begin{equation}
r(2kr^2-2|\mathcal{E}|r^3-\ell^2 r+2B)\left[\frac{dw}{dr}+w^2(r)\right]+
(3kr^2-4|\mathcal{E}|r^3-\ell^2 r+B)w(r)+\ell^2=0.
\end{equation}
It can be verified with Maple, that the above equation admits the following solutions
\begin{equation}
w_\pm(r)=\pm\frac{i\ell}{\sqrt{r(2kr^2-2|\mathcal{E}|r^3-\ell^2 r+2B)}}=
\pm\frac{i\ell/\sqrt{2|\mathcal{E}|}}{\sqrt{r(r-r_0)(r-r_1)(r_2-r)}}.
\end{equation}
At this point the integral in (\ref{ansatzo}) can be easily evaluated with the help of $1.2.34.2$ in \cite{Prudnikov}. More precisely, we find that
\begin{equation}\label{emodulus}
\int_r^{r_2}\frac{du}{\sqrt{u(u-r_0)(u-r_1)(r_2-u)}}=\frac{2}{\sqrt{r_1(r_2-r_0)}}F(\sin{\phi},\kappa),\quad
\sin{\phi}=\sqrt{\frac{r_1(r_2-r)}{(r_2-r_1)r}},\quad
\kappa=\sqrt{\frac{r_0(r_2-r_1)}{r_1(r_2-r_0)}}
\end{equation}
under the assumption that $0<r_0<r_1<r_2$ and $r_1<r<r_2$. Here, $F$ denotes the incomplete elliptic integral of the first kind, $\phi$ is its amplitude and $\kappa$ the elliptic modulus. Hence, two complex linearly independent solutions for (\ref{equah}) are
\begin{equation}
h_{\pm,\mathbb{C}}(r)=\frac{r^{3/2}}{\sqrt{2kr^2-2|\mathcal{E}|r^3-\ell^2 r+2B}}\mbox{exp}\left(\pm\frac{i\sqrt{2}\ell}{\sqrt{|\mathcal{E}|r_1(r_2-r_0)}}F(\sin{\phi},\kappa)\right)
\end{equation}
from which we can set up two real linearly independent solutions as follows
\begin{eqnarray}
h_1(r)&=&\frac{h_{+,\mathbb{C}}(r)+h_{-,\mathbb{C}}(r)}{2}=\frac{r^{3/2}}{\sqrt{2kr^2-2|\mathcal{E}|r^3-\ell^2 r+2B}}\cos{\left(\frac{\sqrt{2}\ell F(\sin{\phi},\kappa)}{\sqrt{|\mathcal{E}|r_1(r_2-r_0)}}\right)},\\
h_2(r)&=&\frac{h_{+,\mathbb{C}}(r)-h_{-,\mathbb{C}}(r)}{2i}=\frac{r^{3/2}}{\sqrt{2kr^2-2|\mathcal{E}|r^3-\ell^2 r+2B}}\sin{\left(\frac{\sqrt{2}\ell F(\sin{\phi},\kappa)}{\sqrt{|\mathcal{E}|r_1(r_2-r_0)}}\right)}.
\end{eqnarray}
Given the function $h(r)$, the corresponding function $g(r)$ can be evaluated by means of (\ref{equaG}) which can be rewritten in the present problem as 
\begin{equation}
g(r)=-\frac{2|\mathcal{E}|}{r^2}(r-r_0)(r-r_1)(r_2-r)\frac{dh}{dr}+\frac{k}{r^3}(r-r_{min})(r-r_{max})h(r).
\end{equation}
By means of the identities
\begin{eqnarray}
\frac{dF(\sin{\phi},\kappa)}{dr}&=&-\frac{1}{2}\sqrt{\frac{r_1(r_2-r_0)}{r(r-r_0)(r-r_1)(r_2-r)}},\\
\frac{\sqrt{2|\mathcal{E}|}}{r^2}(r-r_0)(r-r_1)(r_2-r)\frac{d}{dr}\left(\frac{r^{3/2}}{\sqrt{(r-r_0)(r-r_1)(r_2-r)}}\right)&=&\frac{k(r-r_{min})(r-r_{max})}{r^{3/2}\sqrt{2|\mathcal{E}|(r-r_0)(r-r_1)(r_2-r)}},\label{idII}
\end{eqnarray}
where in deriving (\ref{idII}) we used the following equalities for the the roots of the cubic in (\ref{cubic})
\begin{equation}
r_0+r_1+r_2=\frac{k}{|\mathcal{E}|},\quad
r_0 r_1+r_0 r_2+r_1 r_2=\frac{\ell^2}{|\mathcal{E}|},\quad
r_0 r_1 r_2=\frac{B}{|\mathcal{E}|},
\end{equation}
we find that
\begin{equation}
g_1(r)=-\frac{\ell}{r}\sin{\left(\frac{\sqrt{2}\ell F(\sin{\phi},\kappa)}{\sqrt{|\mathcal{E}|r_1(r_2-r_0)}}\right)},\quad
g_2(r)=\frac{\ell}{r}\cos{\left(\frac{\sqrt{2}\ell F(\sin{\phi},\kappa)}{\sqrt{|\mathcal{E}|r_1(r_2-r_0)}}\right)}.
\end{equation}
Let us consider the pair of functions $h_1(r)$ and $h_2(r)$. With the help of (\ref{Amodulus}) we find that the modulus of the LRL-vector is simply $\mathcal{A}=\ell$ while the trajectory can be computed according to (\ref{equag}) which leads to the equation
\begin{equation}
\sin{\left(\frac{\sqrt{2}\ell F(\sin{\phi},\kappa)}{\sqrt{|\mathcal{E}|r_1(r_2-r_0)}}\right)}=\sin{\left(\varphi+\frac{3}{2}\pi\right)}.
\end{equation}
The above equation can be explicitly solved for $r$ and leads to the following formula for the trajectory
\begin{equation}\label{mastert}
r(\varphi)=\frac{r_1 r_2}{r_1+(r_2-r_1)\mbox{sn}^2{\left(\frac{\sqrt{|\mathcal{E}|r_1(r_2-r_0)}}{2\ell}(\pi-2\varphi),\kappa\right)}},
\end{equation}
where sn is the elliptic sine which belongs to the family of Jacobi elliptic functions \cite{Grad} and $\kappa$ is the elliptic modulus already defined in (\ref{emodulus}). 
\begin{figure}[!ht]\label{quattro}
\centering
    \includegraphics[width=0.3\textwidth]{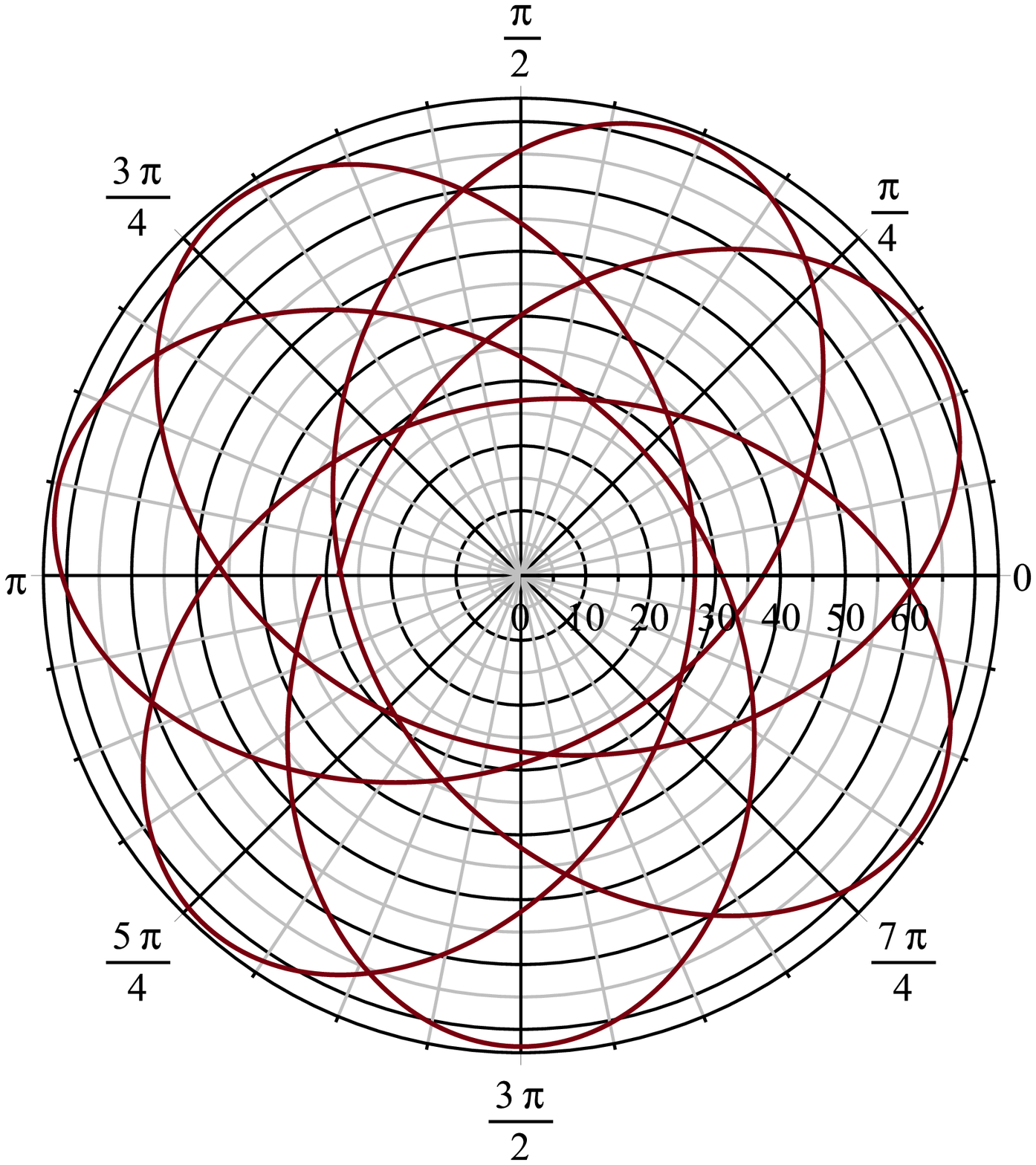}
    \includegraphics[width=0.3\textwidth]{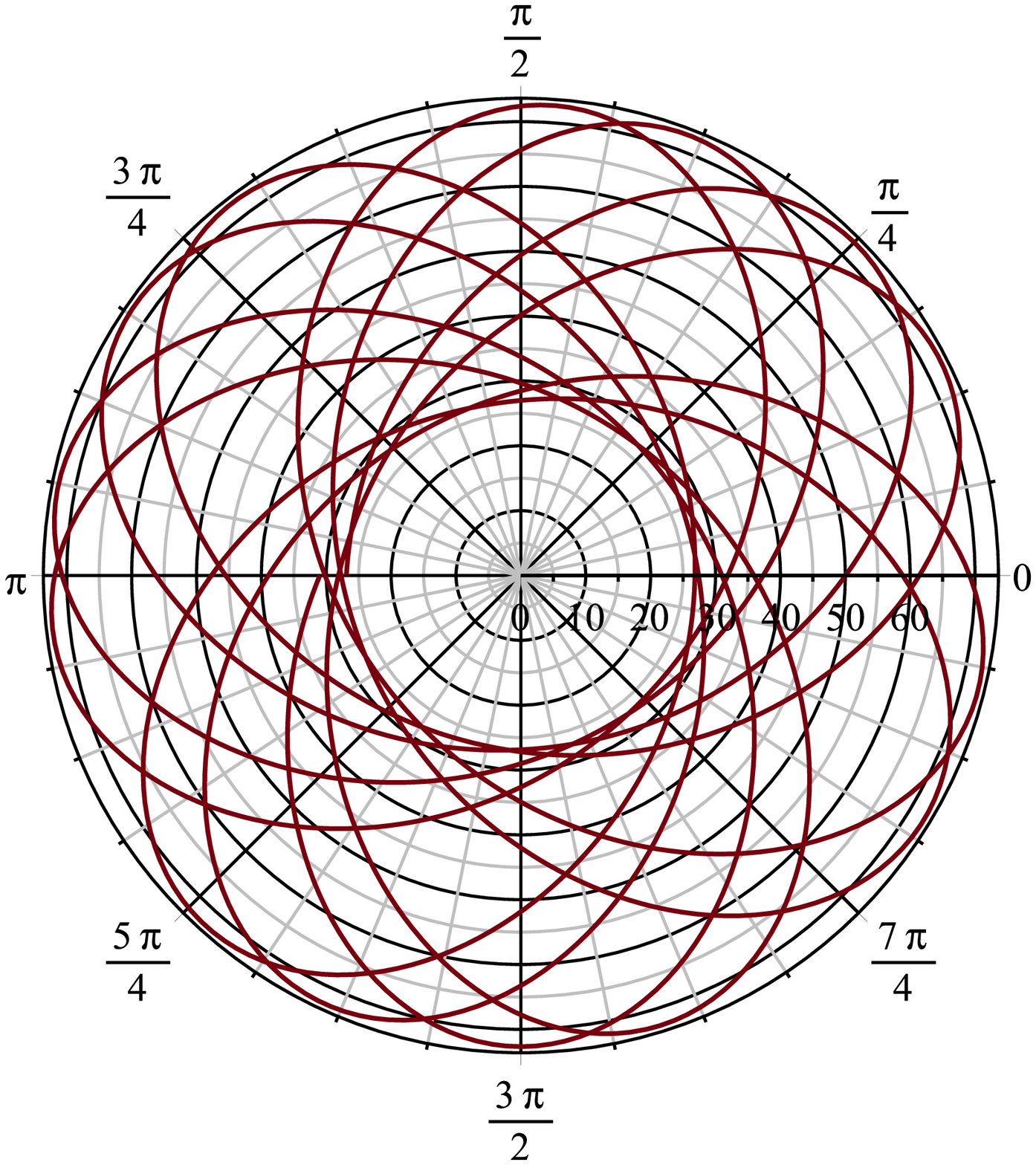}
    \includegraphics[width=0.3\textwidth]{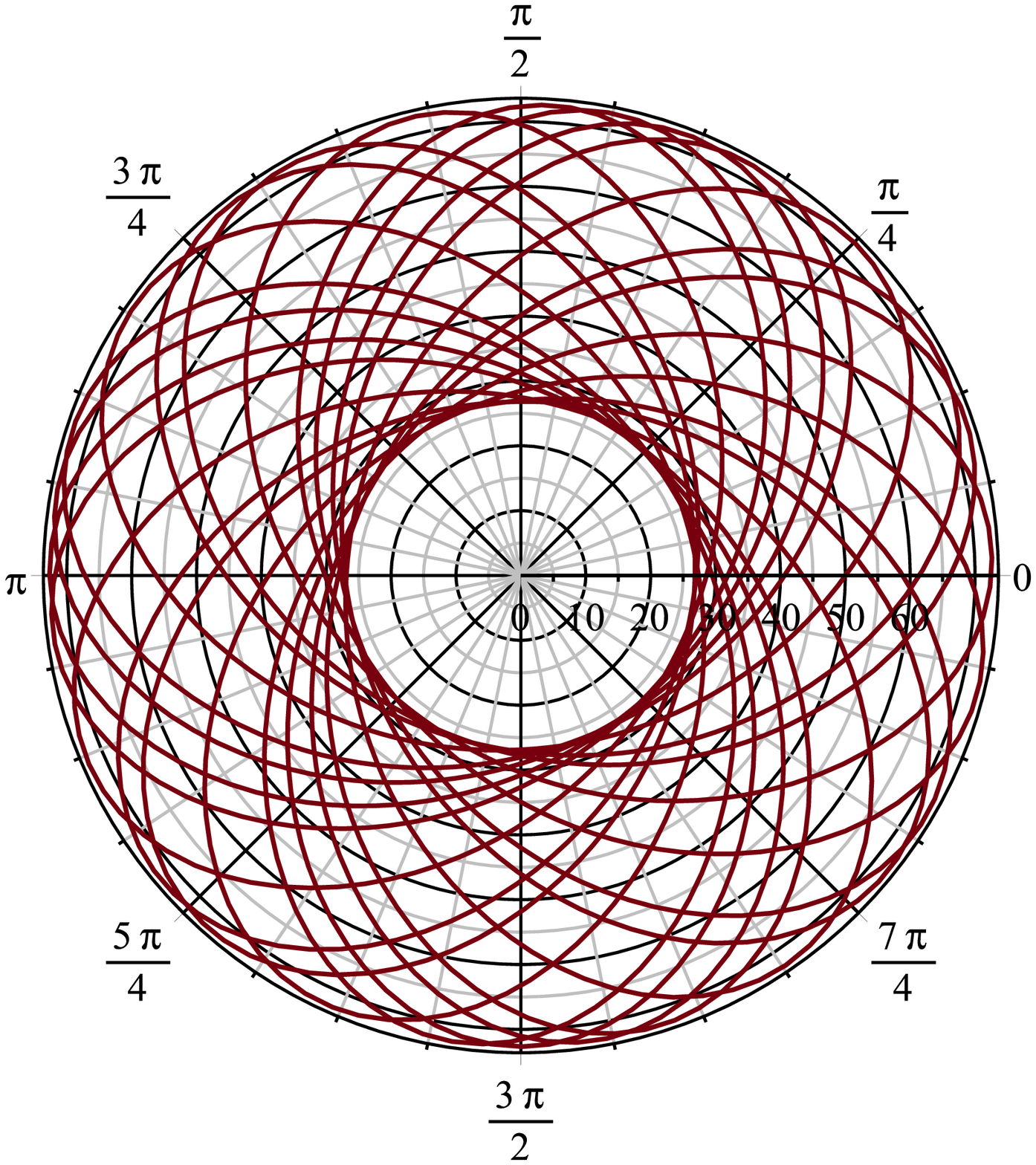}
\caption{\label{figure4}
Plots of the particle trajectory for $\mathcal{E}=10^{-3}$, $k=0.1$, $\ell=2$ and $B=1$ for which $r_0=0.51$, $r_1=26.82$ and $r_2=72.67$. The figure on the far left corresponds to $10$ revolutions around the central star, while the one in the middle and far right show the corresponding trajectories after $20$ and $40$ revolutions, respectively.}
\end{figure}
In Figure~\ref{figure4} we showcased the trajectory described by (\ref{mastert}) for a certain choice of the physical parameters and different numbers of revolutions around the central massive star. We end this part by reminding the reader that similar conclusions and results hold in the case we consider the function $g_2(r)$.
{\bf{The restored expression of the generalised LRL vector for the example studied here is presented below in explicit closed form }}
\begin{equation}
 \boxed{\bm{\mathcal{A}}=rg(r){\widehat{\bf{r}}}+\ell r\dot{r}h(r){\widehat{\bf{r}}}_\bot,\quad\dot{r}=\pm\sqrt{2\left[\mathcal{E}-V_{eff}(r)\right]},\quad V_{eff}(r)=\frac{\ell^2}{2r^2}-\frac{k}{r}-\frac{B}{r^3},\quad k>0,\quad B>0,}
 \end{equation}
\begin{equation}
 \boxed{
 g(r)=\frac{\ell}{r}\left[c_1\sin{\Omega(r)}+c_2\cos{\Omega(r)}\right],\quad
 h(r)=\frac{r^{3/2}}{\sqrt{2kr^2-2|\mathcal{E}|r^3-\ell^2 r+2B}}\left[c_1\cos{\Omega(r)}+c_2\sin{\Omega(r)}\right],
 }
\end{equation}
\begin{equation}
\boxed{
\Omega(r)=\frac{\sqrt{2}\ell F(\sin{\phi},\kappa)}{\sqrt{|\mathcal{E}|r_1(r_2-r_0)}},\quad
\sin{\phi}=\sqrt{\frac{r_1(r_2-r)}{(r_2-r_1)r}},\quad
\kappa=\sqrt{\frac{r_0(r_2-r_1)}{r_1(r_2-r_0)}},
}
\end{equation}
\begin{equation}
\boxed{
r_0=\frac{k}{3|\mathcal{E}|}-2\sqrt{|p|}\cos{\left(\frac{\pi-\psi}{3}\right)},\quad
r_1=\frac{k}{3|\mathcal{E}|}-2\sqrt{|p|}\cos{\left(\frac{\pi+\psi}{3}\right)},\quad
r_2=\frac{k}{3|\mathcal{E}|}+2\sqrt{|p|}\cos{\left(\frac{\psi}{3}\right)},
}
\end{equation}
\begin{equation}
\boxed{
\cos{\psi}=\frac{|q|}{|p|^{3/2}},\quad
p=\frac{3|\mathcal{E}|\ell^2-2k^2}{18\mathcal{E}^2},\quad
q=\frac{9k|\mathcal{E}|\ell^2-54B\mathcal{E}^2-4k^3}{108|\mathcal{E}|^3}.
}
\end{equation}
{\bf{For the definition of the elliptic function $F$ we refer to the glossary in Appendix C.}}

\subsection{The cosmological potential $V(r)=-k/r-\lambda r^2$ with $k>0$.}
We consider two scenarios: $\lambda<0$ and $\lambda>0$ corresponding to the anti-de Sitter and de Sitter cases, respectively. For $\lambda<0$ we focus on the physically interesting case when the particle energy is zero and the effective potential admits two real positive roots. This allows us to understand how the behaviour of the particle trajectory differs from the one analysed for the Kepler potential.

\subsubsection{The anti-de Sitter case}
Let $\lambda=-|\lambda|$. The corresponding effective potential is
\begin{equation}\label{K1}
V_{eff}(r)=\frac{\ell^2}{2r^2}-\frac{k}{r}+|\lambda| r^2=\frac{2|\lambda| r^4-2kr+\ell^2}{2r^2}.
\end{equation}
A simple application of Descartes' rule of signs indicates that if $r<0$, there is no sign change and therefore the potential will not intersect the negative real axis. This observation is relevant for the subsequent discussion. In the case $r>0$, we may have two positive real roots or none. Here, we are interested in the scenario where the effective potential has two positive roots and in such a framework we are going to derive the LRL-vector for the zero energy case. More precisely, according to \cite{Arnon} the quartic $x^4+px^2+qx+R$ admits two distinct real roots and two complex conjugate roots if
\begin{equation}\label{deltaA}
\delta(p,q,R)=256R^3-128p^2 R^2+144pq^2 R+16 p^4R-27q^4-4p^3 q^2<0.
\end{equation}
For the quartic appearing in (\ref{K1}) we have $p=0$, $q=-k/|\lambda|$ and $R=\ell^2/2|\lambda|$ so that the condition above becomes
\begin{equation}
\delta(0,-k/|\lambda|,\ell^2/2|\lambda|)=\frac{32|\lambda|\ell^6-27k^4}{|\lambda|^4}<0
\end{equation}
and leads to the following bound
\begin{equation}\label{bound1}
|\lambda|<\frac{27k^4}{32\ell^6}.
\end{equation}
If we impose $dV_{eff}/dr=0$, we end up with the quartic equation
\begin{equation}\label{quartic_c}
2|\lambda| r^4+kr-\ell^2=0.
\end{equation}
It will display two distinct real roots and two complex conjugate roots if 
\begin{equation}
\delta(0,k/2|\lambda|,-\ell^2/2|\lambda|)=\frac{512|\lambda|\ell^6-27k^4}{16|\lambda|^4}<0
\end{equation}
or equivalently
\begin{equation}\label{bound2}
|\lambda|<\frac{27k^4}{512\ell^6}.
\end{equation}
Clearly, both bounds (\ref{bound1}) and (\ref{bound2}) are simultaneously satisfied if (\ref{bound2}) holds. Moreover, it will turn out that one of the real roots of (\ref{quartic_c}) is always negative. In order to determine the minimum in the effective potential, we observe that the coefficient going with the cubic power in (\ref{quartic_c}) vanishes and therefore, according to \cite{Bron} we first need to apply the variable transformation
\begin{equation}
r=u-\frac{1}{2}\left(\frac{k}{|\lambda|}\right)^{1/3}
\end{equation}
to (\ref{quartic_c}) leading to
\begin{equation}\label{ocap}
u^4-2\left(\frac{k}{|\lambda|}\right)^{1/3}u^3+\frac{3}{2}\left(\frac{k}{|\lambda|}\right)^{2/3}u^2-\frac{\ell^2}{2|\lambda|}-\frac{3}{16}\left(\frac{k}{|\lambda|}\right)^{4/3}=0.
\end{equation}
The roots of this equation can be expressed in terms of one of the real roots of the cubic \cite{Bron}
\begin{equation}\label{1cap}
8y^3-6\left(\frac{k}{|\lambda|}\right)^{2/3}y^2+\left[\frac{4\ell^2}{|\lambda|}+\frac{3}{2}\left(\frac{k}{|\lambda|}\right)^{4/3}\right]y-\frac{1}{8}\left(\frac{k}{|\lambda|}\right)^{2/3}\left[\frac{8\ell^2}{|\lambda|}+3\left(\frac{k}{|\lambda|}\right)^{4/3}\right]=0.
\end{equation}
By means of the transformation
\begin{equation}
y=v+\frac{1}{4}\left(\frac{k}{|\lambda|}\right)^{2/3}
\end{equation}
we obtain the reduced cubic
\begin{equation}
v^3+3pv+2q=0,\quad
p=\frac{\ell^2}{6|\lambda|},\quad
q=-\frac{k^2}{64\lambda^2}.
\end{equation}
Note that $p>0$ ensures that the discriminant $D=q^2+p^3$ is always positive. Since $D>0$, $p>0$ and $q<0$, the cubic above has only one real root which is computed to be \cite{Bron} 
\begin{equation}
v_1=\frac{2\ell}{\sqrt{6|\lambda|}}\sinh{\frac{\phi}{3}},\quad
\sinh{\phi}=\frac{3\sqrt{6}k^2}{32\ell^3\sqrt{|\lambda|}}
\end{equation}
and the corresponding real root of (\ref{1cap}) is
\begin{equation}
y_1=\frac{1}{4}\left(\frac{k}{|\lambda|}\right)^{2/3}+\frac{2\ell}{\sqrt{6|\lambda|}}\sinh{\frac{\phi}{3}}.
\end{equation}
Following \cite{Bron} the roots of (\ref{ocap}) coincide with the roots of the equations
\begin{equation}
\eta^2+\left[A_\pm-2\left(\frac{k}{|\lambda|}\right)^{1/3}\right]\frac{\eta}{2}+y_1\left[1-\frac{2}{A_\pm}\left(\frac{k}{|\lambda|}\right)^{1/3}\right]=0,\quad
A_\pm=\pm\sqrt{8y_1-2\left(\frac{k}{|\lambda|}\right)^{2/3}}=\pm4\sqrt{\frac{\ell}{\sqrt{6|\lambda|}}\sinh{\frac{\phi}{3}}}
\end{equation}
and are
\begin{eqnarray}
u_{+,1}&=&-\frac{1}{4A_+}\left[A_+^2-2\left(\frac{k}{|\lambda|}\right)^{1/3}A_+-\sqrt{\Delta_+}\right],\quad
u_{+,2}=-\frac{1}{4A_+}\left[A_+^2-2\left(\frac{k}{|\lambda|}\right)^{1/3}A_++\sqrt{\Delta_+}\right]\\
u_{-,1}&=&-\frac{1}{4A_-}\left[A_-^2-2\left(\frac{k}{|\lambda|}\right)^{1/3}A_--\sqrt{\Delta_-}\right],\quad
u_{-,2}=-\frac{1}{4A_-}\left[A_-^2-2\left(\frac{k}{|\lambda|}\right)^{1/3}A_-+\sqrt{\Delta_-}\right]
\end{eqnarray}
with
\begin{equation}
\Delta_\pm=A_\pm^4-4A_\pm^3\left(\frac{k}{|\lambda|}\right)^{1/3}+4A_\pm^2\left(\frac{k}{|\lambda|}\right)^{2/3}-16A_\pm^2 y_1+32A_\pm y_1\left(\frac{k}{|\lambda|}\right)^{1/3}.
\end{equation}
Switching back to the variable $r$ gives
\begin{equation}
r_{+,1}=-\frac{A_+}{4}+\frac{\sqrt{\Delta_+}}{4A_+},\quad
r_{+,2}=-\frac{A_+}{4}-\frac{\sqrt{\Delta_+}}{4A_+},\quad
r_{-,1}=-\frac{A_-}{4}+\frac{\sqrt{\Delta_-}}{4A_-},\quad
r_{-,2}=-\frac{A_-}{4}-\frac{\sqrt{\Delta_-}}{4A_-}.
\end{equation}
Further simplifications occur if we observe that the following identities hold true, namely
\begin{equation}
A_\pm^4+4A_\pm^2\left(\frac{k}{|\lambda|}\right)^{2/3}-16A_\pm^2 y_1=-A_\pm^4,\quad
-4A_\pm^3\left(\frac{k}{|\lambda|}\right)^{1/3}+32A_\pm y_1\left(\frac{k}{|\lambda|}\right)^{1/3}=\pm\frac{8k}{|\lambda|}A
\end{equation}
with
\begin{equation}
A=4\sqrt{\frac{\ell}{\sqrt{6|\lambda|}}\sinh{\frac{\phi}{3}}}.
\end{equation}
At this point it is straightforward to verify that the roots of (\ref{quartic_c}) can be nicely expressed as follows
\begin{eqnarray}
r_{+,1}&=&-\frac{A}{4}+\frac{1}{4A}\sqrt{-A^4+\frac{8k}{|\lambda|}A},\quad
r_{+,2}=-\frac{A}{4}-\frac{1}{4A}\sqrt{-A^4+\frac{8k}{|\lambda|}A},\\
r_{-,1}&=&\frac{A}{4}-\frac{i}{4A}\sqrt{A^4+\frac{8k}{|\lambda|}A},\quad
r_{-,2}=\frac{A}{4}+\frac{i}{4A}\sqrt{A^4+\frac{8k}{|\lambda|}A}.
\end{eqnarray}
We immediately observe that $r_{-,1}$ and $r_{-,2}$ are imaginary roots while $r_{+,2}$ is always negative. Hence, we conclude that $r_{+,1}$ represents the position of the minimum in the effective potential. Small note aside the positivity of the quantity $-A^4+8kA/|\lambda|$ is ensured by (\ref{bound2}).  Since we are interested in the zero energy case, the turning points of the particle trajectory are simply computed by finding the positive real roots of the effective potential, here denoted by $r_A$ and $r_B$ with $0<r_A<r_B$. To this purpose we transform the quartic in (\ref{K1}) according to
\begin{equation}
r=\widetilde{u}+\left(\frac{k}{4|\lambda|}\right)^{1/3}
\end{equation}
and we end up with the equation
\begin{equation}\label{omegacap}
\widetilde{u}^4+\left(\frac{16k}{|\lambda|}\right)^{1/3}\widetilde{u}^3+\frac{3}{2}\left(\frac{2k}{|\lambda|}\right)^{2/3}\widetilde{u}^2+\frac{\ell^2}{2|\lambda|}-\frac{3}{16}\left(\frac{2k}{|\lambda|}\right)^{4/3}=0
\end{equation}
whose roots can be expressed in terms of one of the real roots of the cubic
\begin{equation}\label{Icap}
8\widetilde{y}^3-\left(\frac{8k}{|\lambda|}\right)^{2/3}\widetilde{y}^2-\left[\frac{4\ell^2}{|\lambda|}-2\left(\frac{2k}{|\lambda|}\right)^{4/3}\right]\widetilde{y}+\left(\frac{2k}{|\lambda|}\right)^{2/3}\left[\frac{\ell^2}{|\lambda|}-\frac{3}{2^{5/3}}\left(\frac{k}{|\lambda|}\right)^{4/3}\right]=0.
\end{equation}
With the help of the transformation
\begin{equation}
\widetilde{y}=\widetilde{v}+\left(\frac{k}{4|\lambda|}\right)^{2/3}
\end{equation}
we get the reduced cubic
\begin{equation}
\widetilde{v}^3+3\widetilde{p}\widetilde{v}+2\widetilde{q}=0,\quad
\widetilde{p}=-\frac{\ell^2}{6|\lambda|},\quad
\widetilde{q}=-\frac{k^2}{16\lambda^2}.
\end{equation}
Note that the corresponding discriminant
\begin{equation}\label{dt}
\widetilde{D}=\widetilde{q}^2+\widetilde{p}^3=-\frac{32|\lambda|\ell^6-27k^4}{6912\lambda^4}
\end{equation}
is always positive because the numerator in (\ref{dt}) is negative due to the condition (\ref{bound2}). Since $\widetilde{D}>0$, $\widetilde{p}<0$ and $\widetilde{q}<0$, there are only one real root and two complex conjugate roots. The real root is \cite{Bron}
\begin{equation}
\widetilde{v}_1=\frac{2\ell}{\sqrt{6|\lambda|}}\cosh{\frac{\widetilde{\phi}}{3}},\quad
\cosh{\widetilde{\phi}}=\frac{3\sqrt{6}k^2}{32\ell^3\sqrt{|\lambda|}}
\end{equation}
and the corresponding real root of (\ref{Icap}) reads
\begin{equation}
\widetilde{y}_1=\left(\frac{k}{4|\lambda|}\right)^{2/3}+\frac{2\ell}{\sqrt{6|\lambda|}}\cosh{\frac{\widetilde{\phi}}{3}}.
\end{equation}
According to \cite{Bron} the roots of (\ref{omegacap}) coincide with the roots of the equations
\begin{equation}
\eta^2+\left[\widetilde{A}_\pm+\left(\frac{16k}{|\lambda|}\right)^{1/3}\right]\frac{\eta}{2}+\widetilde{y}_1\left[1+\frac{1}{\widetilde{A}_\pm}\left(\frac{16k}{|\lambda|}\right)^{1/3}\right]=0,\quad
\widetilde{A}_\pm=\pm\sqrt{8\widetilde{y}_1-2\left(\frac{2k}{|\lambda|}\right)^{2/3}}=\pm4\sqrt{\frac{\ell}{\sqrt{6|\lambda|}}\cosh{\frac{\widetilde{\phi}}{3}}}
\end{equation}
and are
\begin{eqnarray}
\widetilde{u}_{+,1}&=&-\frac{1}{4\widetilde{A}_+}\left[\widetilde{A}_+^2+2\left(\frac{2k}{|\lambda|}\right)^{1/3}\widetilde{A}_+-\sqrt{\widetilde{\Delta}_+}\right],\quad
\widetilde{u}_{+,2}=-\frac{1}{4\widetilde{A}_+}\left[\widetilde{A}_+^2+2\left(\frac{2k}{|\lambda|}\right)^{1/3}\widetilde{A}_++\sqrt{\widetilde{\Delta}_+}\right]\\
\widetilde{u}_{-,1}&=&-\frac{1}{4\widetilde{A}_-}\left[\widetilde{A}_-^2+2\left(\frac{2k}{|\lambda|}\right)^{1/3}\widetilde{A}_--\sqrt{\widetilde{\Delta}_-}\right],\quad
\widetilde{u}_{-,2}=-\frac{1}{4\widetilde{A}_-}\left[\widetilde{A}_-^2+2\left(\frac{2k}{|\lambda|}\right)^{1/3}\widetilde{A}_-+\sqrt{\widetilde{\Delta}_-}\right],
\end{eqnarray}
where
\begin{equation}
\widetilde{\Delta}_\pm=\widetilde{A}_\pm^4+4\widetilde{A}_\pm^3\left(\frac{2k}{|\lambda|}\right)^{1/3}+4\widetilde{A}_\pm^2\left(\frac{2k}{|\lambda|}\right)^{2/3}-16\widetilde{A}_\pm^2\widetilde{y}_1-32\widetilde{A}_\pm \widetilde{y}_1\left(\frac{2k}{|\lambda|}\right)^{1/3}.
\end{equation}
Transforming back to the variable $r$ gives
\begin{equation}
\widetilde{r}_{+,1}=-\frac{\widetilde{A}_+}{4}+\frac{\sqrt{\widetilde{\Delta}_+}}{4\widetilde{A}_+},\quad
\widetilde{r}_{+,2}=-\frac{\widetilde{A}_+}{4}-\frac{\sqrt{\widetilde{\Delta}_+}}{4\widetilde{A}_+},\quad
\widetilde{r}_{-,1}=-\frac{\widetilde{A}_-}{4}+\frac{\sqrt{\widetilde{\Delta}_-}}{4\widetilde{A}_-},\quad
\widetilde{r}_{-,2}=-\frac{\widetilde{A}_-}{4}-\frac{\sqrt{\widetilde{\Delta}_-}}{4\widetilde{A}_-}.
\end{equation}
Also in this case it is not difficult to verify that the following identities hold true, namely
\begin{equation}
\widetilde{A}_\pm^4+4\widetilde{A}_\pm^2\left(\frac{2k}{|\lambda|}\right)^{2/3}-16\widetilde{A}_\pm^2\widetilde{y}_1=-\widetilde{A}_\pm^4,\quad
4\widetilde{A}_\pm^3\left(\frac{2k}{|\lambda|}\right)^{1/3}-32\widetilde{A}_\pm\widetilde{y}_1\left(\frac{2k}{|\lambda|}\right)^{1/3}=\mp\frac{16k}{|\lambda|}\widetilde{A}
\end{equation}
with
\begin{equation}
\widetilde{A}=4\sqrt{\frac{\ell}{\sqrt{6|\lambda|}}\cosh{\frac{\widetilde{\phi}}{3}}}.
\end{equation}
Finally, it is not difficult to check that the roots of quartic in (\ref{K1})  admit the following representation
\begin{eqnarray}
\widetilde{r}_{+,1}&=&-\frac{\widetilde{A}}{4}+\frac{i}{4\widetilde{A}}\sqrt{\widetilde{A}^4+\frac{16k}{|\lambda|}\widetilde{A}},\quad
\widetilde{r}_{+,2}=-\frac{\widetilde{A}}{4}-\frac{i}{4\widetilde{A}}\sqrt{\widetilde{A}^4+\frac{16k}{|\lambda|}\widetilde{A}},\\
\widetilde{r}_{-,1}&=&\frac{\widetilde{A}}{4}-\frac{1}{4\widetilde{A}}\sqrt{-\widetilde{A}^4+\frac{16k}{|\lambda|}\widetilde{A}},\quad
\widetilde{r}_{-,2}=\frac{\widetilde{A}}{4}+\frac{1}{4\widetilde{A}}\sqrt{-\widetilde{A}^4+\frac{16k}{|\lambda|}\widetilde{A}}.
\end{eqnarray}
Hence, the turning points are $r_A=\widetilde{r}_{-,1}$ and $r_B=\widetilde{r}_{-,2}$. In the zero energy case $\mathcal{E}=0$, the reality condition reads
\begin{equation}
-V_{eff}(r)=2|\lambda|(r-r_A)(r_B-r)\left[(r-a)^2+b^2\right]\geq 0,\quad
a=-\frac{\widetilde{A}}{4},\quad
b=\frac{1}{4\widetilde{A}}\sqrt{\widetilde{A}^4+\frac{16k}{|\lambda|}\widetilde{A}}
\end{equation}
with $r_A\leq r\leq r_B$. To construct the LRL-vector associated to bounded trajectories, we compute the solution of the differential equation (\ref{equah}) with
\begin{equation}
P_1(r)=-\frac{1}{r}+\frac{12|\lambda|r^3-3k}{2|\lambda|r^4-2kr+\ell^2},\quad
P_2(r)=\frac{6|\lambda|r^2}{2|\lambda|r^4-2kr+\ell^2}.
\end{equation}
If we employ the ansatz 
\begin{equation}\label{ansatzqds}
h(r)=\frac{re^{\int_r^{r_B} w(\tau)d\tau}}{\sqrt{2|\lambda|r^4-2kr+\ell^2}},
\end{equation}
we obtain the following nonlinear first order differential equation for the unknown function $w(r)$, namely
\begin{equation}
r^2(2|\lambda|r^4-2kr+\ell^2)\left[\frac{dw}{dr}+w^2(r)\right]+
r(6|\lambda|r^4-3kr+\ell^2)w(r)-\ell^2=0.
\end{equation}
It can be easily checked with Maple that the above equation admits the following solutions
\begin{equation}
w_\pm(r)=\pm\frac{\ell}{r\sqrt{2|\lambda|r^4-2kr+\ell^2}}=\mp\frac{i\ell/\sqrt{2|\lambda|}}{r\sqrt{(r-r_A)(r_B-r)\left[(r-a)^2+b^2\right]}}.
\end{equation}
Hence, two complex linearly independent solutions are
\begin{equation}\label{olivo}
h_{\pm,\mathbb{C}}(r)=\frac{ir}{\sqrt{-2|\lambda|r^4+2kr-\ell^2}}\mbox{exp}\left(\pm\frac{i\ell}{\sqrt{2|\lambda|}}\int_r^{r_B}\frac{d\tau}{\tau\sqrt{(\tau-r_A)(r_B-\tau)\left[(\tau-a)^2+b^2\right]}}\right).
\end{equation}
In order to compute the above integral, we first rewrite it as follows
\begin{equation}
\int_r^{r_B}\frac{d\tau}{\tau\sqrt{(\tau-r_A)(r_B-\tau)\left[(\tau-a)^2+b^2\right]}}=\int_r^{r_B}\frac{R(\tau)}{Y(\tau)}d\tau,\quad R(\tau)=\frac{1}{\tau},\quad
Y(\tau)=\sqrt{S_1(\tau)S_2(\tau)}
\end{equation}
with
\begin{equation}
S_1(\tau)=-\tau^2+(r_A+r_B)\tau-r_A r_B,\quad
S_2(\tau)=\tau^2-2a\tau+a^2+b^2
\end{equation}
and then, we apply a technique outlined in \cite{Lawden}. To this purpose, we introduce the quadratic expression
\begin{equation}
S_1+\widehat{\lambda}S_2=(\widehat{\lambda}-1)\tau^2+2\left(\frac{r_A+r_B}{2}-a\widehat{\lambda}\right)\tau+(a^2+b^2)\widehat{\lambda}-r_A r_B
\end{equation}
which becomes a perfect square if
\begin{equation}
\widehat{D}(\widehat{\lambda})=\left(\frac{r_A+r_B}{2}-a\widehat{\lambda}\right)^2-(\widehat{\lambda}-1)\left[(a^2+b^2)\widehat{\lambda}-r_A r_B\right]=0.
\end{equation}
It is not difficult to verify that the roots of the equation $\widehat{D}(\widehat{\lambda})=0$ are
\begin{equation}
\widehat{\lambda}_1=\frac{3|\lambda|\widetilde{A}^3+2\sqrt{2\lambda^2\widetilde{A}^6+64k^2}}{|\lambda|\widetilde{A}^3+16k},\quad
\widehat{\lambda}_2=\frac{3|\lambda|\widetilde{A}^3-2\sqrt{2\lambda^2\widetilde{A}^6+64k^2}}{|\lambda|\widetilde{A}^3+16k}.
\end{equation}
When $\widehat{D}(\widehat{\lambda})=0$, the corresponding roots of $S_1+\widehat{\lambda}S_2$ have algebraic multiplicity two and can be obtained from the expression
\begin{equation}
\tau=\frac{2a\widehat{\lambda}-(r_A+r_B)}{2(\widehat{\lambda}-1)}.
\end{equation}
More precisely, to $\widehat{\lambda}_1$ and $\widehat{\lambda}_2$ there correspond the roots
\begin{equation}
\tau_1=-\frac{\widetilde{A}\left(2|\lambda|\widetilde{A}^3+\sqrt{2\lambda^2\widetilde{A}^6+64k^2}+8k\right)}{4\left(|\lambda|\widetilde{A}^3+\sqrt{2\lambda^2\widetilde{A}^6+64k^2}-8k\right)},\quad
\tau_2=-\frac{\widetilde{A}\left(2|\lambda|\widetilde{A}^3-\sqrt{2\lambda^2\widetilde{A}^6+64k^2}+8k\right)}{4\left(|\lambda|\widetilde{A}^3-\sqrt{2\lambda^2\widetilde{A}^6+64k^2}-8k\right)}.
\end{equation}
Then,
\begin{equation}
S_1+\widehat{\lambda}_1 S_2=(\widehat{\lambda}_1-1)(\tau-\tau_1)^2,\quad
S_1+\widehat{\lambda}_2 S_2=(\widehat{\lambda}_2-1)(\tau-\tau_2)^2.
\end{equation}
Solving the above system for $S_1$ and $S_2$, these quadratics can be expressed as
\begin{equation}
S_1=A_1(\tau-\tau_1)^2+B_1(\tau-\tau_2)^2,\quad
S_2=A_2(\tau-\tau_1)^2+B_2(\tau-\tau_2)^2
\end{equation}
with
\begin{equation}
A_1=\frac{\widehat{\lambda}_2-\widehat{\lambda}_1\widehat{\lambda}_2}{\widehat{\lambda}_1-\widehat{\lambda}_2},\quad
B_1=\frac{\widehat{\lambda}_1\widehat{\lambda}_2-\widehat{\lambda}_1}{\widehat{\lambda}_1-\widehat{\lambda}_2},\quad
A_2=\frac{\widehat{\lambda}_1-1}{\widehat{\lambda}_1-\widehat{\lambda}_2},\quad
B_2=\frac{1-\widehat{\lambda}_2}{\widehat{\lambda}_1-\widehat{\lambda}_2}
\end{equation}
or equivalently
\begin{eqnarray}
A_1&=&\frac{\sqrt{2}(2\sqrt{2\lambda^2\widetilde{A}^6+64k^2}-3|\lambda|\widetilde{A}^3)(|\lambda|\widetilde{A}^3+\sqrt{2\lambda^2\widetilde{A}^6+64k^2}-8k)}{4(|\lambda|\widetilde{A}^3+16k)\sqrt{\lambda^2\widetilde{A}^6+32k^2}},\\
B_1&=&-\frac{\sqrt{2}(3|\lambda|\widetilde{A}^3+2\sqrt{2\lambda^2\widetilde{A}^6+64k^2})(8k+\sqrt{2\lambda^2\widetilde{A}^6+64k^2}-|\lambda|\widetilde{A}^3)}{4(|\lambda|\widetilde{A}^3+16k)\sqrt{\lambda^2\widetilde{A}^6+32k^2}},\\
A_2&=&\frac{\sqrt{2}(|\lambda|\widetilde{A}^3+\sqrt{2\lambda^2\widetilde{A}^6+64k^2}-8k)}{4\sqrt{\lambda^2\widetilde{A}^6+32k^2}},\quad
B_2=\frac{\sqrt{2}(8k+\sqrt{2\lambda^2\widetilde{A}^6+64k^2}-|\lambda|\widetilde{A}^3)}{4\sqrt{\lambda^2\widetilde{A}^6+32k^2}}.
\end{eqnarray}
It is not difficult to check that all coefficients are positive except $B_1$. Let us introduce the coordinate transformation
\begin{equation}
t=\frac{\tau-\tau_2}{\tau-\tau_1}.
\end{equation}
Such a transformation maps the pair  $(r_A,r_B)$ into $(t_A,t_B)$. A  simple check with Maple shows that the identities $t_A+t_B=0$ with $t_A<0<t_B$ and $A_1/|B_1|=t_A^2=t_B^2$ hold true.  Moreover,
\begin{equation}
S_1=(\tau_2-\tau_1)^2\frac{A_1+B_1 t^2}{(t-1)^2},\quad
S_2=(\tau_2-\tau_1)^2\frac{A_2+B_2 t^2}{(t-1)^2}
\end{equation}
and taking into account that
\begin{equation}
\tau_2-\tau_1=\frac{\sqrt{2\lambda^2\widetilde{A}^6+64k^2}}{2|\lambda|\widetilde{A}^2}>0
\end{equation}
we find that
\begin{equation}\label{inta}
\int_r^{r_B}\frac{R_1(\tau)}{Y(\tau)}d\tau=\frac{1}{\tau_2-\tau_1}\int_{t(r)}^{t_B}\frac{t-1}{\tau_1 t-\tau_2}\frac{dt}{\sqrt{X(t)}},\quad
X(t)=(A_1+B_1 t^2)(A_2+B_2 t^2).
\end{equation}
If we introduce the decomposition
\begin{equation}
\frac{t-1}{\tau_1 t-\tau_2}=\frac{\tau_2-\tau_1 t^2}{\tau_2^2-\tau_1^2 t^2}-\frac{\tau_2-\tau_1}{\tau_2^2-\tau_1^2 t^2}
\end{equation}
and let
\begin{equation}
\widetilde{a}=\sqrt{\frac{A_2}{B_2}},\quad
\widetilde{b}=\sqrt{\frac{A_1}{|B_1|}}=t_B,\quad
\gamma=\left(\frac{\tau_2}{\tau_1}\right)^2=\frac{\left(16k\sqrt{2\lambda^2\widetilde{A}^6+64k^2}-2\lambda^2\widetilde{A}^6-128k^2\right)^2}{4\lambda^4\widetilde{A}^{12}},
\end{equation}
the computation of the integral (\ref{inta}) breaks down into the evaluation of the following three elementary integrals
\begin{equation}\label{inta1}
\sqrt{|B_1|B_2}\int_r^{r_B}\frac{R(\tau)}{Y(\tau)}d\tau=\frac{1}{\tau_1(\tau_2-\tau_1)}\underbrace{\int_{t(r)}^{\widetilde{b}}\frac{dt}{\sqrt{\widetilde{X}(t)}}}_{(I)}
+\frac{\tau_2}{\tau_1^3}\underbrace{\int_{t(r)}^{\widetilde{b}}\frac{dt}{(t^2-\gamma)\sqrt{\widetilde{X}(t)}}}_{(II)}+\frac{1}{\tau_1^2}\underbrace{\int_{t(r)}^{\widetilde{b}}\frac{tdt}{(t^2-\gamma)\sqrt{\widetilde{X}(t)}}}_{(III)}
\end{equation}
with $\widetilde{X}(t)=(\widetilde{a}^2+t^2)(\widetilde{b}^2-t^2)$. The first integral is simply an elliptic integral of the first kind and can be computed by means of 3.152.4 in \cite{Grad}. We obtain
\begin{equation}\label{fin1}
\int_{t(r)}^{\widetilde{b}}\frac{dt}{\sqrt{\widetilde{X}(t)}}=\frac{F\left(\sin{\widetilde{\varphi}},\widetilde{\kappa}\right)}{\sqrt{\widetilde{a}^2+\widetilde{b}^2}},\quad
\widetilde{\varphi}=\arccos{\left(\frac{t(r)}{\widetilde{b}}\right)},\quad
\widetilde{\kappa}=\frac{\widetilde{b}}{\sqrt{\widetilde{a}^2+\widetilde{b}^2}}.
\end{equation}
Concerning the second integral in (\ref{inta1}), 3.157.4 in \cite{Grad} gives
\begin{equation}
\int_{t(r)}^{\widetilde{b}}\frac{dt}{(t^2-\gamma)\sqrt{\widetilde{X}(t)}}=-\frac{\Pi\left(\sin{\widetilde{\varphi}},\widetilde{\xi},\widetilde{\kappa}\right)}{(\gamma-\widetilde{b}^2)\sqrt{\widetilde{a}^2+\widetilde{b}^2}},\quad
\widetilde{\xi}=\frac{\widetilde{b}^2}{\widetilde{b}^2-\gamma},
\end{equation}
where $\Pi$ denotes the elliptic integral of the third kind,  $\widetilde{\xi}$ is the so-called parameter of the aforementioned integral and $\widetilde{\varphi}$ is given as in (\ref{fin1}). Finally, the third integral in (\ref{inta1}) can be easily evaluated by first introducing the transformation $t^2=x$ followed by the Euler substitution $\sqrt{(\widetilde{a}^2+x)(\widetilde{b}^2-x)}=s(x+\widetilde{a}^2)$. In this case, we find that
\begin{equation}
\int_{t(r)}^{\widetilde{b}}\frac{tdt}{(t^2-\gamma)\sqrt{\widetilde{X}(t)}}=-\frac{1}{\sqrt{(\widetilde{a}^2+\gamma)(\gamma-\widetilde{b}^2)}}\arctan{\left(\sqrt{\frac{(\widetilde{a}^2+\gamma)[\widetilde{b}^2-t^2(r)]}{(\gamma-\widetilde{b}^2)[t^2(r)+\widetilde{a}^2]}}\right)}
\end{equation}
with
\begin{equation}
\frac{(\widetilde{a}^2+\gamma)[\widetilde{b}^2-t^2(r)]}{(\gamma-\widetilde{b}^2)[t^2(r)+\widetilde{a}^2]}=\frac{\lambda^2\widetilde{A}^6-8k|\lambda|\widetilde{A}^3+128k^2}{\lambda^2\widetilde{A}^6+8k|\lambda|\widetilde{A}^3+128k^2}\frac{-r^2+(r_A+r_B)r-r_A r_B}{(r-a)^2+b^2}.
\end{equation}
Note that the quantity $\gamma-\widetilde{b}^2$ is always positive and can never vanish. This can be easily seen by the following direct computation
\begin{equation}
\gamma-\widetilde{b}^2=\left(\frac{\tau_2}{\tau_1}\right)^2-\frac{\widetilde{A}_1}{|\widetilde{B}_1|}=\frac{4\mathfrak{Q}_1\sqrt{2\lambda^2\widetilde{A}^6+64k^2}}{\mathfrak{Q}_2^2\mathfrak{Q}_3^2\mathfrak{Q}_4\mathfrak{Q}_5},
\end{equation}
where
\begin{eqnarray}
\mathfrak{Q}_1&=&|\lambda|^5\widetilde{A}^{15}+56k\lambda^4\widetilde{A}^{12}+1280k^2|\lambda|^3\widetilde{A}^9+16384k^3\lambda^2\widetilde{A}^6+131072k^4|\lambda|\widetilde{A}^3+524288k^5,\\
\mathfrak{Q}_2&=&|\lambda|\widetilde{A}^3+\sqrt{2\lambda^2\widetilde{A}^6+64k^2}-8k,\quad
\mathfrak{Q}_3=2|\lambda|\widetilde{A}^3-\sqrt{2\lambda^2\widetilde{A}^6+64k^2}+8k,\\
\mathfrak{Q}_4&=&3|\lambda|\widetilde{A}^3+2\sqrt{2\lambda^2\widetilde{A}^6+64k^2},\quad
\mathfrak{Q}_5=8k+\sqrt{2\lambda^2\widetilde{A}^6+64k^2}-|\lambda|\widetilde{A}^3.
\end{eqnarray}
Clearly, $\mathfrak{Q}_1$ is positive while $\mathfrak{Q}_5>0$ from the trivial observation that $8k+\sqrt{2\lambda^2\widetilde{A}^6+64k^2}>|\lambda|\widetilde{A}^3$. Bringing everything together leads to the final result
\[
\int_r^{r_B}\frac{R(\tau)}{Y(\tau)}d\tau=\frac{1}{\sqrt{|B_1|B_2}}
\left[
\frac{F(\sin{\widetilde{\varphi}},\widetilde{\kappa})}{\tau_1(\tau_2-\tau_1)\sqrt{\widetilde{a}^2+\widetilde{b}^2}}-\frac{\tau_2\Pi(\sin{\widetilde{\varphi}},\widetilde{\xi},\widetilde{\kappa})}{\tau_1^3(\gamma-\widetilde{b}^2)\sqrt{\widetilde{a}^2+\widetilde{b}^2}}\right.
\]
\begin{equation}
\left.
-\frac{1}{\tau_1^2\sqrt{(\widetilde{a}^2+\gamma)(\gamma-\widetilde{b}^2)}}\arctan{\left(\sqrt{\frac{(\widetilde{a}^2+\gamma)[\widetilde{b}^2-t^2(r)]}{(\gamma-\widetilde{b}^2)[t^2(r)+\widetilde{a}^2]}}\right)}
\right].
\end{equation}
Let 
\begin{equation}
\mathcal{I}=\frac{\ell}{\sqrt{2|\lambda|}}\int_r^{r_B}\frac{R(\tau)}{Y(\tau)}d\tau.
\end{equation}
Then, two linearly independent solutions of (\ref{equah}) can be constructed with the help of (\ref{olivo}) as follows
\begin{equation}
h_1(r)=\frac{r\cos{\mathcal{I}}}{\sqrt{-2|\lambda|r^4+2kr-\ell^2}},\quad
h_2(r)=\frac{r\sin{\mathcal{I}}}{\sqrt{-2|\lambda|r^4+2kr-\ell^2}}.
\end{equation}
Given $h(r)$, the corresponding function $g(r)$ can be evaluated by means of (\ref{equaG}) as 
\begin{equation}
g(r)=\frac{2|\lambda|r^4-2kr+\ell^2}{r}\frac{dh}{dr}+\frac{2|\lambda|r^4+kr-\ell^2}{r^2}h(r).
\end{equation}
If we pick for instance $h(r)=h_1(r)$, a lengthy but straightforward computation shows that
\begin{equation}
g_1(r)=-\frac{\ell}{r}\sin{\mathcal{I}}.
\end{equation}
This completes the construction of the LRL-vector. Last but not least, if we consider the pair of functions $(h_1, g_1)$ and $(h_2, g_2)$, it is gratifying to observe that the modulus of the LRL-vector is simply $\mathcal{A}=\ell$.
{\bf{The restored expression of the generalised LRL vector for the example studied here is presented below in explicit closed form }}
\begin{equation}
 \boxed{\bm{\mathcal{A}}=rg(r){\widehat{\bf{r}}}+\ell r\dot{r}h(r){\widehat{\bf{r}}}_\bot,\quad\dot{r}=\pm\sqrt{-2V_{eff}(r)},\quad V_{eff}(r)=\frac{\ell^2}{2r^2}-\frac{k}{r}-\lambda r^2,\quad k>0,\quad\lambda<-\frac{27k^2}{512\ell^6},}
 \end{equation}
\begin{equation}
 \boxed{
 g(r)=\frac{\ell}{r}\left[c_1\sin{\mathcal{I}(r)}+c_2\cos{\mathcal{I}(r)}\right],\quad
 h(r)=\frac{r}{\sqrt{-2|\lambda|r^4+2kr-\ell^2}}\left[c_1\cos{\mathcal{I}(r)}+c_2\sin{\mathcal{I}(r)}\right],
 }
\end{equation}
\begin{equation}
\boxed{
\mathcal{I}(r)=\frac{\ell}{\sqrt{2|\lambda B_1|B_2}}
\left[
\frac{F(\sin{\widetilde{\varphi}},\widetilde{\kappa})}{\tau_1(\tau_2-\tau_1)\sqrt{\widetilde{a}^2+\widetilde{b}^2}}-\frac{\tau_2\Pi(\sin{\widetilde{\varphi}},\widetilde{\xi},\widetilde{\kappa})}{\tau_1^3(\gamma-\widetilde{b}^2)\sqrt{\widetilde{a}^2+\widetilde{b}^2}}
-\frac{\arctan{\Psi(r)}}{\tau_1^2\sqrt{(\widetilde{a}^2+\gamma)(\gamma-\widetilde{b}^2)}}
\right],
}
\end{equation}
\begin{equation}
\boxed{
\Psi(r)=\sqrt{\frac{\lambda^2\widetilde{A}^6-8k|\lambda|\widetilde{A}^3+128k^2}{\lambda^2\widetilde{A}^6+8k|\lambda|\widetilde{A}^3+128k^2}\frac{-r^2+(r_A+r_B)r-r_A r_B}{(r-a)^2+b^2}},\quad r_A=-\frac{\widetilde{A}_-}{4}+\frac{\sqrt{\widetilde{\Delta}_-}}{4\widetilde{A}_-},\quad
r_B=-\frac{\widetilde{A}_-}{4}-\frac{\sqrt{\widetilde{\Delta}_-}}{4\widetilde{A}_-},
}
\end{equation}
\begin{equation}
\boxed{
\widetilde{A}_{-}=-\widetilde{A}=-4\sqrt{\frac{\ell}{\sqrt{6|\lambda|}}\cosh{\frac{\widetilde{\phi}}{3}}},\quad\cosh{\widetilde{\phi}}=\frac{3\sqrt{6}k^2}{32\ell^3\sqrt{|\lambda|}},\quad a=-\frac{\widetilde{A}}{4},\quad
b=\frac{1}{4\widetilde{A}}\sqrt{\widetilde{A}^4+\frac{16k}{|\lambda|}\widetilde{A}}
}
\end{equation}
\begin{equation}
\boxed{
\widetilde{\Delta}_{-}=\widetilde{A}_{-}^4+4\widetilde{A}_{-}^3\left(\frac{2k}{|\lambda|}\right)^{1/3}+4\widetilde{A}_{-}^2\left(\frac{2k}{|\lambda|}\right)^{2/3}-16\widetilde{A}_{-}^2\widetilde{y}_1-32\widetilde{A}_{-} \widetilde{y}_1\left(\frac{2k}{|\lambda|}\right)^{1/3},\quad
\widetilde{y}_1=\left(\frac{k}{4|\lambda|}\right)^{2/3}+\frac{2\ell}{\sqrt{6|\lambda|}}\cosh{\frac{\widetilde{\phi}}{3}}
}
\end{equation}
\begin{equation}
\boxed{
\widetilde{\varphi}=\arccos{\left(\frac{t(r)}{\widetilde{b}}\right)},\quad
\widetilde{\kappa}=\frac{\widetilde{b}}{\sqrt{\widetilde{a}^2+\widetilde{b}^2}},\quad \widetilde{\xi}=\frac{\widetilde{b}^2}{\widetilde{b}^2-\gamma},\quad
t(r)=\frac{r-\tau_2}{r-\tau_1},\quad\widetilde{a}=\sqrt{\frac{A_2}{B_2}},\quad
\widetilde{b}=\sqrt{\frac{A_1}{|B_1|}},\quad\gamma=\left(\frac{\tau_2}{\tau_1}\right)^2,
}
\end{equation}
\begin{equation}
\boxed{
\tau_1=-\frac{\widetilde{A}\left(2|\lambda|\widetilde{A}^3+\sqrt{2\lambda^2\widetilde{A}^6+64k^2}+8k\right)}{4\left(|\lambda|\widetilde{A}^3+\sqrt{2\lambda^2\widetilde{A}^6+64k^2}-8k\right)},\quad
\tau_2=-\frac{\widetilde{A}\left(2|\lambda|\widetilde{A}^3-\sqrt{2\lambda^2\widetilde{A}^6+64k^2}+8k\right)}{4\left(|\lambda|\widetilde{A}^3-\sqrt{2\lambda^2\widetilde{A}^6+64k^2}-8k\right)},
}
\end{equation}
\begin{equation}
\boxed{
A_1=\frac{\sqrt{2}(2\sqrt{2\lambda^2\widetilde{A}^6+64k^2}-3|\lambda|\widetilde{A}^3)(|\lambda|\widetilde{A}^3+\sqrt{2\lambda^2\widetilde{A}^6+64k^2}-8k)}{4(|\lambda|\widetilde{A}^3+16k)\sqrt{\lambda^2\widetilde{A}^6+32k^2}},\quad
A_2=\frac{\sqrt{2}(|\lambda|\widetilde{A}^3+\sqrt{2\lambda^2\widetilde{A}^6+64k^2}-8k)}{4\sqrt{\lambda^2\widetilde{A}^6+32k^2}},
}
\end{equation}
\begin{equation}
\boxed{
B_1=-\frac{\sqrt{2}(3|\lambda|\widetilde{A}^3+2\sqrt{2\lambda^2\widetilde{A}^6+64k^2})(8k+\sqrt{2\lambda^2\widetilde{A}^6+64k^2}-|\lambda|\widetilde{A}^3)}{4(|\lambda|\widetilde{A}^3+16k)\sqrt{\lambda^2\widetilde{A}^6+32k^2}},\quad
B_2=\frac{\sqrt{2}(8k+\sqrt{2\lambda^2\widetilde{A}^6+64k^2}-|\lambda|\widetilde{A}^3)}{4\sqrt{\lambda^2\widetilde{A}^6+32k^2}}
}
\end{equation}
{\bf{For the definition of the elliptic function $F$ and $\Pi$ we refer to the glossary in Appendix C.}}

\subsubsection{The de Sitter case}
\begin{figure}[!ht]\label{ABD1}
\centering
    \includegraphics[width=0.5\textwidth]{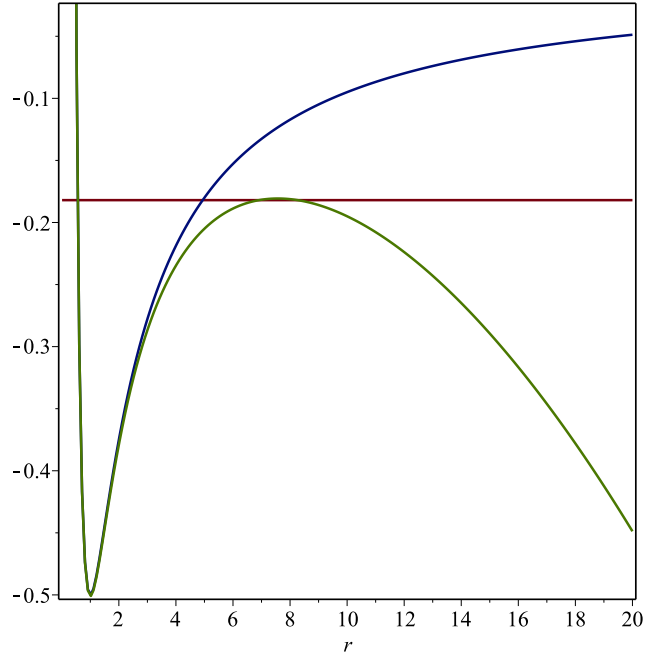}
\caption{\label{figureNM}
Plots of the Kepler (blue) and the de Sitter effective potentials (green) for $k=\ell=1$ and $\lambda=10^{-3}$. The horizontal black line slightly below the maximum in the de Sitter potential corresponds to a choice of the particle energy given by $\mathcal{E}=-0.182$. Regarding the de Sitter potential the tuning points are at $r_1=0.556$ and $r_2=6.909$ while $r_0=-15.734$ and $r_3=8.270$. The minimum and maximum are located at $r_m=1.002$ and $r_M=7.571$, respectively. The de Sitter potential at the maximum is $V_{eff}(r_M)=-0.181$. The turning points in the Kepler potential for $\mathcal{E}=-0.182$ are $r_{1,K}=0.556$ and $r_{2,K}=4.938$.}
\end{figure}

The corresponding effective potential is
\begin{equation}\label{dsV}
V_{eff}(r)=\frac{\ell^2}{2r^2}-\frac{k}{r}-\lambda r^2=\frac{\ell^2-2kr-2\lambda r^4}{2r^2}
\end{equation}
with $\lambda>0$. According to Descartes' rule of sign the quartic in (\ref{dsV}) admits one negative and one positive real root. This signalizes that when the effective potential has a maximum, it must be below the positive $r$-axis. According to \cite{Arnon} the quartic in (\ref{dsV}) has always two distinct real roots and two complex conjugate
roots because the quantity 
\begin{equation}
\delta(0, k/\lambda, -\ell^2/2\lambda)=-\frac{27k^4+32\lambda\ell^6}{\lambda^4}
\end{equation}
already defined in (\ref{deltaA}) is negative for any $\lambda>0$. If we impose $dV_{eff}/dr=0$, we end up with the quartic equation
\begin{equation}\label{kva}
-2\lambda r^4+kr-\ell^2=0. 
\end{equation}
It will display two distinct real roots and two complex conjugate roots if
\begin{equation}
\widetilde{\delta}(0, -k/2\lambda, \ell^2/2\lambda)=\frac{512\lambda\ell^6-27k^4}{16\lambda^4}<0
\end{equation}
or equivalently
\begin{equation}\label{newboundp}
0<\lambda<\frac{27k^4}{512\ell^6}.
\end{equation}
In order to determine the minimum and maximum in the effective potential, we observe that the coefficient going with the cubic power in (\ref{kva}) vanishes and therefore, according to \cite{Bron} we first need to apply the variable transformation
\begin{equation}
r=u+\frac{1}{2}\left(\frac{k}{\lambda}\right)^{1/3}
\end{equation}
to (\ref{kva}) leading to
\begin{equation}\label{o}
u^4+2\left(\frac{k}{\lambda}\right)^{1/3}u^3+\frac{3}{2}\left(\frac{k}{\lambda}\right)^{2/3}u^2+\frac{\ell^2}{2\lambda}-\frac{3}{16}\left(\frac{k}{\lambda}\right)^{4/3}=0.
\end{equation}
The roots of the above equation can be written in terms of one of the real roots of the cubic \cite{Bron}
\begin{equation}\label{1}
8y^3-6\left(\frac{k}{\lambda}\right)^{2/3}y^2-\left[\frac{4\ell^2}{\lambda}-\frac{3}{2}\left(\frac{k}{\lambda}\right)^{4/3}\right]y+\frac{1}{8}\left(\frac{k}{\lambda}\right)^{2/3}\left[\frac{8\ell^2}{\lambda}-3\left(\frac{k}{\lambda}\right)^{4/3}\right]=0.
\end{equation}
Substituting the transformation
\begin{equation}
y=v+\frac{1}{4}\left(\frac{k}{\lambda}\right)^{2/3}
\end{equation}
into (\ref{1}) we end up with the reduced cubic
\begin{equation}
v^3+3pv+2q=0,\quad
p=-\frac{\ell^2}{6\lambda},\quad
q=-\frac{k^2}{64\lambda^2}
\end{equation}
whose discriminant
\begin{equation}
D=q^2+p^3=\frac{27k^4-512\lambda\ell^6}{11059\lambda^4}
\end{equation}
is always negative due to the condition (\ref{newboundp}). Since $D$, $p$ and $q$ are all negative, the cubic (\ref{1}) has only one real root which is computed to be \cite{Bron} 
\begin{equation}
v_1=\frac{2\ell}{\sqrt{6\lambda}}\cosh{\frac{\alpha}{3}},\quad
\cosh{\alpha}=\frac{3\sqrt{6}k^2}{32\ell^3\sqrt{\lambda}}.
\end{equation}
The corresponding real root of (\ref{1}) is
\begin{equation}
y_1=\frac{1}{4}\left(\frac{k}{\lambda}\right)^{2/3}+\frac{2\ell}{\sqrt{6\lambda}}\cosh{\frac{\alpha}{3}}.
\end{equation} 
Moreover, the roots of (\ref{o}) coincide with the roots of the equations \cite{Bron}
\begin{equation}
\eta^2+\left[A_\pm+2\left(\frac{k}{|\lambda|}\right)^{1/3}\right]\frac{\eta}{2}+y_1\left[1+\frac{2}{A_\pm}\left(\frac{k}{\lambda}\right)^{1/3}\right]=0,\quad
A_\pm=\pm\sqrt{8y_1-2\left(\frac{k}{\lambda}\right)^{2/3}}=\pm4\sqrt{\frac{\ell}{\sqrt{6\lambda}}\cosh{\frac{\alpha}{3}}}
\end{equation}
and are
\begin{eqnarray}
u_{+,1}&=&-\frac{1}{4A_+}\left[A_+^2+2\left(\frac{k}{\lambda}\right)^{1/3}A_+-\sqrt{\Delta_+}\right],\quad
u_{+,2}=-\frac{1}{4A_+}\left[A_+^2+2\left(\frac{k}{\lambda}\right)^{1/3}A_++\sqrt{\Delta_+}\right]\\
u_{-,1}&=&-\frac{1}{4A_-}\left[A_-^2+2\left(\frac{k}{\lambda}\right)^{1/3}A_--\sqrt{\Delta_-}\right],\quad
u_{-,2}=-\frac{1}{4A_-}\left[A_-^2+2\left(\frac{k}{\lambda}\right)^{1/3}A_-+\sqrt{\Delta_-}\right]
\end{eqnarray}
with
\begin{equation}
\Delta_\pm=A_\pm^4+4A_\pm^3\left(\frac{k}{\lambda}\right)^{1/3}+4A_\pm^2\left(\frac{k}{\lambda}\right)^{2/3}-16A_\pm^2 y_1+32A_\pm y_1\left(\frac{k}{\lambda}\right)^{1/3}.
\end{equation}
Transforming back to the variable $r$ gives
\begin{equation}
r_{+,1}=-\frac{A_+}{4}+\frac{\sqrt{\Delta_+}}{4A_+},\quad
r_{+,2}=-\frac{A_+}{4}-\frac{\sqrt{\Delta_+}}{4A_+},\quad
r_{-,1}=-\frac{A_-}{4}+\frac{\sqrt{\Delta_-}}{4A_-},\quad
r_{-,2}=-\frac{A_-}{4}-\frac{\sqrt{\Delta_-}}{4A_-}.
\end{equation}
Invoking the following identities 
\begin{equation}
A_\pm^4+4A_\pm^2\left(\frac{k}{\lambda}\right)^{2/3}-16A_\pm^2 y_1=-A_\pm^4,\quad
4A_\pm^3\left(\frac{k}{\lambda}\right)^{1/3}-32A_\pm y_1\left(\frac{k}{\lambda}\right)^{1/3}=\mp\frac{8k}{\lambda}A
\end{equation}
with
\begin{equation}
A=4\sqrt{\frac{\ell}{\sqrt{6\lambda}}\cosh{\frac{\alpha}{3}}}
\end{equation}
leads to the following representation of the roots of (\ref{kva})
\begin{eqnarray}
r_{+,1}&=&-\frac{A}{4}+\frac{i}{4A}\sqrt{A^4+\frac{8k}{\lambda}A},\quad
r_{+,2}=-\frac{A}{4}-\frac{i}{4A}\sqrt{A^4+\frac{8k}{\lambda}A},\\
r_{-,1}&=&\frac{A}{4}-\frac{1}{4A}\sqrt{\frac{8k}{\lambda}A-A^4},\quad
r_{-,2}=\frac{A}{4}+\frac{1}{4A}\sqrt{\frac{8k}{\lambda}A-A^4}.\label{minmax}
\end{eqnarray}
We conclude that $r_{-,1}=r_m$ and $r_{-,2}=r_M$ represent the positions of the minimum and maximum of the effective potential, respectively. Moreover,  the positivity of the quantity $8kA/\lambda-A^4$ is ensured by (\ref{newboundp}). Let us impose the reality condition
\begin{equation}\label{rcc}
\mathcal{E}-V_{eff}(r)=\frac{2\lambda r^4+2\mathcal{E}r^2+2kr-\ell^2}{2r^2}>0,\quad V_{eff}(r_m)<\mathcal{E}<V_{eff}(r_M)<0.
\end{equation}
In this case, the quartic in (\ref{rcc}) admits four distinct real roots $r_0<0<r_1<r_2<r_3$ and 
\begin{equation}
\mathcal{E}-V_{eff}(r)=\frac{\lambda}{r^2}(r-r_0)(r-r_1)(r_2-r)(r_3-r)>0,\quad r_1<r<r_2
\end{equation}
where $r_1$ and $r_2$ are the turning points in the particle trajectory. In order to compute them, we first observe that the coefficient going with the cubic power in the quartic polynomial in (\ref{rcc}) is zero. Hence, according to \cite{Bron} we introduce the transformation
\begin{equation}
r=\widetilde{u}+\left(\frac{|\mathcal{E}|}{6\lambda}\right)^{1/2}
\end{equation}
in the aforementioned quartic and we end up with
\begin{equation}\label{aid}
\widetilde{u}^4+4\left(\frac{|\mathcal{E}|}{6\lambda}\right)^{1/2}\widetilde{u}^3+\left[\frac{k}{\lambda}-\frac{2}{9}\sqrt{6}\left(\frac{|\mathcal{E}|}{\lambda}\right)^{3/2}\right]\widetilde{u}+\frac{k}{\lambda}\left(\frac{|\mathcal{E}|}{6\lambda}\right)^{1/2}-\frac{\ell^2}{2\lambda}-\frac{5\mathcal{E}^2}{36\lambda^2}=0.
\end{equation}
The roots of the above equation can be expressed in terms of one of the real roots of the cubic \cite{Bron}
\begin{equation}
y^3+3py+2q=0,\quad
p=\frac{6\lambda\ell^2-\mathcal{E}^2}{36\lambda^2},\quad
q=\frac{36|\mathcal{E}|\lambda\ell^2+2|\mathcal{E}|^3-27\lambda k^2}{432\lambda^3}.
\end{equation}
Let $y_0$ be such a root. Then, the roots of (\ref{aid}) coincide with the roots of the equation \cite{Bron}
\begin{equation}
\eta^2+\left[\widetilde{B}_\pm+4\left(\frac{|\mathcal{E}|}{6\lambda}\right)^{1/2}\right]\frac{\eta}{2}+y_1+\frac{1}{B_\pm}\left[4y_1\left(\frac{|\mathcal{E}|}{6\lambda}\right)^{1/2}-\frac{k}{\lambda}+\frac{2}{9}\sqrt{6}\left(\frac{|\mathcal{E}|}{\lambda}\right)^{3/2}\right]=0,\quad
\widetilde{B}_\pm=\pm\sqrt{8\widetilde{y}_1+\frac{8|\mathcal{E}|}{3\lambda}}.
\end{equation}
To construct the LRL-vector associated to bounded trajectories, we compute the solution of the differential equation (\ref{equah}) with
\begin{equation}
P_1(r)=-\frac{1}{r}+\frac{12\lambda r^3+6\mathcal{E}r+3k}{2\lambda r^4+2\mathcal{E}r^2+2kr-\ell^2},\quad
P_2(r)=\frac{6\lambda r^2}{2\lambda r^4+2\mathcal{E}r^2+2kr-\ell^2}.
\end{equation}
If we employ the ansatz 
\begin{equation}\label{ansatzds1}
h(r)=\frac{r}{\sqrt{2\lambda r^4+2\mathcal{E}r^2+2kr-\ell^2}}\mbox{exp}\left(\int_{r_1}^{r} w(\tau)d\tau\right),
\end{equation}
we obtain the following nonlinear first order differential equation for the unknown function $w(r)$
\begin{equation}
r^2(2\lambda r^4+2\mathcal{E}r^2+2kr-\ell^2)\left[\frac{dw}{dr}+w^2(r)\right]+
r(6\lambda r^4+4\mathcal{E}r^2+3kr-\ell^2)w(r)+\ell^2=0
\end{equation}
which admits the solutions
\begin{equation}
w_\pm(r)=\pm\frac{i\ell}{r\sqrt{2\lambda r^4+2\mathcal{E}r^2+2kr-\ell^2}}=\pm\frac{i\ell/\sqrt{2\lambda}}{r\sqrt{(r-r_0)(r-r_1)(r_2-r)(r_3-r)}}.
\end{equation}
By means of 3.149.4 in \cite{Grad} we find that
\begin{equation}
\int_{r_1}^r\frac{d\tau}{\tau\sqrt{(\tau-r_0)(\tau-r_1)(r_2-\tau)(r_3-\tau)}}=\frac{2}{r_0 r_1\sqrt{(r_3-r_1)(r_2-r_0)}}\left[r_1 F(\sin{\widehat{\delta}},\widehat{\kappa})-(r_1-r_0)\Pi(\sin{\widehat{\delta}},\widehat{\xi},\widehat{\kappa})\right]
\end{equation}
where $\Pi$ and $F$ denote the incomplete elliptic integral of the third kind and the elliptic integral of the first kind, respectively, and
\begin{equation}\label{ampldS}
\sin{\widehat{\delta}}=\sqrt{\frac{(r_2-r_0)(r-r_1)}{(r_2-r_1)(r-r_0)}},\quad
\widehat{\xi}=\frac{r_0(r_2-r_1)}{r_1(r_2-r_0)},\quad
\widehat{\kappa}=\sqrt{\frac{(r_2-r_1)(r_3-r_0)}{(r_3-r_1)(r_2-r_0)}}.
\end{equation}
By this process we end up with two complex linearly independent solutions
\begin{equation}\label{XidS}
h_{\pm,\mathbb{C}}(r)=\frac{re^{\pi i\widehat{\Xi}(r)}}{\sqrt{2\lambda r^4+2\mathcal{E}r^2+2kr-\ell^2}},\quad
\widehat{\Xi}(r)=\frac{\sqrt{2}\ell\left[r_1 F(\sin{\widehat{\delta}},\widehat{\kappa})-(r_1-r_0)\Pi(\sin{\widehat{\delta}},\widehat{\xi},\widehat{\kappa})\right]}{r_0 r_1\sqrt{\lambda(r_3-r_1)(r_2-r_0)}}
\end{equation}
from which it is straightforward to extract the following linearly independent real solutions
\begin{equation}
h_1(r)=\frac{r\cos{\widehat{\Xi}(r)}}{\sqrt{2\lambda r^4+2\mathcal{E}r^2+2kr-\ell^2}},\quad
h_2(r)=\frac{r\sin{\widehat{\Xi}(r)}}{\sqrt{2\lambda r^4+2\mathcal{E}r^2+2kr-\ell^2}}.
\end{equation}
Once $h(r)$ is known, the corresponding function $g(r)$ can be evaluated by means of (\ref{equaG}) as 
\begin{equation}
g(r)=-\frac{2\lambda}{r} (r-r_0)(r-r_1)(r_2-r)(r_3-r)\frac{dh}{dr}+\frac{\ell^2-2kr-2\lambda r^4}{r^2}h(r).
\end{equation}
If we pick for instance $h(r)=h_1(r)$ and recall the identities
\begin{equation}
\frac{dF(\sin{\widehat{\delta}},\widehat{\kappa})}{dr}=-\frac{1}{2}\sqrt{\frac{(r_2-r_0)(r_3-r_1)}{(r-r_0)(r-r_1)(r_2-r)(r_3-r)}},\quad
\frac{d\Pi(\sin{\widehat{\delta}},\widehat{\xi},\widehat{\kappa})}{dr}=\frac{r_1(r-r_0)}{r(r_1-r_0)}\frac{dF(\sin{\widehat{\delta}},\widehat{\kappa})}{dr}
\end{equation}
a lengthy but straightforward computation shows that
\begin{equation}\label{g1dS}
g_1(r)=\frac{\ell}{r}\sin{\widehat{\Xi}(r)}
\end{equation}
and similarly, $g_2(r)=(\ell/r)\cos{\Xi(r)}$. Also in this case, as a consistency check we verified that the modulus of the LRL-vector is constant, more precisely $\mathcal{A}=\ell$. The above procedure gives the particle trajectory from $r_1$ to $r_2$. In particular, if we pick the function $g_1$, the particle path is described by the implicit representation
\begin{equation}\label{repres1}
\widehat{\Xi}(r)=\frac{\pi}{2}+\varphi.
\end{equation}
In order to get the corresponding part of the trajectory from the outer turning point to innermost, we can use 3.149.5 in \cite{Grad} to evaluate the integral 
\begin{equation}
\int_{r_2}^r\frac{d\tau}{\tau\sqrt{(\tau-r_0)(\tau-r_1)(r_2-\tau)(r_3-\tau)}}=-\frac{2}{r_2 r_3\sqrt{(r_3-r_1)(r_2-r_0)}}\left[r_2 F(\sin{\widetilde{\delta}},\widehat{\kappa})+(r_3-r_2)\Pi(\sin{\widetilde{\delta}},\widetilde{\xi},\widehat{\kappa})\right]
\end{equation}
where $\widehat{\kappa}$ is given as in (\ref{ampldS})  while
\begin{equation}
\sin{\widetilde{\delta}}=\sqrt{\frac{(r_3-r_1)(r_2-r)}{(r_2-r_1)(r_3-r)}},\quad
\widetilde{\xi}=\frac{r_3(r_2-r_1)}{r_2(r_3-r_1)}.
\end{equation}
Proceeding as before we end up with the trajectory equation
\begin{equation}\label{repres2}
\widetilde{\Xi}(r)=\varphi-\frac{\pi}{2},\quad
\widetilde{\Xi}(r)=\frac{\sqrt{2}\ell
\left[r_2 F(\sin{\widetilde{\delta}},\widehat{\kappa})+(r_3-r_2)\Pi(\sin{\widetilde{\delta}},\widetilde{\xi},\widehat{\kappa})\right]}{r_2 r_3\sqrt{\lambda(r_3-r_1)(r_2-r_0)}}.
\end{equation}
In order to plot the particle trajectory and compare it with the corresponding Kepler orbit, we consider the case with $k=\ell=1$ and $\lambda=10^{-3}$. More precisely, we consider the case where the particle energy is slightly below the maximum of the de Sitter effective potential (see Fig.~\ref{figureNM}). This approach allows us to probe into the effect of the cosmological constant over the corresponding Kepler orbit. 
\begin{figure}[!ht]\label{ABD2}
\centering
    \includegraphics[width=0.5\textwidth]{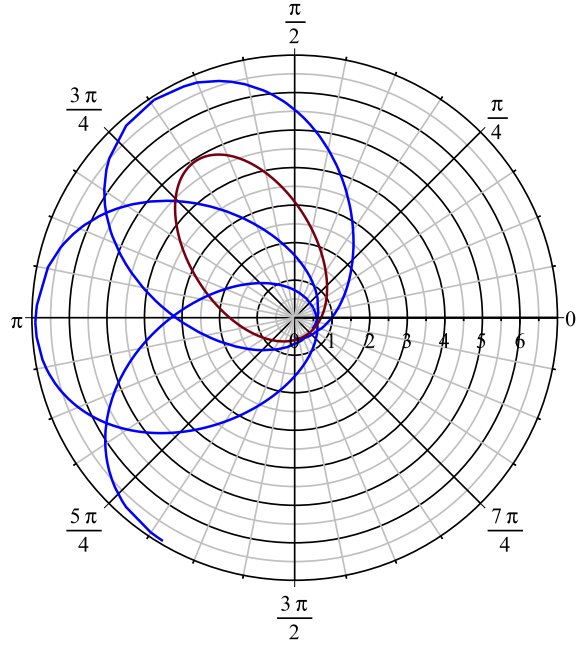}
\caption{\label{figureNMM}
Plots of the particle orbits in the de Sitter (blue) and Kepler (red)
case for $k=\ell=1$, $\lambda=10^{-3}$ and $\mathcal{E}=-0.182$. The
eccentricity of the Kepler orbit is $e=1.168$. The de Sitter
trajectory has been obtained by means of (\ref{repres1}) and
(\ref{repres2}) while the orbit in the Kepler potential is described
by (\ref{traj1}). A superficial inspection shows that the de Sitter
orbits are not really ellipses experiencing a precession, but have more an egg-shape form.}
\end{figure}
As it can be seen in Fig.~\ref{figureNMM}, even though in both the de Sitter and Kepler cases the particle had the same energy, we observe that in the de Sitter case the particle trajectory not only stretches further out in space than the Keplerian one but it also undergoes a precession. We conclude our analysis by considering the special case when the particle energy coincides with the maximum in the effective potential. Even though we already derived exact formulae for the maximum and minimum located at $r=r_m$ and $r=r_M$ (see equation (\ref{minmax})), it is convenient to exploit the fact that $\lambda\approx 10^{-52}~\mbox{m}^{-2}\ll 1$ because it allows to further simplify the treatment of the problem. Imposing that $dV_{eff}/dr=0$ leads to the problem of finding the real roots of the quartic equation $-2\lambda r^4+kr-\ell^2=0$. Looking at $\lambda$ as a small perturbation, it is immediately clear that if $\lambda=0$, a root of the unperturbed quartic is simply located at $\ell^2/k$, i.e. at the minimum of the Kepler problem. Applying a perturbative ansatz for $r_m$, it is straightforward to check that
\begin{equation}
r_m=\frac{\ell^2}{k}+\mathcal{O}(\lambda),\quad
V_{eff}(r_m)=-\frac{k^2}{2\ell^2}+\mathcal{O}(\lambda).
\end{equation}
The process of finding an expansion in $\lambda$ for $r=r_M$ is a little bit trickier. The key point is to observe that in the limit of $\lambda\to 0$ we have $r_M\to\infty$. This signalizes that the first term in the expansion must be proportional to $\lambda$ raised to some negative power. To this purpose, it is useful to introduce the trial rescaling $\delta=\delta(\lambda)$ and set $r=\delta(\lambda)R$ with $R$ being strictly of order one. Moreover, we require that as $\lambda\to 0$, $R$ is neither small nor large. Implementing this substitution into the aforementioned quartic leads to the equation
\begin{equation}\label{scaled_q}
-2\lambda\delta^4 R^4+k\delta R-\ell^2=0.
\end{equation}
Let LHS stand for the left hand side of the equation (\ref{scaled_q}). If $\delta\ll 1$, we would have $\mbox{LHS}=\mbox{small}+\mbox{small}-\ell^2=0$ which is never satisfied because $\ell\neq 0$. If $\delta=1$, $\mbox{LHS}=\mbox{small}+kR-\ell^2$ and we end up with the regular unperturbed root describing the minimum in the Kepler problem. Moreover, when $1\ll\delta\ll\lambda^{-1/3}$, we get $\mbox{LHS}/\delta=\mbox{small}+kR-\mbox{small}=0$. This is impossible because $R$ being of order one does not allow for $R=0$. Let $\delta=\lambda^{-1/3}$. Then, $\mbox{LHS}/\delta=-2R^4+kR-\mbox{small}=0$. This case is admissible and leads to the computation of the first perturbative term for $r_M$. Finally, for $\delta\gg\lambda^{-1/3}$ it results $\mbox{LHS}/\lambda\delta^4=-2R^4+\mbox{small}-\mbox{small}=0$ which can be neglected because $R$ is of order one. Hence, we found the distinguished rescaling $\delta=\lambda^{-1/3}$ and if we try the ansatz $r_M=(k/2\lambda)^{1/3}+\rho_0+\rho_1\lambda^{1/3}+\rho_2\lambda^{2/3}+\mathcal{O}(\lambda)$, it is easy to verify with Maple that
\begin{equation}\label{M_exp}
r_M=\left(\frac{k}{2\lambda}\right)^{1/3}-\frac{\ell^2}{3k}-\frac{2^{4/3}\ell^4}{9k^{7/3}}\lambda^{1/3}-\frac{20}{81}\frac{2^{2/3}\ell^6}{k^{11/3}}\lambda^{2/3}+\mathcal{O}(\lambda)
\end{equation}
for which
\begin{equation}
V_{eff}(r_M)=-3\left(\frac{k}{2}\right)^{2/3}\lambda^{1/3}+\frac{\ell^2}{2}\left(\frac{2}{k}\right)^{2/3}\lambda^{2/3}+\mathcal{O}(\lambda).
\end{equation}
We conclude that closed trajectories are allowed whenever the energy particle is chosen according to the following inequality
\begin{equation}
-\frac{k^2}{2\ell^2}\leq\mathcal{E}\leq -3\left(\frac{k}{2}\right)^{2/3}\lambda^{1/3}.
\end{equation}
As mentioned before, we are interested in the case $\mathcal{E}_M=V_{eff}(r_M)$. In this scenario, a remark is in order concerning the roots $r_0,\cdots,r_3$ of the equation $\mathcal{E}_M-V_{eff}(r)=0$. First of all, the turning point $r_2$   coalesces with $r_M$ and the same happens for $r_3$. Hence, $r_2=r_M=r_3$ and the asymptotic expansion for $r_2$ is given by (\ref{M_exp}). Regarding the turning point $r_1$ and the root $r_0$, it is not difficult to construct perturbative expansions for them. To find $r_1$, we start by noticing that it should be close to the intersection of the Kepler effective potential with the $r$-axis because as $\lambda\to 0$, then $r_M\to\infty$ and $V_{eff}(r_M)\to 0$ and therefore, $\mathcal{E}_M\to 0$ as well. By means of the educated guess $r_1=\ell^2/2k+a_1\lambda^{1/3}+a_2\lambda^{2/3}+\mathcal{O}(\lambda)$ we obtain the following expansion
\begin{equation}
r_1=\frac{\ell^2}{2k}+\frac{3}{8}\frac{2^{1/3}\ell^4}{k^{7/3}}\lambda^{1/3}+\frac{7}{16}\frac{2^{2/3}\ell^6}{k^{11/3}}\lambda^{2/3}+\mathcal{O}(\lambda).
\end{equation}
To construct an expansion for $r_0$, we can get inspiration from the fact that for $k=\ell=1$ and $\lambda=10^{-52}$, it turns out that modulo a sign both $r_M$ and $r_0$ are of order $10^{17}$ and moreover, $r_0\to-\infty$ and $r_M\to\infty$ as $\lambda\to 0$. If we try the ansatz $r_0=r_0^{(0)}\lambda^{-1/3}+r_0^{(1)}+r_0^{(2)}\lambda^{1/3}+r_0^{(3)}\lambda^{2/3}+\mathcal{O}(\lambda)$, we obtain
\begin{equation}\label{r0e}
r_0=-2^{2/3}\left(\frac{k}{\lambda}\right)^{1/3}+\frac{\ell^2}{6k}-\frac{2^{1/3}\ell^4}{216k^{7/3}}\lambda^{1/3}-\frac{5}{3888}\frac{2^{2/3}\ell^6}{k^{11/3}}\lambda^{2/3}+\mathcal{O}(\lambda).
\end{equation}
As a curious side note, the modulus of $r_0$ is practically twice the distance of $r_M$ to the origin of the $r$-axis. This can be easily seen from the following expansion obtained by means of (\ref{M_exp}) and (\ref{r0e}), namely
\begin{equation}
\frac{r_M}{r_0}=-\frac{1}{2}+\frac{2^{1/3}\ell^2}{8k^{4/3}}\lambda^{1/3}+\frac{53}{432}\frac{2^{2/3}\ell^4}{k^{8/3}}\lambda^{2/3}+\mathcal{O}(\lambda).
\end{equation}
Regarding the particle motion we need to distinguish between two cases. If the particle starts at $r=r_M$ without being perturbed, it will remain there since $r_M$ is an unstable equilibrium point in the effective potential. Hence, the trajectory will be a circle of radius $r_M$. In the case $k=\ell=1$ and $\lambda=10^{-52}$, we have $r_M=1.71\cdot 10^{17}~\mbox{m}\approx 5.5~\mbox{pc}\approx 18~\mbox{ly}$. On that other hand, if the particle motion starts at $r=r_1$, the trajectory can be easily derived with the help of corresponding LRL-vector ${\bf{\mathcal{A}}}$. We recall that ${\bf{\mathcal{A}}}\cdot{\bf{r}}=\ell r\cos{\varphi}$ and ${\bf{\mathcal{A}}}\cdot{\bf{r}}=r^2 g(r)$. If we pick for $g$ the expression given by (\ref{g1dS}), we end up with the following implicit representation of the trajectory, namely
\begin{equation}\label{eqmo}
\Xi(r)=\frac{\pi}{2}-\varphi
\end{equation}
with $\Xi(r)$ as given in (\ref{XidS}). Taking into account that $r_2=r_M=r_3$, the modulus, amplitude and characteristic parameter $\widehat{\xi}$ of the elliptic functions entering in the expression of $\Xi(r)$ and represented by (\ref{ampldS}) simplify as follows
\begin{equation}
\widehat{\kappa}=1,\quad
\sin{\widehat{\delta}}=\sqrt{\frac{(r_M+|r_0|)(r-r_1)}{(r_M-r_1)(r+|r_0|)}},\quad
\widehat{\xi}=-\frac{|r_0|(r_M-r_1)}{r_1(r_M+|r_0|)}
\end{equation}
with
\begin{eqnarray}
\sqrt{\frac{r_M+|r_0|}{r_M-r_1}}&=&\sqrt{3}+\frac{2^{1/3}\ell^2}{3^{1/2}k^{4/3}}\lambda^{1/3}+\frac{157\cdot 3^{1/2}}{324}\frac{2^{2/3}\ell^4}{k^{8/3}}\lambda^{2/3}+\mathcal{O}(\lambda),\\
\widehat{\xi}&=&-\frac{2^{5/3}k^{4/3}}{3\ell^2}\lambda^{-1/3}+2+\frac{425}{972}\frac{2^{1/3}\ell^2}{k^{4/3}}\lambda^{1/3}-\frac{39829}{34992}\frac{2^{2/3}\ell^4}{k^{8/3}}\lambda^{2/3}+\mathcal{O}(\lambda).
\end{eqnarray}
Note that for $\lambda\ll 1$, $\widehat{\xi}$ is a large negative number. Furthermore, $\widehat{\delta}=0$ for $r=r_1$ while $\widehat{\delta}=\pi/2$ at $r=r_M$.   Moreover, using the definition of the incomplete elliptic functions of the first and third kind, it is easy to verify that
\begin{equation}
F(\sin{\widehat{\delta}},1)=\arctanh{(\sin{\widehat{\delta}})},\quad
\Pi(\sin{\widehat{\delta}},\widehat{\xi},1)=\frac{\sqrt{|\widehat{\xi}|}}{1+|\widehat{\xi}|}\arctan{(|\widehat{\xi}|\sin{\widehat{\delta}})}+\frac{1}{1+|\widehat{\xi}|}\arctanh{\sin{\widehat{\delta}}}.
\end{equation}
A lengthy but straightforward computation shows that
\begin{equation}
\sqrt{\frac{\lambda}{2}}\frac{\Xi(r)}{\ell}=\frac{1}{r_M\sqrt{|r_0|r_1}}\arctan{\sqrt{\frac{|r_0|(r-r_1)}{r_1(r+|r_0|)}}}+\frac{1}{r_M\sqrt{(r_M+|r_0|)(r_M-r_1)}}\arctanh{\sqrt{\frac{(r_M+|r_0|)(r-r_1)}{(r_M-r_1)(r+|r_0|)}}}
\end{equation}
which replaced into (\ref{eqmo}) gives the following implicit equation for the motion trajectory
\begin{equation}
\frac{1}{\ell}\sqrt{\frac{\lambda}{2}}\left(\frac{\pi}{2}-\varphi\right)=\frac{1}{r_M\sqrt{|r_0|r_1}}\arctan{\sqrt{\frac{|r_0|(r-r_1)}{r_1(r+|r_0|)}}}+\frac{1}{r_M\sqrt{(r_M+|r_0|)(r_M-r_1)}}\arctanh{\sqrt{\frac{(r_M+|r_0|)(r-r_1)}{(r_M-r_1)(r+|r_0|)}}}.
\end{equation}
Note that when the particle starts at $r=r_1$, we have $\varphi=\pi/2$. For the motion trajectory we refer to Fig.~\ref{figureN}.
\begin{figure}[!ht]\label{dSS}
\centering
    \includegraphics[width=0.4\textwidth]{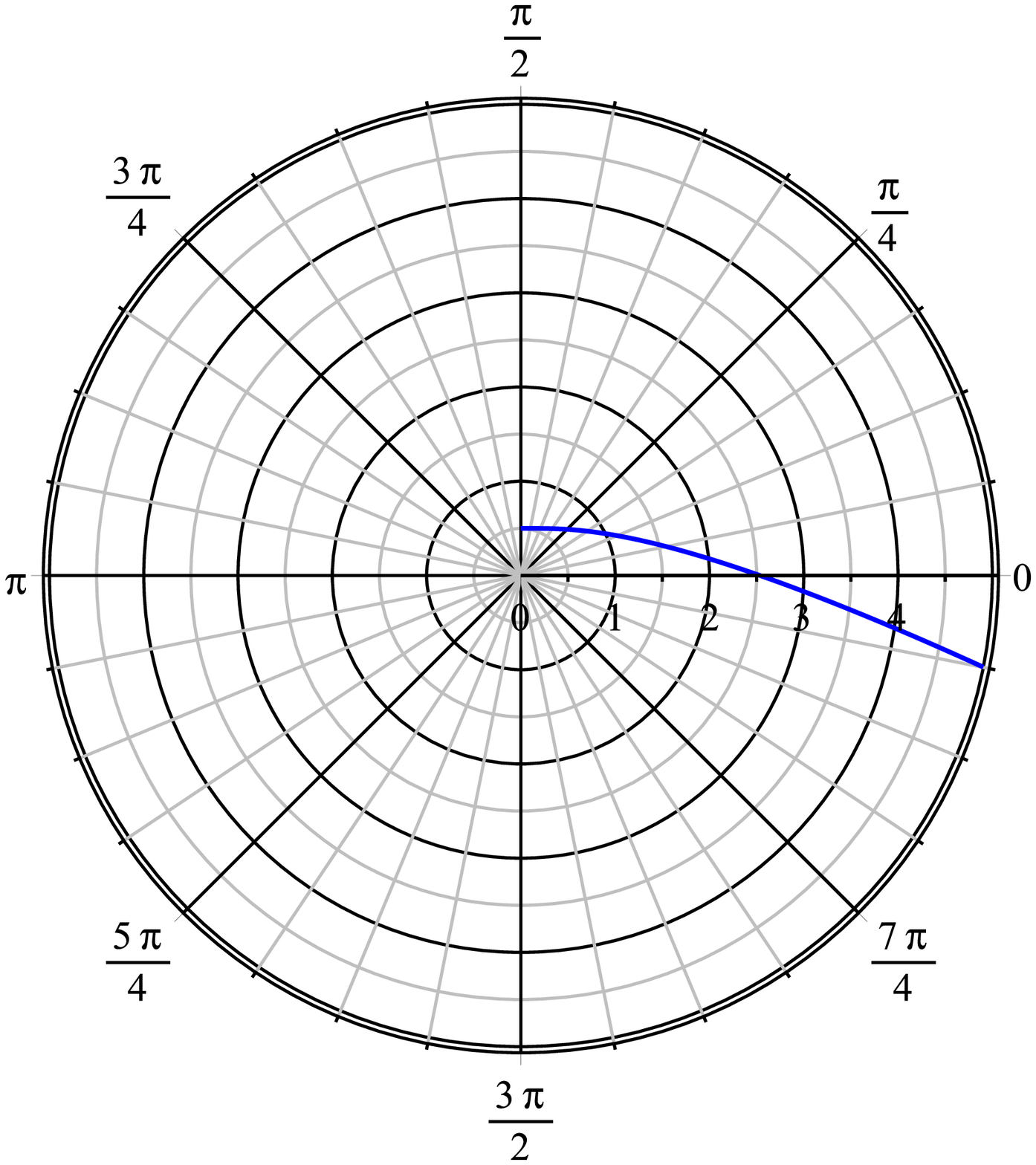}
    \includegraphics[width=0.4\textwidth]{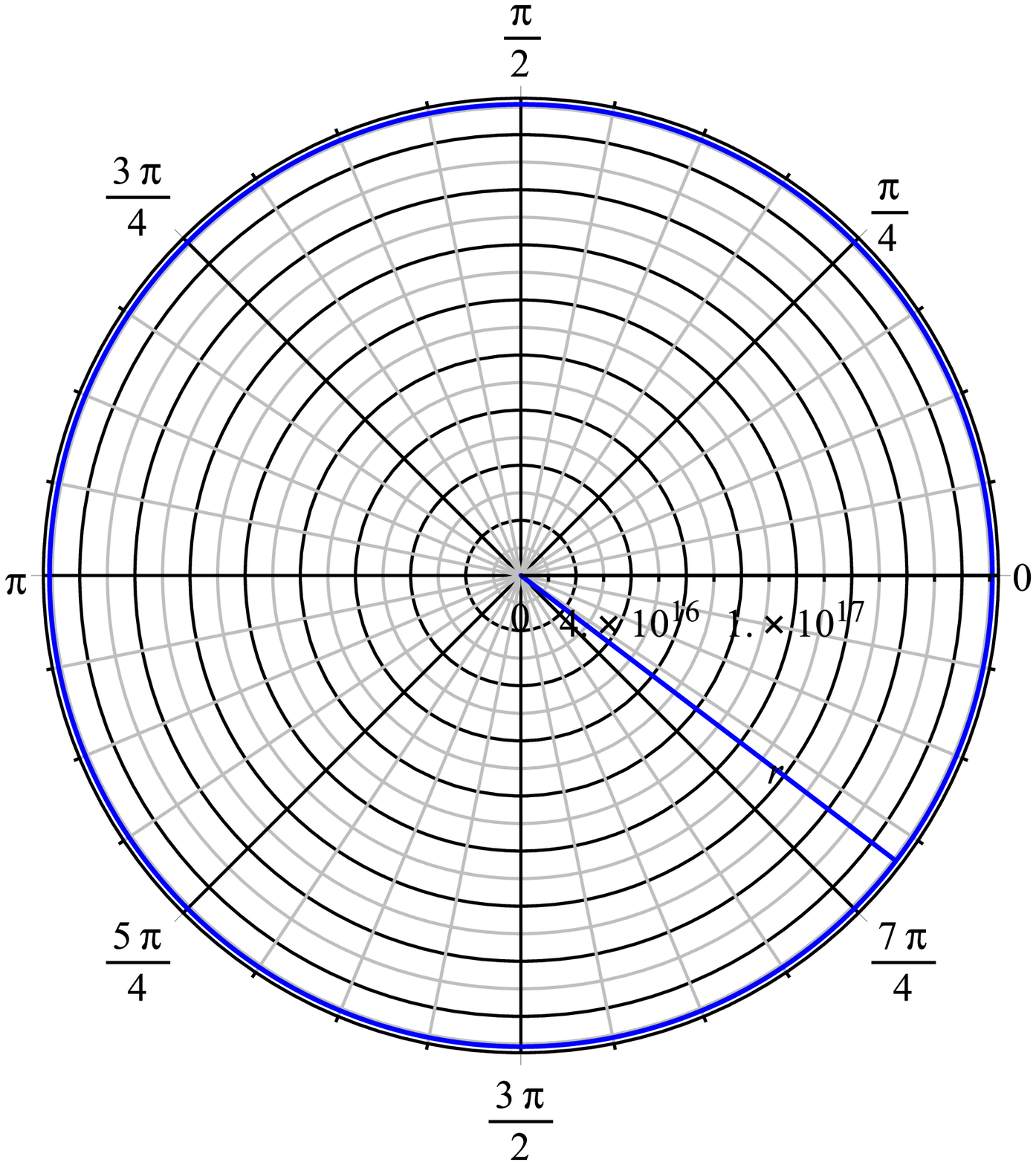}
\caption{\label{figureN}
Plots of the particle trajectories (blue color) with $\ell=k=1$, $\lambda=10^{-52}$ for the case $\mathcal{E}_M=-8.77\cdot 10^{-18}$. In this regime $r_1=0.5$ and $r_M=1.71\cdot 10^{17}$. The figure on the left zooms into the trajectory when the particle starts at $r=r_1$ and moves away from it up to $r=10r_1$. The figure on the right showcases the whole trajectory which consists of two parts: the inner path starts at $r=r_1$ and ends up at  $r=r_M$ where the particle will move around a circle whose radius coincide with the unstable equilibrium point in the effective potential.}
\end{figure}
{\bf{The restored expression of the generalised LRL vector for the example studied here is presented below in explicit closed form }}
\begin{equation}
 \boxed{\bm{\mathcal{A}}=rg(r){\widehat{\bf{r}}}+\ell r\dot{r}h(r){\widehat{\bf{r}}}_\bot,\quad\dot{r}=\pm\sqrt{2\left[\mathcal{E}-V_{eff}(r)\right]},\quad V_{eff}(r)=\frac{\ell^2}{2r^2}-\frac{k}{r}-\lambda r^2,\quad k>0,\quad 0<\lambda<\frac{27k^4}{512\ell^6},}
 \end{equation}
\begin{equation}
\boxed{
V_{eff}(r_{-})<\mathcal{E}<V_{eff}(r_{+}),\quad
r_\pm=\frac{A}{4}\pm\frac{1}{4A}\sqrt{\frac{8k}{\lambda}A-A^4},\quad
A=4\sqrt{\frac{\ell}{\sqrt{6\lambda}}\cosh{\frac{\alpha}{3}}},\quad
\cosh{\alpha}=\frac{3\sqrt{6}k^2}{32\ell^3\sqrt{\lambda}},
}
\end{equation}
\begin{equation}
\boxed{
 g(r)=\frac{\ell}{r}\left[c_1\sin{\widehat{\Xi}(r)}+c_2\cos{\widehat{\Xi}(r)}\right],\quad
 h(r)=\frac{r}{\sqrt{2\lambda r^4+2\mathcal{E} r^2 +2kr-\ell^2}}\left[c_1\cos{\widehat{\Xi}(r)}+c_2\sin{\widehat{\Xi}(r)}\right],
}
\end{equation}
\begin{equation}
\boxed{
\widehat{\Xi}(r)=\frac{\sqrt{2}\ell\left[r_1 F(\sin{\widehat{\delta}},\widehat{\kappa})-(r_1-r_0)\Pi(\sin{\widehat{\delta}},\widehat{\xi},\widehat{\kappa})\right]}{r_0 r_1\sqrt{\lambda(r_3-r_1)(r_2-r_0)}},\quad\sin{\widehat{\delta}}=\sqrt{\frac{(r_2-r_0)(r-r_1)}{(r_2-r_1)(r-r_0)}},\quad
\widehat{\xi}=\frac{r_0(r_2-r_1)}{r_1(r_2-r_0)},
}
\end{equation}
\begin{equation}
\boxed{
\widehat{\kappa}=\sqrt{\frac{(r_2-r_1)(r_3-r_0)}{(r_3-r_1)(r_2-r_0)}},
}
\end{equation}
{\bf{where $r_0,\cdots,r_3$ are the roots of the quartic equation $2\lambda r^4+2\mathcal{E}r^2+2kr-\ell^2=0$. For the definition of the elliptic function $F$ and $\Pi$ we refer to the glossary in Appendix C.}}

\subsection{A general relativistic potential}
The trajectory $x^\kappa(\lambda)$ of a particle immersed in a
gravitational field obeys the geodesic equation \cite{Fliessbach}
\begin{equation}\label{geo1}
\frac{d^2 x^\kappa}{d\lambda^2}+\Gamma^\kappa{}_{\mu\nu}\frac{dx^\mu}{d\lambda}\frac{dx^\nu}{d\lambda}=0
\end{equation}
subject to the constraint
\begin{equation}
g_{\mu\nu}\frac{dx^\mu}{d\lambda}\frac{dx^\nu}{d\lambda}=\left(\frac{ds}{d\lambda}\right)^2=c^2\left(\frac{d\tau}{d\lambda}\right)^2=\left\{
\begin{array}{cc}
c^2 & \mbox{if}~m\neq 0\\
0   & \mbox{if}~m= 0
\end{array}
\right.,
\end{equation}
where $\lambda$ is an affine parameter. In the case of a massive particle with mass $m$, we set $\lambda=\tau$ where $\tau$ denotes the proper time of the particle. For massless particles, $d\tau=0$ and we need to choose a different parametrization. In the presence of a static and spherically symmetric gravitational field represented by the Schwarzschild metric
\begin{equation}\label{metric}
ds^2=B(r)c^2 dt^2-A(r)dr^2-r^2(d\vartheta^2+\sin^2{\vartheta}d\varphi^2),\quad
B(r)=\frac{1}{A(r)}=1-\frac{2G_N M}{c^2 r},
\end{equation}
where $M$ stands for the total mass of the gravitational object, one can cast the radial equation emerging from (\ref{geo1}) coupled to (\ref{metric}) in the form
\begin{equation} \label{GRUeff}
\frac{\dot{r}^2}{2}+V_{eff}(r)=const.
\end{equation}
with effective potential given by
\begin{equation}
V_{eff}(r)=\left\{
\begin{array}{cc}
\frac{\ell^2}{2r^2}-\frac{G_N M}{r}-\frac{G_N M\ell^2}{c^2 r^3} & \mbox{if}~m\neq 0 \label{SPGR}\\
\frac{\ell^2}{2r^2}-\frac{G_N M\ell^2}{c^2 r^3} & \mbox{if}~m= 0
\end{array}
\right..
\end{equation}
and the dots indicating the derivative with respect to the proper
  time $\tau$.  Hence, the vector conservation law we are discussing
 refers to this parameter. At this point, it is straightforward to realize that the potential in (\ref{SPGR}) for a massive particle is a special case of the potential (\ref{GRpotG}) with
\begin{equation}
k=G_N M,\quad
B=\frac{G_N M\ell^2}{c^2}.
\end{equation}
On the other hand the only dynamical input we have been using before
in constructing the LRL-vector is indeed in the form of equation (\ref{GRUeff}). All steps can be then
readily repeated recalling that we have now the proper time in place
of time $t$. Hence, we can construct the corresponding LRL-vector by proceeding as in Section~\ref{tukaj}. To the best of our knowledge, this is the first result indicating how the LRL-vector emerges in General Relativity in the context of a spherically symmetric and static metric and at the same time providing an exact analytic expression for the LRL-vector. Regarding the massless case, it suffices here to mention that the corresponding effective potential admits only a maximum to which there correspond unstable circular orbits.

\subsection{A Laplace-Runge-Lenz vector in Special Theory of Relativity}

Examining the possibility of the existence of a LRL-vector in Special
Theory of Relativity (STR), we mention first that, as it was the case
in the General Theory of Relativity, the conservation law refers to the
proper time $\tau$ related to the time $t$ by $d\tau=(1/\gamma)dt$
with $\gamma=(1-v^2)^{-1/2}$ (here $\mathbf{v}=d\mathbf{r}/dt$ and,
for convenience we set $c=1$). The special relativistic version of Newton' second
law reads \cite{Rafelski}
\begin{equation} \label{N2}
\frac{dp^{\mu}}{d\tau}=F^{\mu},\quad p^{\mu}=m_0 u^{\mu}=m_0\frac{d^2 x^{\mu}}{d\tau^2}
\end{equation}
with $m_0$ the rest mass.  The four-force $F^{\mu}$ is given by the
Newtonian counter-part $\mathbf{K}$ as
\begin{equation} \label{F}
  F^{\mu}=(\gamma \mathbf{K}, \gamma \mathbf{K}\cdot \mathbf{v}).
\end{equation}
A second important remark is now in order. It is necessary to clarify which type of
forces are we allowed to use. We could be tempted to go back to
Newtonian gravith with $\mathbf{K}=-G_Nm_1m_2 \hat{\mathbf{r}}/r^2$.
This, albeit mathematically possible, would be a wrong way to
generalize Newtonian gravity.  Indeed, due to the equivalence principle
gravity's fate lies in its geometrization in the framework of the
General Theory of Relativity. However, this does not apply to the
Coulomb law. Having said this, it suffices to put to zero the magnetic field
$\mathbf{B}$ in the Lorentz force
$\mathbf{K}=q(\mathbf{E}+\mathbf{v} \times \mathbf{B})$ and interpret
$q\mathbf{E}$ as the Coulomb force. Therefore, the $\mathbf{K}$ we will consider is 
\begin{equation} \label{K}
  \mathbf{K}=\frac{k}{r^2}\hat{\mathbf{r}}
\end{equation}
and receives the special relativity generalization in form of
(\ref{F}) and (\ref{N2}). Keep in mind that $k$ is proportional to 
the product of two charges. 
Considering the two fundamental
$1/r^2$-forces in nature there is probably no better place to realize their
distinct fates in modern physics.

With the convention that all dots signify a derivative with respect to
proper time $\tau$ the equations of motion can be cast into the
following three equations containing $r$ and the plane angle $\varphi$
\cite{Greiner}

\begin{eqnarray} 
  \gamma \frac{k}{r^2} &=& m_0(\ddot{r} -r\dot{\varphi}^2),\\
  0&=& m_0(2\dot{r}\dot{\varphi} +r\ddot{\varphi}),\\
  \dot{r}\frac{k}{r^2} &=& m_0\dot{\gamma}.\label{eqnmotion}
\end{eqnarray}
It is now relatively easy to see that the second equation above is equivalent to the conservation of the magnitude
  of the angular momentum $L$, i.e.
  \begin{equation} \label{Lmarek}
    L=m_0r^2\dot{\varphi}=const,
  \end{equation}
  whereas the third equation guarantees that the relativistic energy
  is conserved
  \begin{equation} \label{E}
    \frac{d \mathcal{E}}{d\tau}=\frac{d}{d\tau}(m_0 \gamma +\frac{k}{r})=0.
    \end{equation}
 To get back the non-relativistic limit one would use the expression
 for $\mathcal{E}$ and expand $\gamma$ for small velocities.  We would
 then obtain again $\dot{r}^2/2 +V_{eff}=E=\mathcal{E}-m_0$ where now
 $\dot{r}=dr/dt$.  In the full relativistic case, it is more convenient
 to use the constraint for the four-velocities $u^{\alpha}$, namely
 \begin{equation} \label{veloc}
   u_{\alpha}u^{\alpha}=-1.
 \end{equation}
 Starting with $x^{\mu}=(r\hat{\mathbf{e}}_r, 0, t)$ in plane cylinder coordinates yields
 $u^{\mu}=(\dot{r}\hat{{\bf{r}}}+r\dot{\varphi}\hat{\bm{\varphi}},
 0, \gamma)$ which allows us to calculate
 \begin{equation} \label{veloc2}
   -1=\dot{r}^2+r^2\dot{\phi}^2 -\gamma^2.
 \end{equation}
 The last expression leads to the form $\dot{r}^2/2
 +V_{eff}(r)=const$.   To see that, we use
 \begin{equation} \label{gamma2}
   \gamma=\frac{\mathcal{E}}{m_0}- \frac{k}{m_0r}
\end{equation}
which follows form (\ref{E}) and replace $\dot{\varphi}$ by (\ref{Lmarek}).
In the end, we arrive at the desired result
\begin{equation} \label{Weff}
  \frac{\dot{r}^2}{2}+\frac{a}{2m_0^2 r^2}
  +\frac{b}{r}=\frac{1}{2}\left(\frac{\mathcal{E}^2}{m_0^2}-1 \right)
  \equiv \epsilon=const
\end{equation}
with
\begin{equation} \label{ab}
  a=L^2-k^2,\quad b=\frac{\mathcal{E} k} {m_0^2}.
\end{equation}
With the obvious replacement of the constants we recognize formally the
non-relativistic Coulomb problem. Therefore, keeping in mind the
differences in the constants, the LRL-vector will have the same form
as in the non-relativistic case.  At the end of this sub-section, we
notice that both in general relativistic and the special
relativistic cases it is now easy to compute the trajectory from the
conserved vectors in the same way we did in the non-relativistic domain. The reason is that we have a conservation in
proper time. As a result for an outside observer, the vector will
vary in the usual time coordinate.

\subsection{Newtonian gravity with friction}
So far in deriving the Laplace-Runge-Lenz vector we have followed a specfic method which we used for different potentials.
It instructive to inspect also other approaches which derive the LRL-vector not only for radial potentials and in doing so employ 
a different strategy. We refer here to the excellent review article \cite{Leach} from which the subsequent example is taken.

First, we establish the conservation of the unitary vector $\widehat{\mathbf{L}}$ of the angular momentum 
$\mathbf{L}=|\mathbf{L}|\widehat{\mathbf{L}}=L\widehat{\mathbf{L}}$. We can assert that if
\begin{equation}\label{Mf1}
\dot{\mathbf{L}}+\mathfrak{h}\mathbf{L}=\pmb{0}
\end{equation}
with $\mathfrak{h}$ some function, it follows readily that
\begin{equation} 
\frac{d\widehat{\mathbf{L}}}{dt}=\pmb{0}
\end{equation}
i.e., $\widehat{\mathbf{L}}$ is conserved. One can easily show this using the argument of contradiction. Suppose that 
 $\frac{d \widehat{\mathbf{L}}}{dt} \neq \pmb{0}$. Then according to (\ref{Mf1})
\begin{equation} \label{f3}
\pmb{0}=\dot{\mathbf{L}}+\mathfrak{h}\mathbf{L}=(\dot{L}+\mathfrak{h}L)\widehat{\mathbf{L}} +L\frac{d \widehat{\mathbf{L}}}{dt}.
\end{equation}
Taking the scalar product of this equation with $\frac{d\widehat{\mathbf{L}}}{dt}$ leads to
\begin{equation}\label{interm}
(\dot{L}+\mathfrak{h}L)\widehat{\mathbf{L}}\cdot\frac{d \hat{\mathbf{L}}}{dt} +L\frac{d \widehat{\mathbf{L}}}{dt}\cdot\frac{d \hat{\mathbf{L}}}{dt}=0.
\end{equation}
On the other hand, $\widehat{{\bf{L}}}$ is a unit vector and therefore, $\widehat{{\bf{L}}}\cdot\widehat{{\bf{L}}}=1$ which upon differentiation yields $\widehat{\mathbf{L}}\cdot\frac{d \hat{\mathbf{L}}}{dt}=0$. Hence, if $L \neq 0$ we obtain from (\ref{interm}) that $\frac{d \widehat{\mathbf{L}}}{dt} \cdot \frac{d \widehat{\mathbf{L}}}{dt}=0$ which contradicts our assumption. Therefore,
we can conclude that $\widehat{\mathbf{L}}$ is conserved.

A nice example, which will accompany us through this section, is given by
the equation of motion with arbitrary functions $\mathfrak{f}$ and $\mathfrak{g}$
\begin{equation}\label{Mf5}
\ddot{\mathbf{r}}+\mathfrak{f}\dot{\mathbf{r}}+\mathfrak{g}\mathbf{r}=\pmb{0}
\end{equation}
which obviously contains a velocity dependent force corresponding to friction. 
On the other hand $\mathfrak{g}\mathbf{r}$ could
be simply the Newtonian gravitational potential.
With 
\begin{equation} 
\pmb{\ell}\equiv\frac{\mathbf{L}}{m}=\mathbf{r} \times \dot{\mathbf{r}}
\end{equation}
we obtain $ \dot{\pmb{\ell}}=\mathbf{r} \times \ddot{\mathbf{r}}$ on one hand and 
$\mathbf{r} \times \ddot{\mathbf{r}} +\mathfrak{f}\mathbf{r} \times \dot{\mathbf{r}}=\pmb{0}$ from the equation of motion. 
Therefore, we can assert that 
\begin{equation}\label{Mf7}
\dot{\pmb{\ell}}+\mathfrak{f}\pmb{\ell}=\pmb{0}
\end{equation}
from which we conclude that $\widehat{\pmb{\ell}}$ remains constant.  Indeed, with a force given by $\mathbf{F}=k(r)\widehat{\mathbf{r}}-\lambda 
\dot{\mathbf{r}}$ where $\lambda$ is a positive constant, the equations of motion in polar coordinates can be readily obtained from (\ref{NL}) together with (\ref{rpol}) and (\ref{exprM}). More precisely, we have
\begin{eqnarray}
m(\ddot{r} -r\dot{\varphi}^2)&=& k(r)-\lambda \dot{r},\\
m(2\dot{r}\dot{\varphi} +r\ddot{\phi})&=&-\lambda r\dot{\varphi}.
\end{eqnarray}
Multiplying the second equation by $r$ gives $d(mr^2\dot{\varphi})/dt=-\lambda r^2\dot{\varphi}$. With $L=mr^2\dot{\varphi}$ we arrive at
\begin{equation} \label{f9}
\frac{dL}{dt}=-\frac{\lambda}{m}L
\end{equation}
whose solution is $L=L_0e^{-\lambda t/m}$. Since $\widehat{\pmb{\ell}}$ is conserved, we conclude that $\widehat{\mathbf{L}}=\mathbf{L}/L=\mathbf{L}_0/L_0=const$.

Having established the conservation of the unit angular momentum vector for the system (\ref{Mf5}), we proceed 
to construct two other conserved vectors: the Hamiltonian vector $\mathbf{K}$ and the LRL-vector $\pmb{\mathcal{A}}$
perpendicular to $\widehat{\mathbf{L}}$. With the help of equation (\ref{Mf7}) we rewrite the equation of motion as
\begin{equation} \label{Mf10}
\frac{d}{dt}\left(\frac{\dot{\mathbf{r}}}{\ell}\right)+ \frac{\mathfrak{g}}{\ell}\mathbf{r}=\pmb{0}.
\end{equation}
This form allows us to conclude that the Hamiltonian vector
\begin{equation} \label{Mf11}
\mathbf{K}=\frac{\dot{\mathbf{r}}}{\ell} +\mathbf{u},\quad \mathbf{u}\equiv\int_{t_0}^{t}\frac{\mathfrak{g}}{\ell}\mathbf{r}dt'
\end{equation}
is a constant of motion for the system (\ref{Mf5}). Since both $\mathbf{K}$ and $\widehat{\pmb{\ell}}$ are conserved, it is obvious 
that 
\begin{equation}
{\pmb{\mathcal{A}}}={\bf{K}} \times \widehat{{\pmb{\ell}}}=\frac{\dot{{\bf{r}}}\times\widehat{\pmb{\ell}}}{\ell}+{\bf{u}}\times\pmb{\ell}
\end{equation}
is also conserved. Since
$\mathbf{K}\cdot \widehat{\pmb{\ell}}=0$, we call $\pmb{\mathcal{A}}$ the corresponding Laplace-Runge-Lenz vector.

To arrive at an equation for the trajectory let us first assume that with the help of some function $v(\varphi)$
we are able to write
\begin{equation} \label{Mf13}
\frac{\mathfrak{g}}{\ell}r=v(\varphi) \dot{\varphi}
\end{equation}
and give a  concrete example for $v$ below. If we invoke (\ref{Mf11}) and (\ref{Mf13}), we can now represent the vector $\mathbf{u}$ as
\begin{equation} \label{f14}
\mathbf{u}=\widehat{\mathbf{e}}_x \int_{\varphi_0}^{\varphi}v(\eta)\cos{\eta} d\eta
+\widehat{\mathbf{e}}_y \int_{\varphi_0}^{\varphi}v(\eta)\sin{\eta} d\eta 
\end{equation}
or with $z'(\varphi)=dz/d\varphi$ equivalently as
\begin{equation} \label{f14a }
\mathbf{u}=z'(\varphi)\widehat{\mathbf{e}}_r -z(\varphi)\widehat{\mathbf{e}}_{\phi},\quad z(\varphi)=\int_{\varphi_0}^{\varphi}v(\eta)\sin(\varphi-\eta)d\eta.
\end{equation}
We are now in the position to calculate the trajectory by taking the scalar product
\begin{equation} \label{Mf15}
\mathbf{r} \cdot \pmb{\mathcal{A}}=r\mathcal{A}\cos{(\varphi-\varphi_0)}=1-z(\varphi)r
\end{equation}
which leads to
\begin{equation} \label{f16}
r(\varphi)=\frac{1}{z(\varphi)+\mathcal{A}\cos{(\varphi-\varphi_0)}}
\end{equation}

\begin{figure}[!ht]\label{lastoftheday}
\centering
    \includegraphics[width=0.3\textwidth]{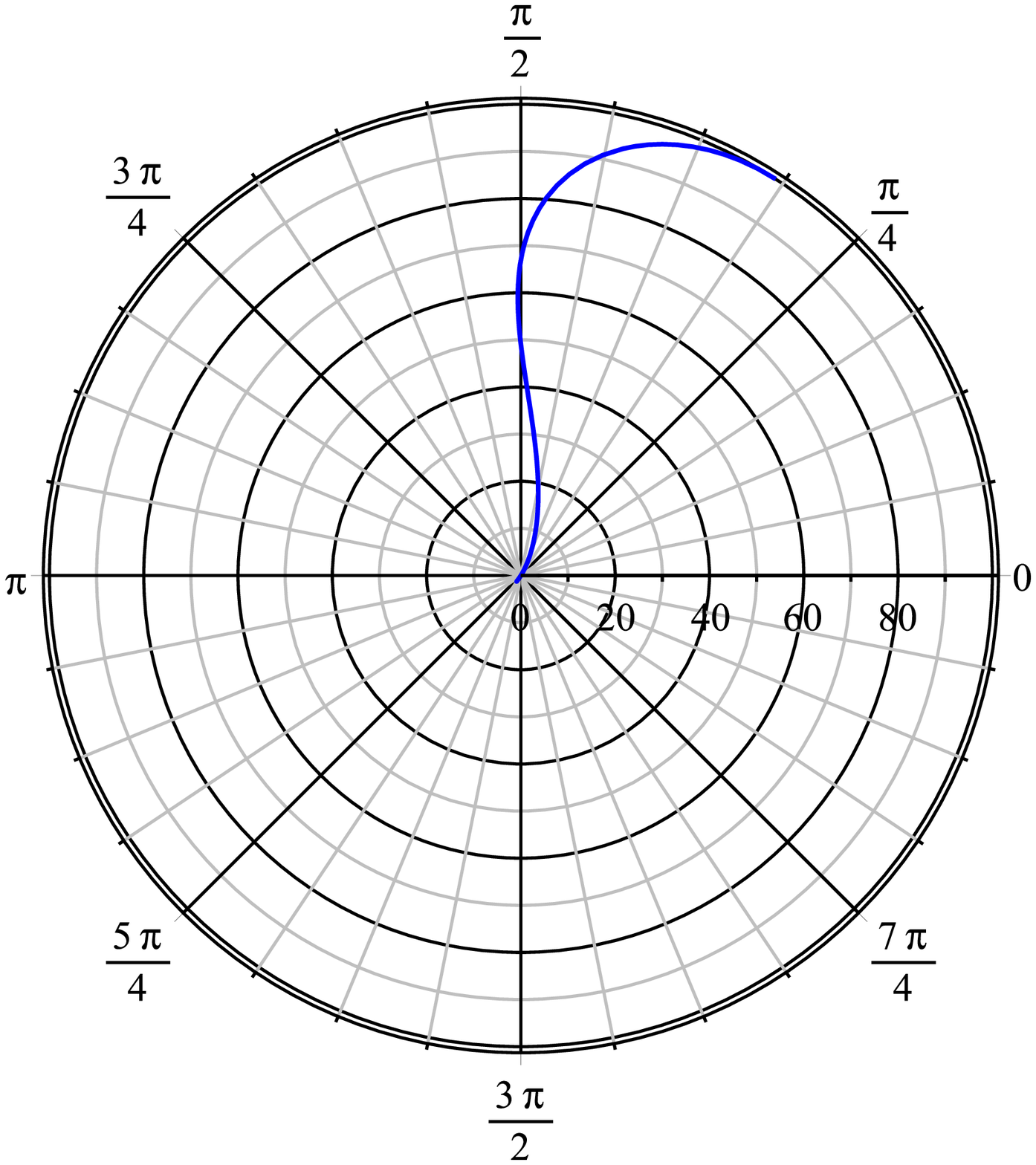}
    \includegraphics[width=0.3\textwidth]{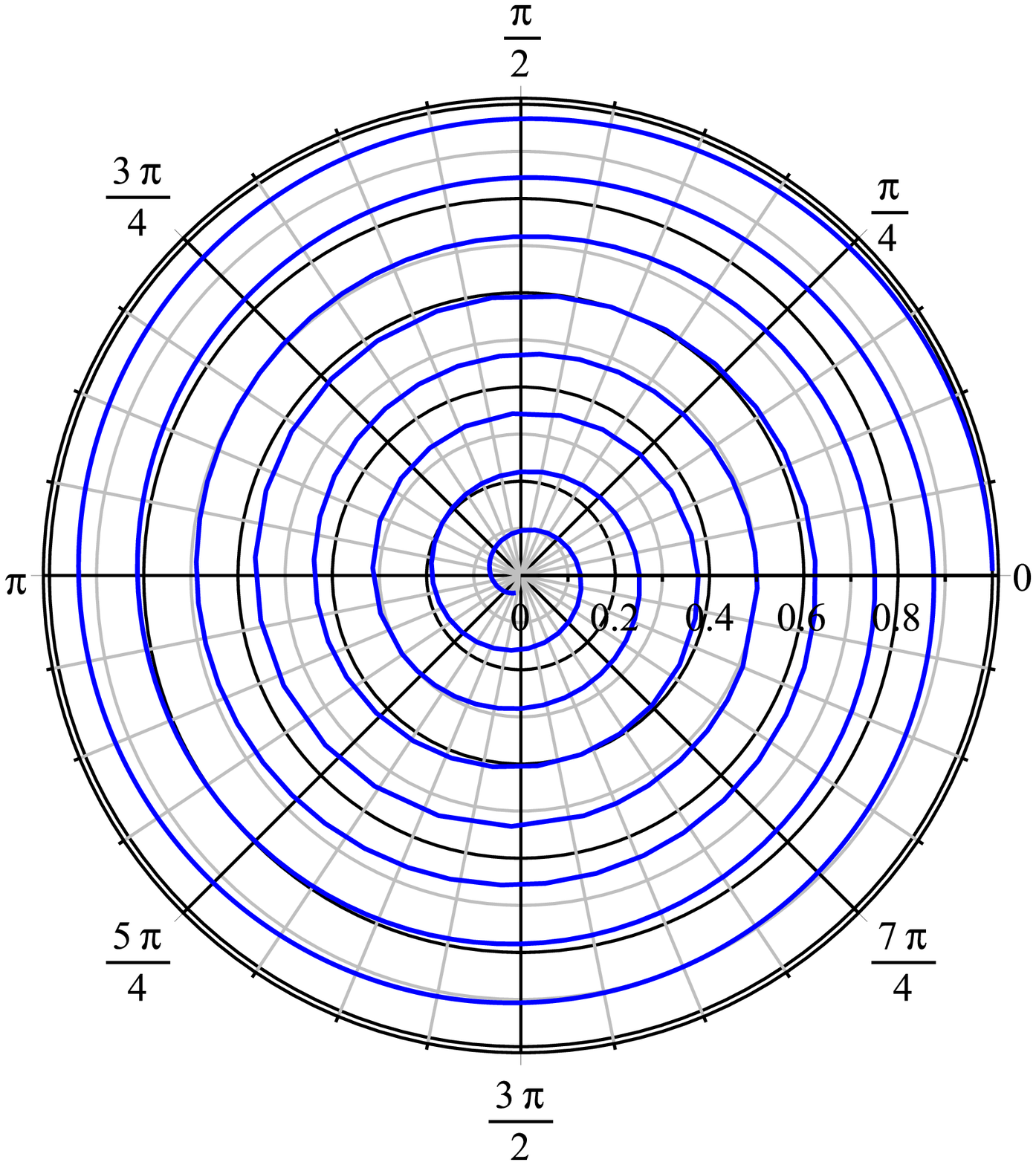}
\caption{\label{figurefriction}
Plots of the particle trajectory according to (\ref{f23}). The figure on the left side corresponds to $\alpha=\mu=\mathcal{A}=1$ and $\xi_0=10^{-2}$ while the one on the right side has been generated by choosing $\alpha=10$, $\mu=10^{-2}$, $\mathcal{A}=1$ and $\xi_0=10^{-2}$. In the first case, the particle crashes into the central mass while in the second case it spirals in towards the central object.}
\end{figure}

It remains to show in an example that equation (\ref{Mf13}) is indeed possible. To this end, we choose in (\ref{Mf5})
\begin{equation} \label{f17}
\mathfrak{f}=\frac{\alpha}{r^2},\quad \mathfrak{g}=\frac{\mu}{r^3}
\end{equation}
with $\alpha$ and $\mu$ constants.
Physically the term $\mathfrak{f}\dot{\mathbf{r}}$ represents friction which falls off quadratically with the distance and $\mathfrak{g}\mathbf{r}$
the gravitational force. Equation (\ref{Mf7}) reads now
\begin{equation} \label{Mf18}
-\frac{\dot{\ell}}{\ell}=\frac{\alpha}{r^2}=\frac{\alpha \dot{\varphi}}{\ell}.
\end{equation}
From the solution
\begin{equation} \label{Mf19}
\ell=\beta -\alpha \varphi
\end{equation}
with $\beta$ an arbitrary integration constant, we can conclude that
\begin{equation} \label{Mf20}
v(\varphi)=\frac{\mathfrak{g}r}{\ell\dot{\varphi}}=\frac{\mu}{\ell r^2\dot{\varphi}}=\frac{\mu}{\ell^2}=\frac{\mu}{(\beta-\alpha \varphi)^2}
\end{equation}
which is of the form (\ref{Mf13}). As a result the function $z$ is now
\begin{equation} \label{f21}
z(\varphi)=\int_{\varphi_0}^{\varphi} \frac{\sin{(\varphi-\eta)}}{(\beta -\alpha \eta)^2} d \eta.
\end{equation}
Defining
\begin{equation} \label{f22}
\xi=\frac{\beta}{\alpha} -\varphi,\quad \xi_0=\frac{\beta}{\alpha} -\varphi_0,
\end{equation}
one can perform the integration to find in terms of the sine and cosine integral functions, denoted by Si and Ci, respectively, the following result
\begin{equation} \label{f23}
z(\varphi)=\frac{\mu}{\alpha^2}\left\{\frac{\sin(\xi-\xi_0)}{\xi_0}-\left[\mbox{Si}(\xi)-\mbox{Si}(\xi_0)\right]\sin{\xi}
-\left[\mbox{Ci}(\xi)-\mbox{Ci}(\xi_0)\right]\cos{\xi}\right\}.
\end{equation} 
Note that the singular behavior in the integral (\ref{f21}) is reflected in the cosine integral function appearing in (\ref{f23}). In fact, Ci$(x)$ diverges logarithmically as $x\to 0^+$. Some sample trajectories described by (\ref{f23}) has been displayed in Fig.~\ref{figurefriction}. This concludes our example to construct the LRL-vector for velocity dependent forces.

\subsection{The Cornell potential $V(r)=-a/r+br$}
The Cornell (or funnel) potential \cite{Eichten1} is a linear superposition of a Coulomb-like potential with a confinement part represented by a spring-like potential where $b$ can be interpreted as a spring tension. Such a potential is used to compute the masses of quarkonium states. Moreover, $r$ is the effective radius of the quarkonium state and $a$, $b$ are parameters that we assume to be positive. The corresponding effective potential is
\begin{equation}\label{quark}
V_{eff}(r)=\frac{\ell^2}{2r^2}-\frac{a}{r}+br.
\end{equation}
Taking into account that
\begin{equation}\label{quark1}
\frac{dV_{eff}}{dr}=\frac{br^3+ar-\ell^2}{r^3}
\end{equation}
together with the fact that the discriminant of the cubic polynomial in (\ref{quark1}) is always positive, it is possible to conclude that such a cubic has only one real root  and it corresponds to a global minimum in the effective potential  (\ref{quark}) given by
\begin{equation}
r_{min}=2\sqrt{\frac{a}{3b}}\sinh{\frac{\psi}{3}},\quad
\sinh{\psi}=\frac{\ell^2}{2b}\left(\frac{3b}{a}\right)^{3/2}.
\end{equation}
As before we introduce the motion reality condition $\mathcal{E}-V_{eff}(r)\geq 0$. From the analysis above we deduce that closed trajectories are possible whenever
$\mathcal{E}> V_{eff}(r_{min})$. Imposing such a condition is equivalent to request that the cubic polynomial appearing in the
expression below
\begin{equation}\label{cubic_q}
\mathcal{E}-V_{eff}(r)=\frac{2\mathcal{E}r^2-2br^3+2ar-\ell^2}{2r^2}=\frac{b}{r}(r-r_0)(r-r_1)(r_2-r)
\end{equation}
admits three real distinct roots here denoted by $r_0<0<r_1<r_2$. The motion reality condition is satisfied between the turning points $r_1$ and $r_2$ and in what follows, we will always consider the case when $r_1<r<r_2$. To find the aforementioned roots, we divide the cubic in (\ref{cubic_q}) by $-2b$ and introduce the variable transformation
\begin{equation}
r=y+\frac{\mathcal{E}}{3b}
\end{equation}
so that the reduced cubic reads
\begin{equation}\label{cubic_q1}
y^3+3p y+2q=0,\quad
p=-\frac{1}{3}\left(\frac{a}{b}+\frac{\mathcal{E}^2}{3b^2}\right),\quad
q=\frac{1}{2}\left(\frac{\ell^2}{2b}-\frac{2\mathcal{E}^3}{27b^3}-\frac{a\mathcal{E}}{3b^2}\right).
\end{equation}
Note that the condition $\mathcal{E}> V_{eff}(r_{min})$ forces the discriminant of (\ref{cubic_q1}) to be negative. After some lengthy but straightforward computations we find that
\begin{equation}
r_0=\frac{\mathcal{E}}{3b}-2\widehat{\rho}\cos{\left(\frac{\widehat{\phi}}{3}\right)},\quad
r_1=\frac{\mathcal{E}}{3b}+2\widehat{\rho}\cos{\left(\frac{\widehat{\phi}+\pi}{3}\right)},\quad
r_2=\frac{\mathcal{E}}{3b}+2\widehat{\rho}\cos{\left(\frac{\widehat{\phi}-\pi}{3}\right)},
\end{equation}
where
\begin{equation}
\cos{\widehat{\phi}}=\frac{q}{\widehat{\rho}^3},\quad\widehat{\rho}=\mbox{sgn}(q)\sqrt{|p|}.
\end{equation}
For instance, in the case $a=b=L=\mathcal{E}=1$ the above formulae give $r_0=-0.85464$, $r_1=0.40303$ and $r_2=1.45161$ in agreements with the roots of the cubic in (\ref{cubic_q}). To construct the LRL-vector associated to bounded trajectories with $r_1<r<r_2$, we compute the solution of the differential equation (\ref{equah}) with
\begin{equation}
P_1(r)=-\frac{1}{r}+\frac{9br^2-6\mathcal{E} r-3a}{2br^3-2\mathcal{E}r^2-2ar-3a},\quad
P_2(r)=\frac{2br}{2br^3-2\mathcal{E}r^2-2ar-3a}.
\end{equation}
If we employ the ansatz 
\begin{equation}\label{ansatzq}
h(r)=\frac{re^{\int_r^{r_2} w(u)du}}{\sqrt{2\mathcal{E} r^2-2br^3+2ar-\ell^2}},
\end{equation}
we obtain the following nonlinear first order differential equation for the unknown function $w(r)$, namely
\begin{equation}
r^2(2\mathcal{E} r^2-2br^3+2ar-\ell^2)\left[\frac{dw}{dr}+w^2(r)\right]+
r(4\mathcal{E} r^2-5br^3+3ar-\ell^2)w(r)+\ell^2=0.
\end{equation}
It can be easily checked with Maple that the above equation admits the following solutions
\begin{equation}
w_\pm(r)=\pm\frac{i\ell}{r\sqrt{2\mathcal{E} r^2-2br^3+2ar-\ell^2}}=
\pm\frac{i\ell/\sqrt{2b}}{r\sqrt{(r-r_0)(r-r_1)(r_2-r)}}.
\end{equation}
The integral in (\ref{ansatzq}) can be evaluated with the help of $1.2.27.5$ in \cite{Prudnikov}. More precisely, we find that
\begin{equation}\label{emodulusq}
\int_r^{r_2}\frac{du}{u\sqrt{(u-r_0)(u-r_1)(r_2-u)}}=\frac{2\Pi(\sin{\phi},\xi,\kappa)}{r_2\sqrt{r_2-r_0}},\quad
\sin{\phi}=\sqrt{\frac{r_2-r}{r_2-r_1}},\quad
\kappa=\sqrt{\frac{r_2-r_1}{r_2-r_0}},\quad 
\xi=1-\frac{r_1}{r_2}
\end{equation}
under the assumption that $0<r_0<r_1<r_2$ and $r_1<r<r_2$. Here, the symbol $\Pi$ denotes the incomplete elliptic integral of the third kind and $\xi$ is called the parameter of the aforementioned integral. At this point, two linearly independent solutions of (\ref{equah}) can be constructed as follows
\begin{eqnarray}
h_1(r)&=&\frac{r}{\sqrt{2\mathcal{E}r^2-2br^3+2ar-\ell^2}}\cos{\left(\frac{\sqrt{2}\ell\Pi(\sin{\phi},\xi,\kappa)}{r_2 \sqrt{b(r_2-r_0)}}\right)},\\
h_2(r)&=&\frac{r}{\sqrt{2\mathcal{E}r^2-2br^3+2ar-\ell^2}}\sin{\left(\frac{\sqrt{2}\ell\Pi(\sin{\phi},\xi,\kappa)}{r_2\sqrt{b(r_2-r_0)}}\right)}.
\end{eqnarray}
Given the function $h(r)$, the corresponding function $g(r)$ can be evaluated by means of (\ref{equaG}) as 
\begin{equation}
g(r)=-\frac{2b}{r}(r-r_0)(r-r_1)(r-r_2)\frac{dh}{dr}+\frac{br^3+ar-\ell^2}{r^2}h(r).
\end{equation}
By means of the identities
\begin{eqnarray}
&&\frac{d\Pi(\sin{\phi},\xi,\kappa)}{dr}=-\frac{r_2\sqrt{r_2-r_0}}{2r\sqrt{(r-r_0)(r-r_1)(r-r_2)}},\\
&&\frac{\sqrt{2b}}{r}(r-r_0)(r-r_1)(r-r_2)\frac{d}{dr}\left(\frac{r}{\sqrt{(r-r_0)(r-r_1)(r-r_2)}}\right)=\frac{br^3+ar-\ell^2}{r\sqrt{2b(r-r_0)(r-r_1)(r-r_2)}},\label{idIIcf}
\end{eqnarray}
where in deriving (\ref{idIIcf}) we used the following equalities for the the roots of the cubic appearing in (\ref{cubic_q})
\begin{equation}
r_0 r_1+r_0 r_2+r_1 r_2 =-\frac{a}{b},\quad
r_0 r_1 r_{2}=-\frac{\ell^2}{2b},
\end{equation}
we find that
\begin{equation}
g_1(r)=-\frac{\ell}{r}\sin{\left(\frac{\sqrt{2}\ell\Pi(\sin{\phi},\xi,\kappa)}{r_2\sqrt{b(r_2-r_0)}}\right)},\quad
g_2(r)=\frac{\ell}{r}\cos{\left(\frac{\sqrt{2}\ell\Pi(\sin{\phi},\xi,\kappa)}{r_2\sqrt{b(r_2-r_0)}}\right)}.
\end{equation}
If we consider the pair of functions $(h_1,g_1)$ and $(h_2,g_2)$, it is gratifying to observe that the modulus of the LRL-vector is simply $\mathcal{A}=\ell$. {\bf{The restored expression of the generalised LRL vector for the example studied here is presented below in explicit closed form }}
\begin{equation}
 \boxed{\bm{\mathcal{A}}=rg(r){\widehat{\bf{r}}}+\ell r\dot{r}h(r){\widehat{\bf{r}}}_\bot,\quad\dot{r}=\pm\sqrt{2\left[\mathcal{E}-V_{eff}(r)\right]},\quad V_{eff}(r)=\frac{\ell^2}{2r^2}-\frac{a}{r}+br,\quad a,b>0,}
 \end{equation}
\begin{equation}
\boxed{
\mathcal{E}> V_{eff}(r_{min}),\quad
r_{min}=2\sqrt{\frac{a}{3b}}\sinh{\frac{\psi}{3}},\quad
\sinh{\psi}=\frac{\ell^2}{2b}\left(\frac{3b}{a}\right)^{3/2},
}
\end{equation}
\begin{equation}
\boxed{
 g(r)=\frac{\ell}{r}\left[c_1\sin{\Sigma(r)}+c_2\cos{\Sigma(r)}\right],\quad
 h(r)=\frac{r\left[c_1\cos{\Sigma(r)}+c_2\sin{\Sigma(r)}\right]}{\sqrt{2\mathcal{E} r^2-2br^3+2ar-\ell^2}},\quad
 \Sigma(r)=\left(\frac{\sqrt{2}\ell\Pi(\sin{\phi},\xi,\kappa)}{r_2 \sqrt{b(r_2-r_0)}}\right)
}
\end{equation}
\begin{equation}
\boxed{
\sin{\phi}=\sqrt{\frac{r_2-r}{r_2-r_1}},\quad
\kappa=\sqrt{\frac{r_2-r_1}{r_2-r_0}},\quad 
\xi=1-\frac{r_1}{r_2},\quad
r_1=\frac{\mathcal{E}}{3b}+2\widehat{\rho}\cos{\left(\frac{\widehat{\phi}+\pi}{3}\right)},\quad
r_2=\frac{\mathcal{E}}{3b}+2\widehat{\rho}\cos{\left(\frac{\widehat{\phi}-\pi}{3}\right)},
}
\end{equation}
\begin{equation}
\boxed{
\cos{\widehat{\phi}}=\frac{q}{\widehat{\rho}^3},\quad\widehat{\rho}=\mbox{sgn}(q)\sqrt{|p|},\quad
p=-\frac{1}{3}\left(\frac{a}{b}+\frac{\mathcal{E}^2}{3b^2}\right),\quad
q=\frac{1}{2}\left(\frac{\ell^2}{2b}-\frac{2\mathcal{E}^3}{27b^3}-\frac{a\mathcal{E}}{3b^2}\right).
}
\end{equation}
{\bf{For the definition of the elliptic function $\Pi$ we refer to the glossary in Appendix C.}}

\section{Conclusions and outlook}\label{Conclusions}
In this work, we have taken up the challenge to construct explicitly a
conserved vector in the plane of motion given a spherically symmetric
potential $V(r)$.  We can call such a vector the generalized
Laplace-Runge-Lenz vector (or simply the LRL-vector) being originally
constructed for the $1/r$-potential. To achieve our goals, we set up
an ansatz which led to a system of differential equations. In choosing the cases
which we considered explicitly, we concentrated on examples where the
differential equations could be solved explicitly in terms of
transcendental functions. We also paid attention to the phenomenological
relevance of our choices which included several gravitational examples
either as a result of an approximation or the non-relativistic
reduction of General Relativity like the (anti) de Sitter
case. Relying on the LRL-vectors we plotted the corresponding
trajectories with the help of Maple.

Our approach allowed also the construction of a conserved LRL-vector in
General Relativity in the case of spherical symmetry where the time
$t$ is replaced by the proper time $\tau$.

It is known that the existence of a conserved LRL-vector for the Coulomb
problem can be used to find the eigenvalues of the correponding
Hamilton operator in quantum mechanics \cite{Bohm}. An obvious question is 
how other LRL vectors, e.g., for the confining potential constructed in
the text,  can be employed to find the eigenvalues algebraically probably with the help of a closed algebra of operators.
{\bf{Such a full undertaking for new LRL-vectors would go well beyond the scope of the present work which focuses to demonstrate explicit forms of new conserved vectors. But it makes sense to briefly outline a program. The usual starting point on the way
to a LRL-vector in quantum mechanics is the Poisson algebra of angular momentum vector component $L_i$, the components of the 
LRL-vector $A_i$ and the Hamiltonian $H$. This algebra finds its corresponding counterpart in quantum mechanics in terms of 
commutators. Firstly, it should be noted that a general theorem regarding conserved quantities, as mentioned in \cite{Hall}, implies that  
$\{A_i, H\}=\{L_i, H\}=0$. Secondly, the transformation properties of the LRL- vector under rotation imply that ${A_i, L_j}=\epsilon_{ijk}A_k$, as guaranteed by a second general theorem in \cite{Hall}. However, the proof for the result
$\{A_i, A_j\}=-2mH \epsilon_{ijk}L_k$ is already lengthy and rather technical for the standard LRL vector in the 
case of $1/r$-potential. The next step to obtain a closed algebra is to calculate the same Poisson brackets for the rescaled
vector $D_i=A_i/\sqrt{2mH}$ and to proceed to quantum mechanics. We plan to pursue this program for the new LRL-vectors in future  publications.}}

\appendix
\section{The Schwarzschild-de Sitter and Schwarzschild-anti de Sitter metrics}
There are, of course, many interesting macroscopic potentials worth
some detailed investigation. Out of these, potentials connected to
gravity have a special role as they will govern the orbits of
astrophysical objects at scales of the solar system and beyond. In this appendix, we present the non-relativistic limit of the Schwarzschild-de Sitter
(also called Kottler metric) and Schwarzschild-anti de Sitter metric connected closely
to recent cosmological discoveries.

The discovery of the accelerated stage of the expansion of the
universe \cite{maccel} forced scientists to reconsider the standard
Einstein equations. One of the simplest models to encompass the cosmic
acceleration is to include in the Einstein equations a positive
cosmological constant $\Lambda$.
\begin{equation} \label{eq:2} 
G_{\mu \nu}=
R_{\mu \nu} -\frac{1}{2}Rg_{\mu \nu} 
+ \Lambda  g_{\mu \nu}. 
\end{equation}
It was Einstein himself who
introduced this constant to obtain a static (unstable, as we know now)
universe and who after the discovery of the expansion rejected it
also. It is therefore somewhat curious to see this constant being
re-introduced  in order to explain an accelerated stage of this
expansion, often in the framework of the cosmological concordance model
called the $\Lambda$-CDM ($\Lambda$-Cold
Dark Matter) \cite{mCDM}. Indeed, starting
with the
Friedmann-Robertson-Walker metric which requires the isotropy and
homogeneity of the underlying space-time
\begin{equation} \label{eq:3}
ds^2=-dt^2 + a^2(t)R_0^2\left(\frac{dr^2}{1-kr^2} +r^2d\Omega\right),
\end{equation}
where $a=R/R_0$, $R_0$ is the value today, and $k$ denotes the spatial curvature
which is zero in our universe, the Einstein equations are now reduced to
differential equations (Friedmann equations) for $a$.
More explicitly involving the Hubble parameter $H$, we have, 
\begin{equation} \label{eq:4}
H^2\equiv \left(\frac{\stackrel{\cdot}{a}}{a}\right)^2=\frac{8\pi G_N}{3}\rho 
+\frac{\Lambda}{3}
-\frac{k}{a^2R_0^2},\quad k=\pm 1, 0
\end{equation}
and a second differential equation of the form
\begin{equation} \label{eq:5}
\frac{\ddot{a}}{a}=-\frac{4 \pi G_N}{3}\left[\rho + 3p(\rho)\right] + \frac{\Lambda}{3},
\end{equation}
where $p$ denotes the pressure. It is now clear that the acceleration $\ddot{a}$ can, in principle, be
bigger than zero provided $\Lambda > 0$. It is instructive to realize
with what length scales we are dealing in the cosmological
scenario.  Using $\hbar=c=1$  the Hubble length $H_0^{-1}$ is of the
order of Gpc (Giga parsec). On the other hand, using the critical density
\begin{equation} \label{eq:12}
\rho_{crit}=\frac{3H_0^2}{8\pi G_N}=\frac{3}{8\pi}H_0^2 m_{pl}^2,
\end{equation}
where $H_0$ is the present value of the Hubble parameter, and energy
density associated with the cosmological constant
\begin{equation} \label{eq:13}
\rho_{vac}=\frac{\Lambda}{8\pi G_N},
\end{equation}
it is possible to write the following useful relation
\begin{equation} \label{eq:14}
\Lambda=3 \left(\frac{\rho_{vac}}{\rho_{crit}}\right)H_0^2
\end{equation}
with the observed value of $\rho_{vac} \simeq 0.6\rho_{crit}$ \cite{mPlanck}.  This
allows us to define a length scale of $\Lambda$ as
\begin{equation} \label{eq:15}
r_{\Lambda}=\frac{1}{\sqrt{\Lambda}}=\frac{1}{\sqrt{3}}\left(\frac{\rho_{vac}}{\rho_{crit}}\right)^{-1/2}H_0^{-1}.
\end{equation}
This equation tells us that $r_{\Lambda}$ is practically the Hubble
length.  Since $\Lambda$ enters the Einstein tensor, it
will affect, in principle,  any calculation where Einstein equations
are used.  On the other hand, it appears the cosmological dimensions
associated with $\Lambda$ will restrict its phenomenological
usefulness to cosmology only.  In other words, intuitively we might
neglect $\Lambda$ while considering local properties of matter at
scales much smaller then the present Hubble radius like stars and
galaxies.

To detect effects due to the cosmological constant, let us probe into
the properties of the Schwarzschild-de Sitter metric which in the case of a spherically symmetric object with mass $M$ is given by 
\begin{equation} \label{eq:8}
ds^2=-{e}^{\nu (r)}dt^2 + {e}^{-\nu (r)}dr^2 +r^2 d\theta^2 +r^2 \sin^2{\theta} d\phi^2
\end{equation}
with
\begin{equation} \label{eq:9}
g_{00}={e}^{\nu (r)} = 1-\frac{2r_s}{r}-\frac{r^2}{3(r_{\Lambda})^2},\quad
r_s \equiv G_N M,\quad r_{\Lambda} \equiv \frac{1}{\sqrt{\Lambda}}.
\end{equation}
Two length scales appear: the Schwarzschild radius $2r_s$ and $r_{\Lambda}$. With the connection of the $g_{00}$ component of the metric to the
gravitational potential $\Phi$ \cite{mWeinberg}, i.e. 
\begin{equation}\label{eq:10}
g_{00} \simeq -(1+2\Phi),
\end {equation}
we obtain the Newtonian limit in the form 
\begin{equation} \label{eq:11}
\Phi (r) =-\frac{r_s}{r} -\frac{1}{6}\left(\frac{r}
{r_{\Lambda}}\right)^2.
\end{equation}
The first term is the standard Newtonian potential. Its form will change when we consider
a non-spherically symmetric mass distribution. But the second term 
will remain as it is.
Indeed, it is an external force
which for a positive cosmological constant plays the role of a repulsive external force
with the interpretation that two points in space separate.
The Galilean spacetime gets replaced by Newton-Hooke  
space-time \cite{mine1} in which two space points go apart due
to the cosmological constant (this is the part of the cosmological expansion). 
The equation of motion for a massive particle with proper time $\tau$ in the  Schwarzschild-de Sitter metric can be written elegantly as 
\begin{equation} \label{eq:63}
\frac{1}{2}\left(\frac{d r}{d\tau}\right)^{2}+U_{eff}=\frac{1}{2}\left( \mathcal{E}^{2}+
\frac{L^{2}\Lambda}{3} -1 \right)\equiv C={\rm constant},
\end{equation}
where $\mathcal{E}$ and $L$ 
are conserved quantities (the analogues of energy and angular momentum in classical mechanics) defined by
\begin{equation} \label{eq:64}
\mathcal{E}= {e}^{\nu(r)}\frac{d t}{d\tau},\quad
L = r^2 \frac{d\varphi}{d\tau},
\end{equation}
where $\varphi$ 
is the azimuthal angle and $U_{eff}$  is defined by \cite{Balaguera-2006}
\begin{equation} \label{eq:65}
U_{eff}(r)=-\frac{r_s}{r}-\frac{1}{6}\left(\frac{r}{r_\Lambda}\right)^2+\frac{L^{2}}{2r^{2}}-\frac{r_s L^{2}}{2r^{3}}\,,
\end{equation}
which is the analogue of an effective potential in classical mechanics. This form of the equation of motion is, of course, equivalent
to the geodesic equation of motion from which it has been derived.  The corresponding Newtonian limit holds if the potential is weak, i.e.
\begin{equation} \label{eq:24}
|\Phi(r)| \ll 1.
\end{equation}From the fact that
$|\Phi(\sqrt{6}r_{\Lambda})|=1+r_s/\sqrt{6}r_{\Lambda}$ and $|\Phi(r_s)|=1+(r_s/\sqrt{6}r_{\lambda})^2$
it is clear that we must satisfy
\begin{equation} \label{eq:27}
\sqrt{6}r_{\Lambda} \gg r \gg r_s.
\end{equation}
In the non-relativistic limit we can neglect the $1/r^3$ term in (\ref{eq:65})
and consider only
\begin{equation} \label{eq:65x}
V_{eff}(r)=-\frac{r_s}{r}-\frac{1}{6}\frac{r^{2}}{(
r_{\Lambda})^{2}}+\frac{l^{2}}{2r^{2}}
\end{equation}
There are three length scales involved in the potential:
$r_{\Lambda}$, $r_s$ and $r_l=l$.  Because of $r_{\Lambda} \gg r_s$
the zero and the local minimum of the effective potential will be
dominated by $r_s$ and $r_l$ (making the physics at small distances
almost Newtonian) whereas at large distances the cosmological constant
will contribute. To see this, we tentatively put $l=r_l=0$ and obtain
a local maximum at
\begin{equation} \label{max}
r_{max}=(3r_s r_{\Lambda}^2)^{1/3}.
\end{equation}

The mixing of a small scale with a large one makes $r_{max}$ relevant
at astrophysical scales. For instance, for the sun the maximum is
roughly at $70$ pc. Provided the accelerated stage of the universe is explained
by a positive cosmological constant, the same constant will also affect the local behaviour of stars. It is nice to see how cosmology can connect local gravitational phenomena.

A small comparison with the pure Newtonian case is due. First we notice that
the local maximum has a physical meaning. Indeed, in contrast to the
Newtonian physics there is now a ``last bound state''  associated with
the local maximum. Secondly, starting at large distances bigger
than $r_{\max}$ and at sufficiently small negative energies (say, well
below the local minimum) the particle will scatter off  ``the wall'' created by the term proportional to $-\Lambda r^2$.  These scattering will occur also at
large distances in contrast to the Newtonian case where the  back scattering
will happen at the centre. These are, of course, predictions of the
Schwarzschild-de Sitter metric, but they have never been put to
test. One reason is that even if  $r_{max}$ is of astrophysical
relevance it is still larger than the length scale associated with the
objects under consideration. For instance, for the sun it is larger than
the size of the solar system.  However, already at distances of few light years we
will encounter many other stars and the problem is not anymore a two
body one. It could be then that the effect manifests itself in some
kind of star clustering with the radius of $r_{max}$.

Finally, let us briefly review qualitatively the case of a negative
cosmological constant. All our formulae above are valid by taking
$\Lambda \to -\Lambda$.  The scattering states at large distances and
negative energies are ``replaced'' by infinite number of bound states
for positive energies.  Any object might be then in gravitational interaction
with any other object giving rise to a bound state.  This might merit
a deeper examination albeit it is already somewhat clear that it is
not very physical.

\section{Power and inverse power law potentials in gravity}
We might ask ourselves the question in which gravitational scenarios
the relevant potential will depend only on $r$. This is not
necessarily always connected to a spherical symmetry, but rather to an
axial symmetry in the equatorial plane $\theta =\pi/2$.  Provided we
can expand the potential $\Phi(r, \theta)$ in terms of Legendre
polynomials, i.e.
\begin{equation} \label{legendre}
  \Phi(r, \theta)=\sum_{n=0}^\infty \Phi_n(r) P_n(\cos \theta).
\end{equation}
The constant angle $\theta=\pi/2$ is a solution of the equations of
motion. To see that, it suffices to notice that the force in the
$\mathbf{\hat{e}}_{\theta} $ direction, i.e. $F_{\theta}$, is
$r^{-1}\partial\Phi/\partial\theta$ and hence equal to
$\sum_{n=0}^\infty (\Phi_n dP_n/d \cos\theta) (-\sin \theta)$.  On the
other hand, the acceleration in the same direction is
$a_{\theta}=r\dot{\theta} +2\dot{r}\ddot{\theta}$.

To see what important potentials arise in this context, we start with
the oblatness of the sun, discuss very briefly the gravitational
potential of a ring and use this result to infer on perturbative
potentials of a planet caused by the surrounding remaining bodies
(planets).  In doing so we will skip many details and refer the reader
to the book of Fitzpatrick \cite{Fitzpatrick}. 

Consider a slightly deformed spheroid. A convenient parametrization of
its shape is given by
\begin{equation} \label{spheroid}
  R_{\theta}=R\left[1 -\frac{2}{3} \epsilon P_2(\cos \theta)\right]
\end{equation}
with $|\epsilon|$ being a small number. In case it is positive our
surface will be oblate and in case that $\epsilon$ is smaller than
zero, a prolate spheroid will emerge. With $R_p=R_{\theta}(\theta=0)$ 
and $R_e=R_{\theta}(\theta=\pi/2)$ we get the intuitively simple
result $\epsilon=(R_e-R_p)/R$.  To proceed further, we mention that the
projections $\Phi_n$ can be conveniently
written as
\begin{eqnarray} \label{projections}
  \Phi_n(r)&=&-\frac{2\pi G_n}{r^{n+1}} \int_0^r \xi^{n+2}   \int_0^{\pi} 
P_n(\cos \phi) \sin \phi \rho(\xi,
  \cos\phi)d\xi d\phi \nonumber \\
&-&2\pi G_N  r^{n+1} \int_r^{\infty}r \xi^{1-n}   \int_0^{\pi} 
P_n(\cos \phi) \sin \phi \rho(\xi,
  \cos\phi)d\xi d\phi.
  \end{eqnarray}
A slight rearrangement gives
\begin{equation} \label{2ndform}
\Phi(r, \theta)=\frac{G_N M}{R}\sum_{n=0}^\infty J_n\left(\frac{R}{r}\right)P_n(\cos \theta)
\end{equation}
with $J_n$ being now
\begin{equation} \label{Jn}
J_n=-\frac{3}{2}\int_0^{2\pi}P_n(\cos \theta) \int_0^{R_{\theta}}(\theta) \frac{r^{2+n}}{r^{3+n}}dr \sin \theta d\theta
\end{equation}
given here for a constant density. To the leading order in $\epsilon$ we obtain $J_0=-1$ and $J_2=2\epsilon/5$.
The final potential reads
\begin{equation}\label{finalpot}
\Phi(r,\theta)= -\frac{G_N M}{r} +J_2\frac{G_N M R^2}{r^3}P_2(\cos{\theta}).
\end{equation}
This expression is a rather well known potential for a weakly deformed sphere and has some relevance for the solar dynamics.
The $J_2$ expression is known as the sun's quadrupole moment and the astronomers agree on a non-zero value of this moment due to the
sun's rotation. The new term proportional to $1/r^3$ will cause a perihelion motion of the planet (say, Mercury).  The motion of Mercury's perihelion is then due to the quadrupole moment, the perturbative potential of the other planets and General Relativity. 
The present value $J_2 \simeq 2 \times 10^{-7}$ \cite{Rozlet} is too small to stir up a contradiction with General Relativity. 
But bigger value of the order of $10^{-6}$ were advocated previously \cite{Campbell} 
which would spoil the agreement with General Relativity.  

The next gravitational potential we would like to mention is the one of a ring with radius $a$.  Using a constant density 
$\rho=\rho_0 a^{-1}\delta (\cos \theta)\delta(r-a)$ with $\rho_0=M/2\pi a$ we obtain for $r > a$
\begin{equation} \label{ring1}
\Phi(r)=-\frac{G_N M}{r}\sum_{n=0}^\infty P^2_n(0)\left(\frac{a}{r}\right)^{n+1} =-\frac{G_nM}{r}\left[1 +\frac{1}{4}\left(\frac{a}{r}\right)^2 + \frac{9}{64}
\left(\frac{a}{r}\right)^4 +...\right]
\end{equation}
and for $r<a$ 
\begin{equation} \label{ring2}
\Phi(r)=-\frac{G_N M}{a}\sum_{n=0}^\infty P^2_n(0)\left(\frac{r}{a}\right)^{n+1} =-\frac{G_nM}{a}\left[1 +\frac{1}{4}\left(\frac{r}{a}\right)^2 + \frac{9}{64}\left(\frac{r}{a}\right)^4 +...\right].
\end{equation}
The usefulness of this potential is not to have an astrophysical ring, but reveals its power in the art of approximation which goes back to Carl Friedrich Gau\ss who
assumed that the averaged interaction over an orbit of a planet $i$ with another planet $j$ is well represented by taking the potential of a ring. Hence, we will a sum of terms
found in equations (\ref{ring1}) and (\ref{ring2}) where $a_j$ is the average radius of the planet $j$. This makes the inter-planetary interaction pertubative.  To summarize this, 
we write the potential for a planet $i$ as
\begin{equation} \label{pottotal}
\Phi_i(r)=-\frac{G_N M}{r} -\sum_{k=1}^\infty P^2_{2k}(0)\left[\sum_{j < i}\frac{G_N m_j}{a_j}\left(\frac{a_j}{r}\right)^{2k+1} + \sum_{j > i}\frac{G_N m_j}{a_j}\left(\frac{r}{a_j}\right)^{2k}
\right]
\end{equation}
with the convention that $i=1$ represents Mercury for which in leading order we get the Newtonian potential plus a term proportional to $r^2$. Formally, this resembles the non-relativistic limit of the Schwarzschuild- de Sitter potential from General Relativity. For the last planet the leading perturbative term is proportional to $1/r^3$. One can put to test
Gau\ss's method by calculating the perihelion precession and the results are in good agreement with observations. We refer the reader to the book \cite{Fitzpatrick} where more details are given.  

\section{Glossary}\label{glossario}
For the reader convenience we list the special functions used in the present work.
\begin{itemize}
\item
$F(\sin{\phi},\kappa)$ is the incomplete elliptic integral of the first kind and is defined according to \cite{Abra}
\begin{equation}\label{IEI1K}
F(\sin{\phi},\kappa)=\int_0^{\sin{\phi}}\frac{dt}{\sqrt{1-t^2}\sqrt{1-\kappa^2 t^2}},
\end{equation}
where $\phi$ denotes the amplitude and $\kappa$ is the elliptic modulus. Note that $-\frac{\pi}{2}<\phi<\frac{\pi}{2}$ while $0<k^2<1$. Another equivalent definition of the same integral is the following
\begin{equation}\label{ultima}
F(\phi,\kappa)=\int_0^\phi\frac{d\vartheta}{\sqrt{1-\kappa^2\sin^2{\vartheta}}}
\end{equation}
and can be transformed into (\ref{IEI1K}) by the change of variable $t=\sin{\vartheta}$.
\item
$\Pi(\sin{\phi},\xi,\kappa)$ is the incomplete elliptic integral of the third kind which is defined as \cite{Abra}
\begin{equation}
\Pi(\sin{\phi},\xi,\kappa)=\int_0^{\sin{\phi}}\frac{dt}{(1-\xi t^2)\sqrt{1-t^2}\sqrt{1-\kappa^2 t^2}},
\end{equation}
where $\phi$ denotes the amplitude, $\kappa$ is the elliptic modulus and $\xi$ is a constant called the elliptic characteristic or simply the characteristic parameter.
\item
The sine Jacobi elliptic function $\mbox{sn}{(u,\kappa)}$ emerges from the process of inversion of the elliptic integral of the first kind $u=F(\phi,\kappa)$ defined as in (\ref{ultima}). If we define the Jacobi amplitude $\phi$ as 
\begin{equation}
\phi=F^{-1}(\phi,\kappa)=\mbox{am}(u,\kappa),
\end{equation}
then the sine Jacobi elliptic function is simply
\begin{equation}
\mbox{sn}(u,\kappa)=\sin{(\mbox{am}(u,\kappa))}.
\end{equation}
\item
The so-called sine integral is defined as follows \cite{Abra}
\begin{equation}
\mbox{Si}(z)=\int_0^z\frac{\sin{t}}{t} dt.
\end{equation}
Clearly, $\mbox{Si}(0)=0$ while for real argument $x\to+\infty$ we have \cite{Spiegel}
\begin{equation}
\mbox{Si}(x)\sim\frac{\pi}{2}-\frac{\sin{x}}{x}\sum_{n=0}^\infty(-1)^n\frac{(2n+1)!}{x^{2n+1}}-\frac{\cos{x}}{x}\sum_{n=0}^\infty(-1)^n\frac{(2n)!}{x^{2n}}
\end{equation}
with $\sim$ denoting asymptotically equivalence. Note that $S(x)\to\pi/2$ as $x\to+\infty$. 
\item
The cosine integral is specified according to \cite{Abra}
\begin{equation}
\mbox{Ci(z)}=-\int_{z}^\infty\frac{\cos{t}}{t}dt=\gamma+\ln{z}+\int_0^z\frac{\cos{t}-1}{t}dt,\quad |\mbox{arg}z|<\pi,
\end{equation}
where $\gamma$ is the Euler-Mascheroni constant. Note that the above function diverges logarithmically as $z\to 0$. Moreover, for real argument $x\to+\infty$ the Cosine integral admits the asymptotic expansion \cite{Spiegel} 
\begin{equation}
\mbox{Ci}(x)\sim\frac{\cos{x}}{x}\sum_{n=0}^\infty(-1)^n\frac{(2n+1)!}{x^{2n+1}}-\frac{\sin{x}}{x}\sum_{n=0}^\infty(-1)^n\frac{(2n)!}{x^{2n}}.
\end{equation}
\end{itemize}

\bigskip

{\bf{Data accessibility}} This article does not use data.

\end{document}